\documentclass[superscriptaddress,
reprint,aps,amssymb,nofootinbib,floatfix]{revtex4-2}

\usepackage{graphicx}
\usepackage{amsmath}
\usepackage[svgnames]{xcolor} %OMG - use SVG, not DVI or ODP or PS or POOP
\usepackage[colorlinks=True, linkcolor=SlateBlue,
            citecolor=SteelBlue,urlcolor=BlueViolet]{hyperref}
\usepackage[capitalise]{cleveref}

\usepackage{aas_macros}

\newcommand{\imag}{\text{i}}
\newcommand{\out}{\text{o}}

\newcommand{\tot}{\text{t}}
\newcommand{\vect}[1]{\boldsymbol{#1}}
\newcommand{\uvect}[1]{\vect{\hat{#1}}}

\newcommand{\chieff}{\ensuremath{\chi_{\rm eff}}}

\begin{document}

\author{Hang Yu}\thanks{hangyu@caltech.edu}
\affiliation{TAPIR, Walter Burke Institute for Theoretical Physics, Mailcode 350-17 California Institute of Technology, Pasadena, CA 91125, USA}

\author{Sizheng Ma}
\affiliation{TAPIR, Walter Burke Institute for Theoretical Physics, Mailcode 350-17 California Institute of Technology, Pasadena, CA 91125, USA}

\author{Matthew Giesler}
\affiliation{TAPIR, Walter Burke Institute for Theoretical Physics, Mailcode 350-17 California Institute of Technology, Pasadena, CA 91125, USA}

\author{Yanbei Chen}
\affiliation{TAPIR, Walter Burke Institute for Theoretical Physics, Mailcode 350-17 California Institute of Technology, Pasadena, CA 91125, USA}

\title{Spin and Eccentricity Evolution in Triple Systems: from the Lidov-Kozai Interaction to the Final Merger of the Inner Binary}

\begin{abstract}
  We study the spin and eccentricity evolution of black-hole (BH) binaries that are perturbed by tertiary masses and experience the Lidov-Kozai (LK) excitation. We focus on three aspects. Firstly, we study the spin-orbit alignment of the inner binary following the approach outlined by Antonini et al.~\cite{Antonini:18} and Liu and Lai~\cite{Liu:18}, yet allowing the spins to have random initial orientations. We confirm the existence of a dynamical attractor that drives the spin-orbit angle at the end of the LK evolution to a value given by the initial angle between the spin and the outer orbital angular momentum (instead of to a specific value of the effective spin). Secondly, we follow the (inner) binary's evolution further to the merger to study the final spin-spin alignment. We generalize the effective potential theory to include orbital eccentricity, which allows us to efficiently evolve the system in the early inspiral stages. We further find that the spin-spin and spin-orbit alignments are correlated and the correlation is determined by the initial spin-orbit angle. For systems with the spin vectors initially in the orbital plane, the final spins strongly disfavor an aligned configuration and could thus lead to a greater value of the GW recoil than a uniform spin-spin alignment would predict. Lastly, we study the maximum eccentricity excitation that can be achieved during the LK process, including the effects of gravitational-wave radiation. We find that when the tertiary mass is a super-massive BH and the inner binary is massive, then even with the maximum LK excitation, the residual eccentricity is typically less than 0.1 when the binary's orbital frequency reaches $10\,{\rm Hz}$, and a decihertz detector would be necessary to follow such a system's orbital evolution. 
\end{abstract}

\maketitle

\section{Introduction}

It has been suggested that a significant amount of binary black-hole (BH) mergers detectable by Advanced LIGO (aLIGO;~\cite{LSC:15}) and Advanced Virgo (aVirgo;~\cite{Acernese:15}) may happen in galactic nuclei~\cite{Antonini:16, Liu:19, Fragione:19} or in surrounding gas disks~\cite{Bartos:17, Tagawa:19, McKernan:19}. The recent announcement by the Zwicky Transient Facility~\cite{Graham:20} further strengthens this possibility. In Ref.~\cite{Graham:20}, the authors report a plausible electromagnetic counterpart to a candidate binary BH merger in the accretion disk of an active galactic nucleus,  associating it with aLIGO/aVirgo's gravitational-wave (GW) event GW190521~\cite{GW190521a, GW190521b}.

The deep gravitational potential well in a galactic nucleus enables the possibility of finding mergers involving second-generation (or even higher generation) BHs, i.e., BHs that are themselves products of previous merger events~\cite{Gerosa:17, Rodriguez:19, Gerosa:19b}. Such a high-generation BH may be produced by frequent stellar interactions thanks to the dense stellar environment~\cite{OLeary:09}. Alternative, if there are gas disks around the SMBH, then migration traps may form and cause massive objects to accumulate and collide with each other~\cite{Bellovary:16}. A high-generation BH may be massive, potentially exceeding the upper mass gap set by pair-instability supernovae
~\cite{Woosley:07}. Moreover, such a BH likely possesses significant spin angular momentum, inherited from the residual orbital angular momentum (AM) of its progenitor binary~\cite{Barack:04, Campanelli:06, Gonzalez:07, Berti:07, GWTC1}. This is in contrast to BHs born from stellar evolution, in which case small spins are expected~\cite{Fuller:19, GWTC1}. Ref.~\cite{Graham:20} suggests that the GW190521 event may have a total mass of $\sim 100 M_\odot$ and at least one component is significantly spinning,\footnote{During the preparation of this work, the LIGO parameter estimation on the GW190521 event was not ready and therefore parameters suggested Ref.~\cite{Graham:20} were used. LIGO later reported a more massive binary with component masses of $(M_1, M_2)=(85\,M_\odot, 66\,M_\odot)$ and both components may have potentially significant spin. More importantly, there is a potentially significant spin component in the orbital plane~\cite{GW190521a, GW190521b}.  These parameters further strengthens the possibility of a dynamical origin of the system.} two characteristics consistent with BHs with dynamical origins as expected in galactic nuclei.

Meanwhile, as a super-massive BH (SMBH) typically resides in the galactic nucleus~\cite{Kormendy:13}, binaries in the nucleus might be further perturbed by the SMBH via, e.g, the Lidov-Kozai (LK) mechanism~\cite{Lidov:62, Kozai:62}. In this picture, the SMBH acts as a tertiary perturber that causes the inner binary to oscillate in its orbital inclination and eccentricity. As the pericenter separation decreases with increasing eccentricity, the GW radiation becomes increasingly more efficient. This allows the binaries to merge more quickly and on timescales shorter than, e.g., the age of the Universe or other survival timescales set by local environments.

In fact, the LK mechanism has been considered to be an important channel producing the mergers of binary BHs and belongs to the family of dynamical formation channels (see, e.g., Ref.~\cite{Mandel:18} for a review of different formation scenarios). Different authors have investigated this problem in different context, ranging from galactic nuclei (e.g.,~\cite{Antonini:12, VanLandingham:16, Petrovich:17,  Harmers:18, Liu:19, Fragione:19}), to dense stellar clusters (e.g.,~\cite{Miller:02, Wen:03}), to isolated field stars (e.g.,~\cite{Liu:17, Silsbee:17, Antonini:17, Antonini:18, Liu:18, Rodriguez:18b, Liu:19a, Liu:19b}). 

While most of the references above focus on the merger window (i.e., the parameter space of initial conditions that could lead to successful LK-induced mergers) and the event rates, a few authors~ \cite{Antonini:18, Liu:18, Rodriguez:18b} suggest another interesting aspect of the LK mechanism, namely, its effect on the evolution of the spin vectors in the inner binary. More specifically, Refs.~\cite{Antonini:18, Liu:18, Rodriguez:18b} all report a dynamical attractor that drives each component's spin into the orbital plane at the end of the LK evolution. Consequently, the effective spin parameter [the mass-weighted sum of the component spins along the direction of the orbital AM; see Eq.~(\ref{eq:chieff})] of the inner binary is attracted towards zero. 
However, Refs.~\cite{Antonini:18, Liu:18, Rodriguez:18b} assumed a special initial condition where the spin vectors are aligned with the inner orbit AM vector. This is a reasonable assumption to make for triple systems in the field, where such an alignment might be expected from stellar evolution~\cite{Kalogera:00, Corsaro:17}. It is unclear, however, whether such a condition still holds for binaries formed near an SMBH whose components are more likely to have dynamical origins. This motivates us to study, under more generic initial conditions, how the LK process affects the evolution of the inner binary's spin-orbit alignment. 
This is particularly relevant to GW190521, as significant spin may be expected~\cite{Graham:20, GW190521a, GW190521b},
and would improve our understanding of a more generic class of mergers driven by the LK mechanism.

In addition to the spin-orbit alignment, the spin-spin alignment is also of particular interest in this study. Previous studies suggest that the post-Newtonian (PN) spin evolution may play a significant role in shaping the final distribution of this angle (e.g., Refs.~\cite{Schnittman:04, Kesden:10, Berti:12, Gerosa:13, Gerosa:15, Gerosa:17b, Gerosa:18, Gerosa:19}). While this is not a leading-order post-Newtonian (PN) effect in the inspiral stage, the spin-spin alignment nonetheless affects the GW radiation during the final merger-ringdown stage, and plays a crucial role in determining the GW recoil (also known as the GW kick;~\cite{Campanelli:07, Kesden:10, Berti:12}). Properly modeling this final stage is particularly important for a system like GW190521, which is both intrinsically massive and appearing more massive in the detector frame due to the large cosmological redshift, because the signal information content captured in the LIGO band is dominated by the merger-ringdown stage~\cite{Veitch:15}. This is in contrast to the majority of previous LIGO detections, which typically appear with a detector-frame total mass of $< 100\, M_\odot$, where the signal-to-noise is dominated by the inspiral stage.
% Here we will extend the study by investigating the correlations between the spin-spin alignment with various other physical parameters, including both quantities measured at the merger, and the initial conditions provided by the LK interaction.

Consequently, in this study, we also investigate the evolution of the spin-spin alignment. Particularly, how different initial conditions such as orbital eccentricity and the initial spin-orbit alignment affect the final orientation of the spin vectors. Since in the final evolution stages, the binary effectively decouples from the tertiary perturber, the LK process simply serves as a way of providing the initial condition. Thus, our result has broader applications to other formation channels, provided one properly substitutes in the initial conditions suitable for the formation channel of interest.

The eccentricity is yet another interesting aspect that we explore in this study, as it usually bears unique signatures of a binary's formation channel~\cite{Miller:02b, Wen:03, OLeary:06, Seto:16, Nishizawa:16, Breivik:16, Nishizawa:17, Giesler:2017uyu, Rodriguez:18, Samsing:18, Romero-Shaw:19}, and it is anticipated to be detectable by future space-based GW observatories in the millihertz and decihertz bands such as LISA~\cite{Amaro-Seoane:17}, TianQin~\cite{Luo:16}, and TianGO~\cite{Kuns:19}. 
This motivates investigating the limiting eccentricity that can be excited by the LK mechanism and the observational consequences for future space-based and ground GW detectors.

The rest of the paper is organized as follows. In Sec.~\ref{sec:formalism} we outline the basic formalism of the problem. In the remainder of Sec.~\ref{sec:LK_evol}, we apply the formalism to studying the spin evolution during the LK evolution. Our approach is similar to Ref.~\cite{Liu:18} but with a key extension in the form of sampling the initial spins isotropically. In Sec.~\ref{sec:bin_evol} we further evolve the systems after the LK excitation, which specify the binary initial conditions, and follow the binary's evolution onward to the final merger. This is done by first generalizing the precession-averaged evolution for circular orbits proposed by Ref.~\cite{Kesden:15} to allow for orbital eccentricity in Sec.~\ref{sec:eff_spin_potential}. We study the final spin distributions in Sec.~\ref{sec:fin_spin_dist} and its relation to GW kicks in Sec.~\ref{sec:kick}. We then consider the limiting eccentricity excitation by the LK mechanism in Sec.~\ref{sec:max_ecc}. Lastly, we summarize our results in Sec.~\ref{sec:conclusion}. Throughout this paper we use geometrical units with $G=c=1$.

\section{Evolution of the spin-orbit alignment during the Lidov-Kozai oscillation}
\label{sec:LK_evol}
In this Section we study the dynamics of an inner binary (consisting of masses $M_1$ and $M_2$ with $M_1\geq M_2$ in an orbit with semi-major axis $a_\imag$) perturbed by a tertiary mass $M_3$ that is in an outer orbit with semi-major axis $a_\out$ via the Lidov-Kozai (LK) oscillation. 

Our focus is to examine how the spin vectors of the inner binary evolve with respect to the inner orbital AM. Specifically, we want to examine if the attraction towards $\chieff=0$ reported in Refs.~\cite{Antonini:18, Liu:18, Rodriguez:18b} still holds if we randomize the initial spin orientation. According to Ref.~\cite{Liu:18}, the attraction is most significant for triple systems that experience multiple ``clean'' LK cycles. In other words, the interaction is dominated by the quadrupole interaction potential. The octupole effects are naturally small when the tertiary mass is an SMBH, because the condition $a_\out \gg a_\imag$ is required in order for the triple to be dynamically stable~\cite{Mardling:01, Liu:19}. Consequently, we truncate the LK interaction at the quadrupole order in this work. %Consequently, we will also restrict our study to the ``clean'' problem, and defer various corrections (see, e.g., Refs.~\cite{Petrovich:17, Liu:19a, Liu:19}) to a future study. 

Given the complications of the environment near an SMBH, we do not attempt to make any predictions on the LK-induced event rates in this study. %\matt{this single sentence paragraph seems a little out of place...}\hang{I agree, yet I feel we would want to remind the readers at the beginning that we only want to study the spin evolution here. Predicting the event rates would require making many assumptions and we are not so ambitious to do that here. }

In Sec.~\ref{sec:formalism} we review the basic formalism of the standard LK problem and in Sec.~\ref{sec:analytical_approx} we provide some analytical solutions under the simplifications that the interaction is truncated at the quadrupole order and the GW decay is neglected. Additional corrections due to an SMBH are discussed in Sec.~\ref{sec:smbh_effects}. We present our numerical simulations in Sec.~\ref{sec:num_LK}. Our study in this Section closely follows Ref.~\cite{Liu:18} (see also Refs.~\cite{Antonini:18, Rodriguez:18b}), with a key modification, namely, that we allow the initial orientations of the spin vectors to be drawn isotropically, rather than fixing them along the direction of the AM of the inner orbit. As evident in Sec.~\ref{sec:num_LK}, this has a significant consequence on the final distribution of $\chieff$. 

\subsection{Formalism}
\label{sec:formalism}
 %To do so, we follow the formalism outlined in Ref.~\cite{Liu:18}.
We start our discussion here by presenting the key equations of the ``standard'' LK interactions. Corrections due to an SMBH are discussed in Sec.~\ref{sec:smbh_effects}.

The secular evolution of the inner orbit can be specified by 4 vectors, $\vect{L_{\rm i}}$, $\vect{e_{\rm i}}$, $\vect{S}_{1}$, and $\vect{S}_{2}$, corresponding to the orbital AM of the inner orbit, the eccentricity vector\footnote{It has a direction pointing from the apocenter to the pericenter and its amplitude is equal to the eccentricity. This is equivalent to the Laplace-Runge-Lenz vector divided by $M\mu^2$. }, and the spin vectors associated with masses $M_1$ and $M_2$, respectively. These vectors are further specified by a set of ordinary differential equations as 
\begin{align}
    &\frac{d \vect{L}_{\rm i}}{dt} = 
    \frac{d \vect{L}_{\rm i}}{dt}\Big{|}_{\rm LK} 
    + \frac{d \vect{L}_{\rm i}}{dt}\Big{|}_{\rm GW} 
    + \frac{d \vect{L}_{\rm i}}{dt}\Big{|}_{\rm dS} 
    + \frac{d \vect{L}_{\rm i}}{dt}\Big{|}_{\rm LT}, \\
    &\frac{d \vect{e}_{\rm i}}{dt} = 
    \frac{d \vect{e}_{\rm i}}{dt}\Big{|}_{\rm LK}
    + \frac{d \vect{e}_{\rm i}}{dt}\Big{|}_{\rm GR}
    + \frac{d \vect{e}_{\rm i}}{dt}\Big{|}_{\rm GW} 
    + \frac{d \vect{e}_{\rm i}}{dt}\Big{|}_{\rm dS} 
    + \frac{d \vect{e}_{\rm i}}{dt}\Big{|}_{\rm LT},\\
    &\frac{d\vect{S}_{1,2}}{dt} = 
    \frac{d\vect{S}_{1,2}}{dt}\Big{|}_{\rm dS} 
    +\frac{d\vect{S}_{1,2}}{dt}\Big{|}_{\rm LT},
\end{align}
where in the subscripts we have used ``LK'', ``GR'', ``GW'', ``dS'', and ``LT'' to respectively stand for the Lidov-Kozai (LK) interaction, the (conservative) general-relativistic apsidal precession, the (dissipative) GW radiation, the de Sitter, and the Lense-Thirring precessions. When coupled to the outer orbit via the LK mechanism, the above set of equations gives the complete description of the system's dynamics. Next, we examine each of these terms more closely.

We start with the LK interaction, which together with the Keplerian motion of the inner and outer orbit (i.e., all the Newtonian parts), can be jointly described by a Hamiltonian (see, e.g., Ref.~\cite{Harrington:68}; see also Ref.~\cite{Naoz:16} for a more recent review) of the form 
\begin{equation}
    \mathcal{H} = \frac{1}{2}\mu_\imag |\vect{\dot{r}}_\imag|^2 + \frac{1}{2}\mu_\out |\vect{\dot{r}}_\out|^2 - \frac{M_1 M_2}{r_\imag} - \frac{M_\tot M_3}{r_\out} +\Phi_{\rm LK}.
\end{equation}
Here, $\vect{r}_\imag=r_\imag \uvect{r}_\imag$ and $\vect{r}_\out=r_\out\uvect{r}_\out$ are the inner and outer orbital separations, respectively, while the hats denote unit vectors. %and we have used $\uvect{q}$ to represent the unit vector along the direction of an arbitrary vector $\vect{q}$.
We have also defined $\mu_\imag = M_1M_2/M_\tot$ and $\mu_\out=M_\tot M_3/(M_\tot+M_3)$, the reduced masses of the inner and outer orbits, respectively, where $M_\tot = M_1 + M_2$ is the total mass of the inner orbit. For conciseness, we sometimes drop the subscript ``$\imag$'' for quantities describing the inner orbit. To avoid any confusion, quantities related to the outer orbit retain the subscript ``$\out$'' throughout this paper. 

The quantity $\Phi_{\rm LK}$ describes the tidal potential of the tertiary mass expanded around the center of mass of the inner orbit and it is given by
\begin{align}
    \Phi_{\rm LK} =& -M_1M_2M_3\sum_{l=2} \frac{M_1^{l-1} + (-1)^{l}M_2^{l-1}}{M_\tot^l}\nonumber \\
    &\times\frac{r_\imag^l}{r_\out^{l+1}} P_l\left(\uvect{r}_\imag\cdot\uvect{r}_\out\right),\label{eq:phi_LK_full}
\end{align}
where in the second line, $P_l$ is the Legendre polynomial of degree $l$.
Note that the octupole term is significantly suppressed when $M_3$ is an SMBH as dynamical stability~\cite{Mardling:01, Liu:19} requires $r_\out/r_\imag \ll 1$ (the system we focus on in Sec.~\ref{sec:num_LK} has $a_\out/a_\imag \simeq 10^{-4}$, about  $100-10^3$ times smaller than what is allowed for triples in the field with $M_{3}\sim M_{1}$). More importantly, as our goal is to study the spin attractor under ``clean'' LK interactions~\cite{Liu:18}, we focus solely on the leading order quadrupole ($l=2$) term. 
%Note that the octupole ($l=3$) term $\propto (M_1-M_2)/M_\tot$, and thus it is suppressed if we focus on nearly equal-mass systems with mass ratio $q\equiv M_2/M_1 \simeq 1$. As in this paper we will study systems with comparable masses, and more importantly, our goal is the study the spin attractor under ``clean'' LK interactions~\cite{Liu:18}, we will thus focus on the leading order quadrupole ($l=2$) term. 

To efficiently evolve the system, one typically uses the orbital-averaged (i.e., the  secular) version of the interaction potential $\Phi_{\rm LK}$. Specifically, one may average over both the inner and outer orbits (i.e., the double-averaged, or DA, approximation), which leads to 
\begin{align}
    &\langle\langle\Phi_{\rm LK}\rangle\rangle\Big{|}_{l=2} = \frac{\mu M_3 a^2}{8a_\out^3(1-e_\out)^{3/2}} \nonumber \\
    & \times \left[1-6e^2 - 3(1-e^2)\left(\uvect{L}\cdot\uvect{L}_\out\right)^2 + 15e^2\left(\uvect{e}\cdot\uvect{L}_\out\right)^2\right],
\end{align}
where $\vect{L}_\out$ and $\vect{e}_\out$ are the orbital angular momentum and eccentricity vectors
%\matt{should these vectors be formatted as such? Like in the equation above?} 
of the outer orbit, and they can be jointly evolved with the inner orbit's quantities to solve for the dynamics of the hierarchical triple system. Due to the LK interaction, the inner eccentricity and mutual orbital inclination oscillates at a characteristic rate $\Omega_{\rm LK}$, given by
\begin{equation}
    \Omega_{\rm LK} = \frac{M_3}{M_{\rm t}}\left(\frac{a}{a_\out\sqrt{1-e_\out^2}}\right)^3\sqrt{\frac{M_{\rm t}}{a^3}}.
    \label{eq:omega_LK}
\end{equation}
When the inner orbit's eccentricity $e$ is near its maximum with $e\simeq 1$, the eccentricity varies on a timescale $\tau_{\rm LK}$ given by~\cite{Anderson:16}\footnote{For future convenience, we do not define $\tau_{\rm LK}$ as $1/\Omega_{\rm LK}$. Instead, we define $\tau_{\rm LK}= \sqrt{1-e^2}/\Omega_{\rm LK}$. }
\begin{equation}
    \tau_{\rm LK} = \frac{M_\tot}{M_3}\left(\frac{a_\out \sqrt{1-e_\out^2}}{a}\right)^3 \sqrt{\frac{a^3(1-e^2)}{M_\tot}}.
    \label{eq:tau_lk}
\end{equation}
If this timescale is longer than the period of the outer orbit, we are safely in the DA regime. Otherwise, one should only average over the inner orbit (the single-averaged, or SA, approximation), leading to 
\begin{align}
    &\langle\Phi_{\rm LK}\rangle\Big{|}_{l=2} = \frac{\mu M_3 a^2}{4r_\out^3} \nonumber \\
    &\times\left[-1 + 6e^2 + 3(1-e^2)\left(\uvect{L}\cdot\uvect{r}_\out\right)^2 -15\left(\uvect{e}\cdot\uvect{r}_\out\right)^2\right]. 
\end{align}

Once the Hamiltonian is specified, one can easily obtain the equations of motions for both the inner and outer orbits. The explicit forms are provided in Appx.~\ref{sec:explicit_LK_eqs} (See also Ref.~\cite{Liu:15} for the DA case and Ref.~\cite{Liu:18} for the SA case). 

As the LK oscillation excites a large eccentricity in the inner orbit, it greatly reduces the instantaneous GW decay timescale $\tau_{\rm gw}$, defined by 
\begin{equation}
    \tau_{\rm gw} \equiv \frac{a}{|\dot{a}|} 
    = \frac{5}{64}\frac{a^4}{\mu M_{\rm t}^2}\frac{(1-e^2)^{7/2}}{\left(1+\frac{73}{24}e^2 + \frac{37}{96}e^4\right)}. 
    \label{eq:tau_gw}
\end{equation}
Hence, an initially widely separated system may be able to merge in a reasonable amount of time due to GW radiation when $(1-e^2)\ll 1$. As pointed out by Ref.~\cite{Liu:18} (see also Ref.~\cite{Liu:17}), the total LK-induced merger time can be approximated by 
\begin{equation}
    \tau_{\rm m} = \tau_{\rm gw}|_{e=0}\left(1-e_{\rm max}^2\right)^{3},\label{eq:tau_m}
\end{equation}
where $e_{\rm max}$ is the maximum eccentricity reached during the LK cycle [which is further explored in Eq.~(\ref{eq:j_min}) and Sec.~\ref{sec:max_ecc}]. 

%\sma{I would move Eq. 1-3 here, so that you first discuss the orbital part, then the precession part. }\hang{Eqs. 1-3 are the full dynamics. the rest are describing how to compute each term, from LK to GW to GR. }
To incorporate the GW decay, we have
\begin{align}
    &\frac{d\vect{L}}{dt}\Big{|}_{\rm GW} = -\frac{32}{5}\frac{\mu^2M_{\rm t}^{5/2}}{a^{7/2}}\frac{\left(1+\frac{7}{8}e^2\right)}{\left(1-e^2\right)^2}\uvect{L}, \label{eq:dLdt_gw}\\
    &\frac{d\vect{e}}{dt}\Big{|}_{\rm GW} = 
    -\frac{304}{15}\frac{\mu M_{\rm t}^2}{a^4}\frac{\left(1+\frac{121}{304}e^2\right)}{\left(1-e^2\right)^{5/2}}\vect{e}. \label{eq:dedt_gw}
\end{align}
Note that the above equations preserve the relation that
\begin{equation}
    L = \mu \sqrt{M_{\rm t}a\left(1-e^2\right)}. 
    \label{eq:L_vs_a_e}
\end{equation}

In addition to the dissipative decay, GR also induces a conservative apsidal precession as 
\begin{equation}
    \frac{d \vect{e}}{dt}\Big{|}_{\rm GR} = \vect{\Omega}_{\rm GR} \times \vect{e},
\end{equation}
where 
\begin{equation}
    \vect{\Omega}_{\rm GR} = \frac{3M_{\rm t}}{a(1-e^2)}\Omega_{\rm orb}\uvect{L}, 
\end{equation}
with $\Omega_{\rm orb}=\sqrt{M_t/a^3}$. 

In order to study the evolution of spin orientation, we further incorporate the de Sitter (1.5 PN) and Lense-Thirring (2 PN) precessions according to Ref.~ \cite{Barker:75}, as well as the quadrupole-monopole interaction according to Ref.~\cite{Racine:08}
\begin{align}
    & \frac{d\vect{S}_1}{dt} = \left(\vect{\Omega}_{\rm dS}^{(S_1)} + \vect{\Omega}_{\rm LT}^{(S_1)} + \vect{\Omega}_{\rm QM}^{(S_1)}\right)\times \vect{S}_{1},
    \label{eq:dS1_v_dt}
\end{align}
and similarly for $\vect{S}_2$. These also induce back-reactions on the orbit (denoted with a subscript ``br'') as
\begin{align}
    \frac{d\vect{L}}{dt}\Big{|}_{\rm dS+LT+QM} &= \left(\vect{\Omega}_{\rm dS, br}^{(S_1)}
    +\vect{\Omega}_{\rm dS, br}^{(S_2)}
    +\vect{\Omega}_{\rm LT, br}\right.\nonumber \\
    &\left. + \vect{\Omega}_{\rm QM, br}^{(S_1)}+ \vect{\Omega}_{\rm QM, br}^{(S_2)}\right)\times\vect{L} \label{eq:dL_v_dt}\\
    \frac{d\vect{e}}{dt}\Big{|}_{\rm dS+LT+QM} &= \left(\vect{\Omega}_{\rm dS, br}^{(S_1)}
    +\vect{\Omega}_{\rm dS, br}^{(S_2)}
    +\vect{\Omega}_{\rm LT, br} \right.\nonumber \\
    &\left.+\vect{\Omega}_{\rm QM, br}^{(S_1)}+ \vect{\Omega}_{\rm QM, br}^{(S_2)}\right)\times\vect{e}.
\end{align}
The different $\vect{\Omega}$'s are given by
\begin{align}
    & \vect{\Omega}_{\rm dS}^{(S_1)} = \frac{3\left(M_2 + \mu/3\right)}{2a\left(1-e^2\right)}\Omega_{\rm orb} \uvect{L}=\frac{(4+3M_2/M_1)L}{2a^3(1-e^2)^{3/2}}\uvect{L}, \label{eq:omega_dS}\\
    & \vect{\Omega}_{\rm LT}^{(S_1)} = \frac{S_2}{2a^3\left(1-e^2\right)^{3/2}}\left[\uvect{S}_2 - 3\left(\uvect{L}\cdot\uvect{S}_2\right)\uvect{L}\right],\label{eq:omega_LT}\\
    & \vect{\Omega}_{\rm QM}^{(S_1)} =
    \frac{S_1}{2a^3\left(1-e^2\right)^{3/2}}\frac{M_2}{M_1}\left[\uvect{S}_1 - 3\left(\uvect{L}\cdot\uvect{S}_1\right)\uvect{L}\right], \label{eq:omega_QM}\\
    & \vect{\Omega}_{\rm dS, br}^{(S_1)} = \frac{S_1\left(4+3{M_2}/{M_1}\right)}{2 a^3\left(1-e^2\right)^{3/2}}\left[\uvect{S}_1 - 3\left(\uvect{L}\cdot\uvect{S}_1\right)\uvect{L}\right], \label{eq:omega_dSbr}\\
    & \vect{\Omega}_{\rm LT, br} = -\frac{3S_1S_2}{2 a^3\left(1-e^2\right)^{3/2}L} \nonumber \\
    &\quad\quad \times\left\{\left(\uvect{L}\cdot\uvect{S}_1\right)\uvect{S}_2 +\left(\uvect{L}\cdot\uvect{S}_2\right)\uvect{S}_1 \nonumber \right. \\
    &\quad \quad\quad  +\left.\left[\uvect{S}_1\cdot\uvect{S}_2 - 5\left(\uvect{L}\cdot\uvect{S}_1\right)\left(\uvect{L}\cdot\uvect{S}_2\right)\right]\uvect{L}\right\}, \\
    &\vect{\Omega}_{\rm QM, br}^{(S_1)} = -\frac{3S_1^2}{4a^3(1-e^2)^{3/2}L}\frac{M_2}{M_1}\nonumber \\
    &\quad \quad \times  \left\{2\left(\uvect{L}\cdot\uvect{S_1}\right)\uvect{S_1}  +\left[1-5\left(\uvect{L}\cdot\uvect{S}_1\right)^2\right]\uvect{L}\right\}
\end{align}
Quantities with a superscript of $(S_2)$ can be obtained from those with $(S_1)$ by switching subscripts $(1\leftrightarrow 2)$. 

\subsection{Analytical approximations to conservative systems}
\label{sec:analytical_approx}
The above set of differential equations describe the  dynamics of the triple system and can be solved numerically. Nonetheless, it is also instructive to consider the analytical solutions of the system under certain approximations. Specifically, if one ignores the GW decay and truncates the interaction potential at the quadrupole order [Eq.~(\ref{eq:phi_LK_full})], then the maximum eccentricity of the inner orbit $e_{\rm max}$ can be obtained as a function of the initial (which we define as the moment when the system is nearly circular) inclination $I^{(0)}$ (i.e., the angle between $\vect{L}_\imag$ and $\vect{L}_\out$)  as \cite{Miller:02, Liu:15, Anderson:17}
\begin{align}
    &\frac{3(j_{\rm min}+1)}{8j_{\rm min}}\left[
    \epsilon_{\rm br}^2j_{\rm min}^4
    -\left(3+4\epsilon_{\rm br}\cos I^{(0)} + \frac{9}{4}\epsilon_{\rm br}^2\right)j_{\rm min}^2 \nonumber \right. \\
  & \quad\quad \left. 
  + 5\left(\cos I^{(0)} + \frac{\epsilon_{\rm br}}{2}\right) \right] + \epsilon_{\rm GR} = 0,
  \label{eq:j_min}
\end{align}
where $j_{\rm min}\equiv\sqrt{1-e_{\rm max}^2}$ and 
\begin{align}
    &\epsilon_{\rm GR} = 
    3\left(\frac{M_{\rm t}}{a}\right)
    \left(\frac{M_t}{M_3}\right)
    \left(\frac{a_\out\sqrt{1-e^2_\out}}{a}\right)^3, \\
    &\epsilon_{\rm br} =\frac{L(e=0)}{L_\out}=
    \left.\frac{\mu}{\mu_\out}\right.
    \left[\frac{M_{\rm t} }{(M_{\rm t}+M_3)}\frac{a}{a_\out\left(1-e_\out^2\right)}\right]^{1/2}.
\end{align}

The limiting eccentricity $\tilde{e}_{\rm lim}=\max \left\{e_{\rm max}\left[I^{(0)}\right]\right\}$ is obtained when 
\begin{equation}
    \cos I_{\rm lim}^{(0)} = \frac{\epsilon_{br}}{2}\left(\frac{4}{5}\tilde{j}_{\rm lim}^2-1\right),
    \label{eq:inclination_4_max_e}
\end{equation}
with $\tilde{j}_{\rm lim}\equiv\sqrt{1-\tilde{e}^2_{\rm lim}}$, by solving 
\begin{equation}
    \frac{3}{8}\tilde{j}_{\rm lim}\left(\tilde{j}_{\rm lim} +1 \right)\left[-3+\frac{\epsilon_{\rm br}^2}{4}\left(\frac{4}{5}\tilde{j}_{\rm lim}^2-1\right)\right] + \epsilon_{\rm GR} = 0.
\end{equation}
Under the limit that the back-reaction factor $\epsilon_{\rm br}\ll 1$ and $1-\tilde{e}\ll 1$, we can simplify the equation as 
\begin{align}
    1-\tilde{e}_{\rm lim}\simeq&  1.9\times10^{-5}\left(\frac{M_{\rm t}}{150\,M_\odot}\right)^4 \left(\frac{a}{3\,{\rm AU}}\right)^{-8}\nonumber \\
    \times&  \left(\frac{M_{3}}{10^9\,M_\odot}\right)^{-2} 
    \left(\frac{a_\out \sqrt{1-e_\out^2}}{0.06\,{\rm pc}}\right)^6, 
    \label{eq:e_lim_cons}
\end{align}
and the limiting merger timescale associated with $\tilde{e}_{\rm lim}$ is given by [Eq.~(\ref{eq:tau_m})]
\begin{align}
    \tilde{\tau}_{\rm m,lim} \simeq& 2.3\times10^{1}\,{\rm yr} \nonumber \\
   \times & \left(\frac{M_{\rm t}}{100\,M_\odot}\right)^{10}\left(\frac{\mu }{25\,M_\odot}\right)^{-1}\left(\frac{a}{3\,{\rm AU}}\right)^{-20}\nonumber \\
    \times& \left(\frac{M_{3}}{10^9\,M_\odot}\right)^{-6} 
    \left(\frac{a_\out \sqrt{1-e_\out^2}}{0.06\,{\rm pc}}\right)^{18}. 
    \label{eq:tau_m_lim_cons}
\end{align}

\begin{figure}[tb]
  \centering
  \includegraphics[width=\columnwidth]{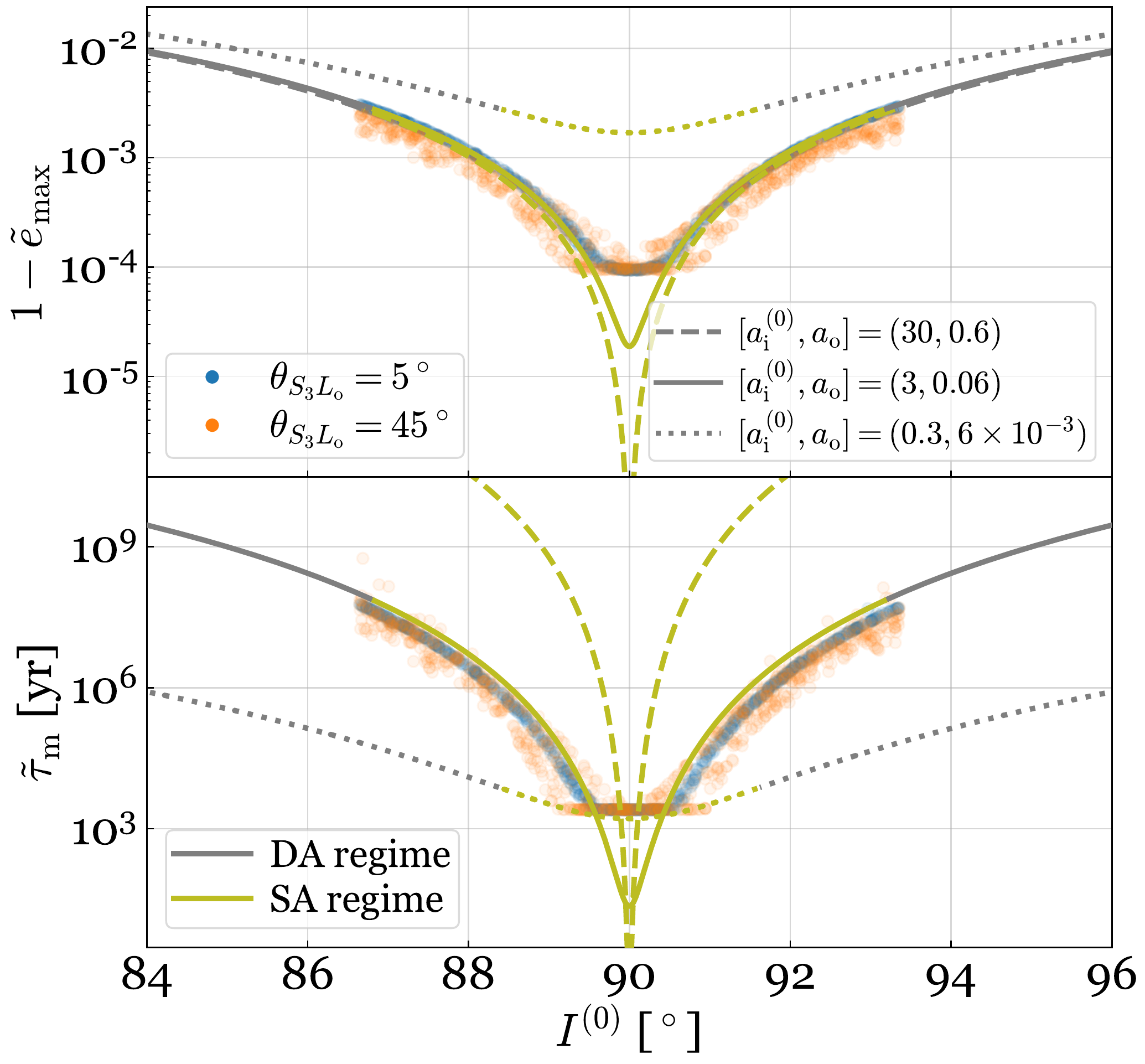}
\caption{Top panel: the maximum eccentricity that can be excited during the LK oscillation as a function of the initial inclination $I^{(0)}$ under the conservative (i.e., no GW radiation), quadrupole approximation [Eq.~(\ref{eq:j_min})]. Bottom panel: the corresponding merger timescale [Eq.~(\ref{eq:tau_m})]. Here we have focused on a system with $(M_1, M_2, M_3)=(55, 45, 10^9)M_\odot$ and three different sets of  $\left[a^{(0)}_\imag, a_\out\right]$ (indicated by different line styles). We have fixed the eccentricity of the outer orbit to be $e_\out=0$ for the cases. Also shown in the dots are the results obtained from numerical simulations including SMBH corrections (blue for $\theta_{S_3L_\out}=5^\circ$ and orange for $\theta_{S_3L_\out}=45^\circ$, where $\theta_{S_3L_\out}$ is the inclination of $\vect{L}_\out$ with respect to $\vect{S}_3$). The pile-up of eccentricity at $1-e_{\rm max}\simeq 10^{-4}$ and the merger time at $\tau_{\rm m}\simeq 3 \times10^{3}\,{\rm yr}$ are explained in Sec.~\ref{sec:max_ecc}.}
\label{fig:max_ecc_Tm_vs_I0}
\end{figure}

%\sma{What's the relation between Eq. 10 and 29?}
%\hang{For each $I^{(0)}$, there is an $\tilde{e}_{\rm max}$ and corresponding merger timescale given by Eq. 10. Eq. 29 is the limiting merger timescale among all the $I^{(0)}$. }
We show a few representative curves of the maximum eccentricity under the conservative approximation, and the corresponding merger timescale calculated according to Eq.~(\ref{eq:tau_m}) in Fig.~\ref{fig:max_ecc_Tm_vs_I0}. Here we assume the triple system has masses of $(M_1, M_2, M_3)=(55, 45, 10^9) M_{\odot}$. We denote the initial semi-major axes of the inner and outer orbits as $a_\imag^{(0)}$ and $a_\out$, respectively, and use three different line styles to represent three sets of separations (we use dashed, solid, and dotted lines for $[a_\imag^{(0)}, a_\out]=\{30\,{\rm AU}, 0.6\,{\rm pc}\},\ \{3\,{\rm AU}, 0.06\,{\rm pc}\},\ \{0.3\,{\rm AU}, 6\times10^{-3}\,{\rm pc}\}$, respectively). Lastly, we use the color grey (olive) to represent systems that are in the DA (SA) regime. Note that the maximum eccentricity varies with respect to $a_\imag^{(0)}$ even if we keep the ratio $a_\imag^{(0)}/a_\out$ a constant. 

Note that the derivation so far is for a \emph{conservative} system only, and we use a tilde symbol to denote the associated quantities. We revisit the limiting eccentricity in Sec.~\ref{sec:max_ecc} to take into account the effect of GW radiation. 

% \begin{figure}[tb]
%   \centering
%   \includegraphics[width=\columnwidth]{merg_win.pdf}
% \caption{Merger window ($\tau_{\rm m}<10\,{\rm Gyr}$) for systems with different $a_\imag^{(0)}$ and $a_\out$. }
% \label{fig:merg_win}
% \end{figure}

\subsection{Effects associated with an SMBH}
\label{sec:smbh_effects}
In addition to the ``standard'' LK equations presented in Sec.~\ref{sec:formalism}, there are additional corrections that may be important when the tertiary perturber is an SMBH~\cite{Liu:19}. In this section, we discuss these effects.

One of the most significant  effects associated with an SMBH is that $\vect{L}_\out$ and $\vect{e}_\out$ may experience a 1.5-PN precession around $\vect{S}_3$ (the spin vector of $M_3$) as\footnote{This is in analog to how $\vect{L}_\imag$ precesses around $\vect{S}_{1}$ (and $\vect{S}_2$). See Eq.~(\ref{eq:omega_dSbr}). Note that whereas $L_\imag/S_1\gg 1$, we have $L_\out/S_3\simeq 5\times 10^{-4}$ if $S_3\simeq M_3^2$, and consequently, the precession of $\vect{S}_3$ around $\vect{L}_\out$ [analog of Eq.~(\ref{eq:omega_dS})] can be safely ignored.}  
\begin{align}
    \frac{d\vect{L}_\out}{dt}\Big{|}_{S_3L_\out} &= \Omega_{S_3L_\out} \uvect{S}_{3}\times \vect{L}_\out, \label{eq:dLo_dt_S3}\\
    \frac{d\vect{e}_\out}{dt}\Big{|}_{S_3L_\out} &= \Omega_{S_3L_\out}\left[\uvect{S}_3-3\left(\uvect{L}_\out \cdot \uvect{S}_3\right)\uvect{L}_\out\right]\times \vect{e}_\out,
    % \frac{d\vect{e}_\out}{dt}\Big{|}_{S_3L_\out} &= \Omega_{S_3L_\out} \uvect{S}_{3}\times \vect{e}_\out\nonumber \\
    % &-3\Omega_{S_3L_\out}\left(\uvect{L}_\out\cdot\uvect{S}_3\right) \uvect{L}_{\out}\times \vect{e}_\out,
\end{align}
where the precession rate is given by
\begin{align}
    \Omega_{S_3L_\out} =& \frac{S_3\left(4+3M_{\rm t}/M_{\rm 3}\right)}{2 a_\out^3 (1-e_\out^2)^{3/2}},\nonumber \\
    &\simeq 3.7\times10^{-3}\Omega_{\rm LK}\left(\frac{S_3}{M_3^2}\right)\left(\frac{M_3}{10^9\,M_\odot}\right)^{-1}\nonumber \\
    \times&\left(\frac{M_{\rm t}}{100\,M_\odot}\right)^{1/2}\left(\frac{a}{3\,{\rm AU}}\right)^{-3/2}. 
\end{align}
Note that in the second line we measure $\Omega_{S_3L_\out}$ in terms of LK precession rate, $\Omega_{\rm LK}$ [Eq.~(\ref{eq:omega_LK})], to compare the relative importance of the two effects. As we focus on inner binaries that are less compact than those studied in Ref.~\cite{Liu:19}, this effect is less significant in our case.

Similarly, $\vect{S}_3$ also causes $\vect{L}_1$, $\vect{S}_1$, and $\vect{S}_2$ to precess around it [in analog to Eq.~(\ref{eq:omega_LT})] as
\begin{align}
    \frac{d \vect{L}_\imag}{dt}\Big{|}_{S_3L_\imag} = \Omega_{S_3L_\imag} \left[\uvect{S}_3 - 3\left(\uvect{L}_\out \cdot \uvect{S}_3\right)\uvect{L}_\out\right]\times \vect{L}_\imag. \label{eq:dLi_dt_S3}
\end{align}
The equations for $\vect{S}_1$ and $\vect{S}_2$ can be easily obtained by replacing $\vect{L}$ by $\vect{S}_{1(2)}$. The three vectors precess at the same rate,
\begin{equation}
    \Omega_{S_3L_\imag} = \frac{S_3}{2a_\out^3\left(1-e_\out^2\right)^{3/2}} \simeq \frac{1}{4}\Omega_{S_3L_\out}.
\end{equation}
Therefore, this effect does not directly alter the angle between $\vect{S}_1$ and $\vect{L}_\imag$. 

Nevertheless, the combination of the above two effects introduces extra variations on the directions of $\vect{L}_\out$ and $\vect{L}_\imag$ relative to each other, which enables a greater eccentricity excitation at a given initial inclination $I^{(0)}$ and typically broadens the LK merger window. Similar effects can also be generated by a non-spherical mass distribution of the ambient star cluster~\cite{Petrovich:17, Harmers:18}, or in the context of field stars, by a quadruple system~\cite{Harmers:17, Liu:19a}.

We demonstrate the significance of this effect numerically in Fig.~\ref{fig:max_ecc_Tm_vs_I0} with the dot markers. When the angle between $\vect{S}_3$ and $\vect{L}_\out$, $\theta_{S_3 L_\out}$, is small (blue dots with $\theta_{S_3 L_\out}=5^\circ$; the azimuthal angle between the two vectors is set randomly), the eccentricity and merger time matches well the analytical approximation [Eq.~(\ref{eq:j_min})].\footnote{The numerically found merger times are slightly shorter than Eq.~(\ref{eq:tau_m}) as Eq.~(\ref{eq:tau_m}) is only a semi-analytical approximation that captures the key scalings. Also note that the eccentricity piles up at $1-e_{\rm max}\simeq 10^{-4}$ and does not reach the limiting values computed in Eq.~(\ref{eq:e_lim_cons}), similarly for the merger time. This is explained in Sec.~\ref{sec:max_ecc} when we take into account the GW radiation. } Indeed, if $L_\out$ is parallel to $\vect{S}_3$, Eq.~(\ref{eq:dLo_dt_S3}) vanishes while Eq.~(\ref{eq:dLi_dt_S3}) reduces to an extra precession of $L_\imag$ around $L_\out$ without providing additional changes in the nutation. On the other hand, when the misalignment is significant (orange dots with $\theta_{S_3L_\out}=45^\circ$), we see more scattering of the numerical results. A greater eccentricity allows a binary to merge in a smaller number of LK cycles. It is thus expected to degrade the dynamical attractor, which we examine in more detail in Sec.~\ref{sec:num_LK}. 

Additionally, both $S_1$ (and $S_2$) and $\vect{L}_\imag$ experience de-Sitter (or a de-Sitter-like) precession around $\vect{L}_\out$ [in analog to Eq.~(\ref{eq:omega_dS})]. 
\begin{align}
    &\frac{d\vect{S}_1}{dt}\Big{|}_{L_\out S_1} = \Omega_{L_\out S_1}\uvect{L}_\out \times\vect{S}_1, \\
    &\frac{d\vect{L}_\imag}{dt}\Big{|}_{L_\out L_\imag} = \Omega_{L_\out L_\imag} \uvect{L}_\out \times \vect{L}_\imag,
\end{align}
where the precession rates are 
\begin{align}
    \Omega_{L_\out S_1} =& \Omega_{L_\out L_\imag} = \frac{3\left(M_3 + \mu_\out/3\right)}{2a_\out (1-e_\out^2)}\sqrt{\frac{M_3}{a_\out^3}},\nonumber \\
    =&0.12\Omega_{\rm LK}
    \left(\frac{M_3}{10^9\,M_\odot}\right)^{1/2}
    \left[\frac{a_\out(1-e_\out^2)}{0.06\,{\rm pc}}\right]^{1/2} \nonumber \\
    \times&\left(\frac{M_{\rm t}}{150\,M_\odot}\right)^{1/2}
    \left(\frac{a}{3\,{\rm AU}}\right)^{-3/2}. 
\end{align}
Note that this effect does not directly affect the angle between $\vect{S}_1$ and $\vect{L}_\imag$, which is the focus of our study here. Thus, despite that $\Omega_{L_\out L_i} > \Omega_{S_3 L_\out}$, it is subdominant compared to the extra precessions around $\vect{S}_3$.

\subsection{Numerical simulations}
\label{sec:num_LK}
Having outlined the set of equations we evolve and their approximate, analytical solutions, we now examine the full numerical evolution of a population of triple systems undergoing the LK excitation. Here we directly integrate the differential equations outlined in Sec.~\ref{sec:formalism} and Appx.~\ref{sec:explicit_LK_eqs} using an explicit Runge-Kutta method of order 5(4)~\cite{Dormand:80}. We developed our own code in \texttt{PYTHON} using standard \texttt{NumPy}~\cite{Harris:20} and \texttt{SciPy}~\cite{Virtanen:20} packages and optimized using \texttt{NUMBA}~\cite{Lam:15}.\footnote{The code is available from the corresponding author on reasonable requests.}

Motivated by Ref.~\cite{Graham:20}, we consider a relatively massive inner binary with masses $(M_1, M_2)=(55\,M_\odot, 45\,M_\odot)$ and initial separation $a_{\rm i}^{(0)}=3\,{\rm AU}$. The tertiary perturber is assumed to be an SMBH of mass $M_3=10^9\,M_\odot$ with separation $a_{\rm o} = 0.06\,{\rm pc}$. The outer orbit is further assumed to be circular. %The spin magnitude of the two BHs in the inner binaries are fixed to $\chi_1 = \chi_2 = 0.7$, where 
Additionally, we assume the two BHs of the inner binaries each have significant spins, i.e. $\chi_1 = \chi_2 = 0.7$, where 
\begin{equation}
    \chi_{1,2} \equiv \frac{S_{1,2}}{M_{1,2}^2}.
\end{equation}
When Lense-Thirring precessions around $\vect{S}_3$ are included (Sec.~\ref{sec:smbh_effects}), we fix $S_3=M_3^2$ or $\chi_3=1$ to maximize its potential consequences. 
We remind the reader that we are focused on studying the spin distribution under the LK interaction, similar to the study of Refs.~\cite{Antonini:18, Liu:18, Rodriguez:18b}, but with a key difference in that we allow the initial direction of the spin vectors to be isotropic and random (independent of the inner orbital plane's orientation), as one may expect if the binary has a dynamical origin as suggested by Refs.~\cite{Graham:20, GW190521a, GW190521b}. We do not attempt to make any predictions on the event rate in this study. 
%Other effects, such as the perturbation due to non-spherical galactic nuclei~\cite{Petrovich:17} and the GR effects associated with $M_3$~\cite{Liu:19} are deferred to future studies. 

To get a population, we uniformly sample the initial inclination of the inner orbit $I^{(0)}$. 
%\sma{how about azimuthal angle?}\hang{The azimuthal angle does not affect the secular equations.} (i.e., the angle between $\vect{L}_\imag$ and $\vect{L}_\out$). 
%For the SA equations, we further randomize the inital orbital phase of the outer orbit. 
Here, the initial instant is defined when the inner orbit is nearly circular with $e_\imag^{(0)}=10^{-3}$. The value of $I^{(0)}$ then determines the merger timescale $\tau_{\rm m}$ [see, Eqs.~(\ref{eq:tau_m}) and (\ref{eq:j_min})]. Although a natural choice is to only retain systems with
%$\tau_{\rm m}<\text{the age of the Universe}\sim 10\,{\rm Gyr}$,
$\tau_{\rm m}\lesssim10\,{\rm Gyr}$ (the approximate age of the Universe),
we note that an inner binary in a dense stellar environment like a galactic nucleus may not be able to survive for such a time. For example, the binary may evaporate due to dynamical interactions with environmental stars on a timescale~\cite{Binney:87}
\begin{align}
&\tau_{\rm ev} \simeq 1\times10^{7}\,{\rm yr}\left(\frac{M_t}{100\,M_\odot}\right)\left(\frac{a_\imag}{3\,{\rm AU}}\right)^{-1} \nonumber\\
&\times\left(\frac{\sigma_\star}{350\,{\rm km\,s^{-1} }}\right)
\left(\frac{m_\star}{10\,M_\odot}\right)^{-1}
\left(\frac{\rho_\star}{10^7\, M_\odot\,{\rm pc}^{-3}}\right)^{-1},
\label{eq:tau_evap}
\end{align}
where $\sigma_\star$ and $\rho_\star$ are the local velocity dispersion and stellar mass density, and $m_\star$ is the mass of a typical object in the local environment. Another potentially limiting timescale is the two-body relaxation timescale~\cite{Spitzer:87}, 
\begin{align}
    &\tau_{\rm 2b} \simeq 5\times10^8\,{\rm yr} 
    \left(\frac{\sigma_\star}{350\,{\rm km\,s^{-1}}}\right)^3\nonumber \\
    &\times\left(\frac{m_\star}{10\,M_\odot}\right)^{-1}
    \left(\frac{\rho}{10^7\, M_\odot\,{\rm pc}^{-3}}\right)^{-1}.
\end{align}
We point interested readers to Ref.~\cite{Antonini:12} and references therein for detailed discussions on different timescales that may be relevant. Here we simply choose a merger window of $\tau_{\rm m}< 10^{8}\,{\rm yr}$ for systems evolved using the DA equations. Despite seeming somewhat arbitrary, our choice is justified, as once $\tau_{\rm m}>\text{(a few)}\times \tilde{\tau}_{\rm m, lim}$,\footnote{In fact, $\tau_{\rm m}$ should be compared to the minimum of $\tilde{\tau}_{\rm m, lim}$ and $\tau_{\rm m, lim}$; see Sec.~\ref{sec:max_ecc} and Eq.~(\ref{eq:tau_m_diss})}
the distribution is insensitive to $\tau_{\rm m}$. 

To compare the effect of orbital averaging, we evolve the triple system using both the DA and SA equations. For the DA systems, we select systems that have $\tau_{\rm m}<10^8\,{\rm yr}$ as argued above. As the SA equations are more computationally expensive, we consider only those with $\tau_{\rm m}<3\times10^7\,{\rm yr}$ (see Fig.~\ref{fig:max_ecc_Tm_vs_I0}). In total, we simulate 2000 (1800) DA  (SA) systems.  

We terminate the three-body interaction when the inner semi-major axis shrinks by a factor of 10, $a_\imag = a_\imag^{(0)}/10$. At this point, $\tau_{\rm LK}\gg \tau_{\rm gw}$ and the inner binary is well decoupled from the tertiary perturber. In the remainder of this section, we focus on examining the properties of the inner binary after decoupling from the third body. The properties of the binary once it enters the LIGO band are studied in detail in Sec.~\ref{sec:bin_evol}.

We examine two cases. First, we examine results obtained under the ``clean'' LK without various SMBH effects as described in Sec.~\ref{sec:smbh_effects} (this also corresponds to the case where $\vect{L}_\out$ is parallel to $\vect{S}_3$). The second is with SMBH corrections, using the DA approximation. In the second case, we focus on two representative values of $\theta_{S_3L_\out}$, a small value of $\theta_{S_3L_\out}=5^\circ$ and a larger value of $\theta_{S_3L_\out}=45^\circ$, while the azimuthal angle between $\vect{L}_\out$ and $\vect{S}_3$ is sampled uniformly. 

To summarize, in our numerical simulations we fix the masses of the triple to $(M_1, M_2, M_3)=(55, 45, 10^9)\,M_\odot$, the spin magnitude of each component to $(\chi_1, \chi_2)=(0.7, 0.7)$, and the initial separations to $(a_\imag^{(0)}, a_\out)=(3\,{\rm AU}, 0.06\,{\rm pc})$. The quantities we randomize are the orientation of $\vect{S}_1$ and $\vect{S}_2$ 
(isotropically), as well as the initial inclination of the inner orbit with respect to the outer one, $I^{(0)}$ (uniform in angle). When considering corrections due to effects associated with the central SMBH, we fix $S_3=M_3^2$ and consider two representative angles between $\vect{L}_\out$ and $\vect{S}_3$ ($\theta_{S_3L_\out}=5^\circ \text{ or }45^\circ$). We further select only systems with $\tau_{\rm m}<10^8\,{\rm yr}$ ($3\times 10^7\,{\rm yr}$) to be evolved using the DA (SA) equations until $a_\imag=a_\imag^{(0)}/10=0.3\,{\rm AU}$. In total we simulate 2000 (1800) realizations with the DA (SA) equations.  The focus of our study here is to understand how the LK excitation affects the inner orbit's  spin-orbit alignment and the distribution of the effective spin, $\chieff$, defined as 
\begin{equation}
    \chieff = \frac{M_1 \vect{\chi}_1\cdot \uvect{L} + M_2 \vect{\chi}_2\cdot\uvect{L}}{M_1 + M_2}.
    \label{eq:chieff}
\end{equation}

\begin{figure}[tb]
  \centering
  \includegraphics[width=\columnwidth]{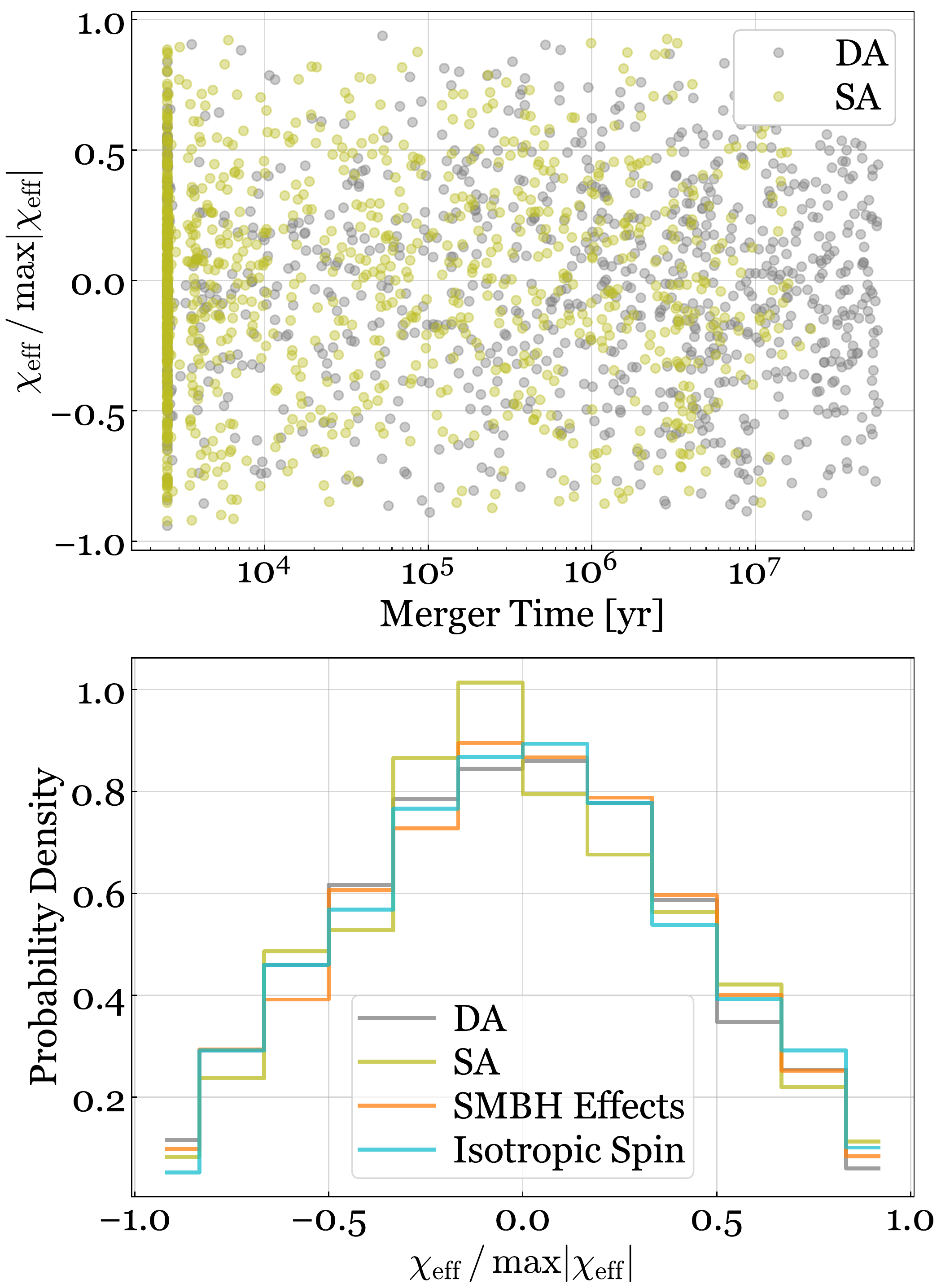}
\caption{Top panel: effective spin $\chieff$ distribution as a function of the LK induced merger time $\tau_{\rm m}$. The grey and olive dots represent systems evolved with the DA and SA equations, respectively. SMBH effects are ignored in this case. Bottom panel: the distribution of the effective spins for systems that experience multiple LK cycles before the eventual merger (i.e., with $\tau_{\rm m}\gtrsim 5\times 10^{3}\,{\rm yrs}$). For reference, the cyan trace corresponds to the initial distribution of $\chieff$ with isotropically oriented spins. Lastly, the orange trace corresponds to the distribution of $\chieff$ with SMBH effects incorporated for $\theta_{S_3L_\out}=45^\circ$. Note that in both panels we express the effective spin as $\chieff$ normalized by the maximum effective spin allowed in the simulations, namely, max$|\chieff| = 0.7$. To generate the distribution, we use 1605 DA runs, 1112 SA runs, and 1287 runs including the SMBH effects after the cut $\tau_{\rm m}\gtrsim 5\times 10^{3}\,{\rm yrs}$ cut. }
\label{fig:chi_eff_dist}
\end{figure}

In the top panel of Fig.~\ref{fig:chi_eff_dist} we present a scatter plot of $\chieff$ as a function of the merger time.\footnote{Note that in the top panel, there is a cluster of points piled up at the vertical line of $\tau_{\rm m}\simeq 2.5\times10^{3}$. This is $\sim 100$ times longer than the limiting merger time one would expect for a conservative system as shown in Eq.~(\ref{eq:tau_m_lim_cons}). This is due to the fact that the limiting eccentricity can be smaller than the prediction of Eq.~(\ref{eq:e_lim_cons}) if the inner orbit decays rapidly due to GW radiation. This is discussed further in Sec.~\ref{sec:max_ecc}.}
We use grey (olive) dots to represent systems evolved using the DA (SA) equations. With randomized initial spin directions, we do not see $\chieff$ attracted toward 0, even for systems that experience multiple ``clean'' LK cycles with merger times greater than $5\times10^3\,{\rm yr}$ and without being perturbed by various SMBH effects. Rather, the effective spin has a distribution consistent with that expected from an isotropic spin direction, as shown in the lower panel of Fig.~\ref{fig:chi_eff_dist}.

\begin{figure}[tb]
  \centering
  \includegraphics[width=\columnwidth]{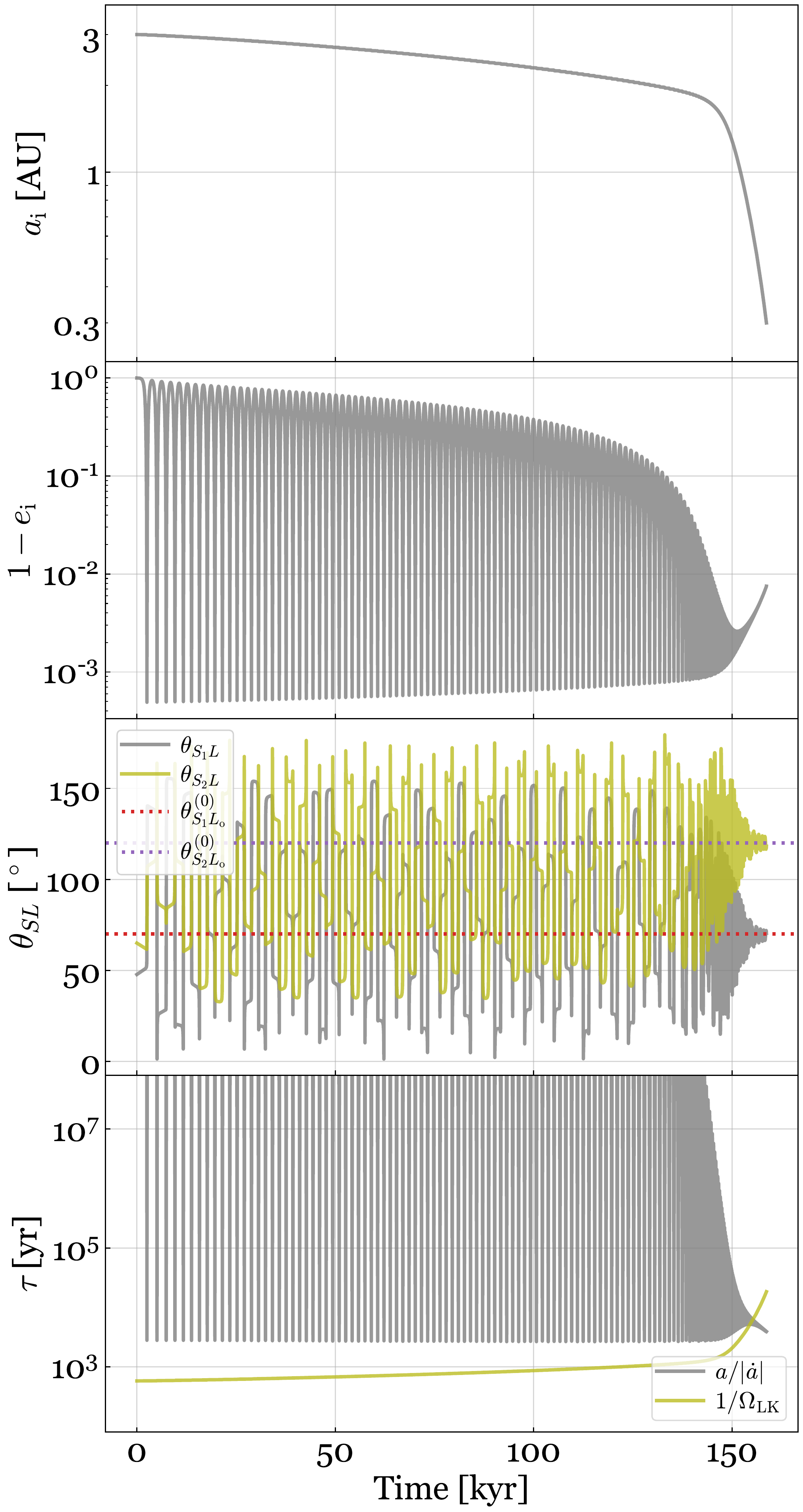}
  \caption{A representative case of an inner binary's evolution during the LK induced oscillations with $(M_1, M_2, M_3) = (55, 45, 10^9)\,M_{\odot}$, $a_\imag^{(0)}=3\,{\rm AU}$, $a_\out=0.06\,{\rm pc}$ and $I^{(0)}=88.7^{\circ}$. %Potential precession around $S_3$ is not included in this plot.
    From the top to bottom, we show the semi-major axis $a_\imag$, the eccentricity $e_\imag$,  the spin-orbit alignment $\theta_{SL}$ of the inner orbit, and the comparison of the GW decay timescale and the inverse of LK frequency [Eqs.~(\ref{eq:tau_gw}) and (\ref{eq:omega_LK})] respectively. In the third panel, the dotted lines correspond to the initial angles between the spin vectors and the outer orbit's orbital angular momentum $L_\out$.}
\label{fig:LK_evol_sample}
\end{figure}

Nevertheless, there still exists a dynamical attractor of the spin orientation. This is illustrated in Fig.~\ref{fig:LK_evol_sample} where we present a sample evolution track of the inner binary under multiple LK cycles (without SMBH effects). From the top to bottom, we show, respectively, the semi-major axis, the eccentricity, and the spin-orbit alignment of the inner orbit, $\theta_{S_{1(2)}L}$. We see that at the end of the LK evolution, the angles between the spin vectors and the inner orbital angular momentum, $\theta_{S_{1(2)}L}$, converge to fixed values, which correspond to the anglea between the initial spin vectors and the AM of the outer orbit, $\theta_{S_{1(2)}L_\out}^{(0)}$. 

\begin{figure}[tb]
  \centering
  \includegraphics[width=\columnwidth]{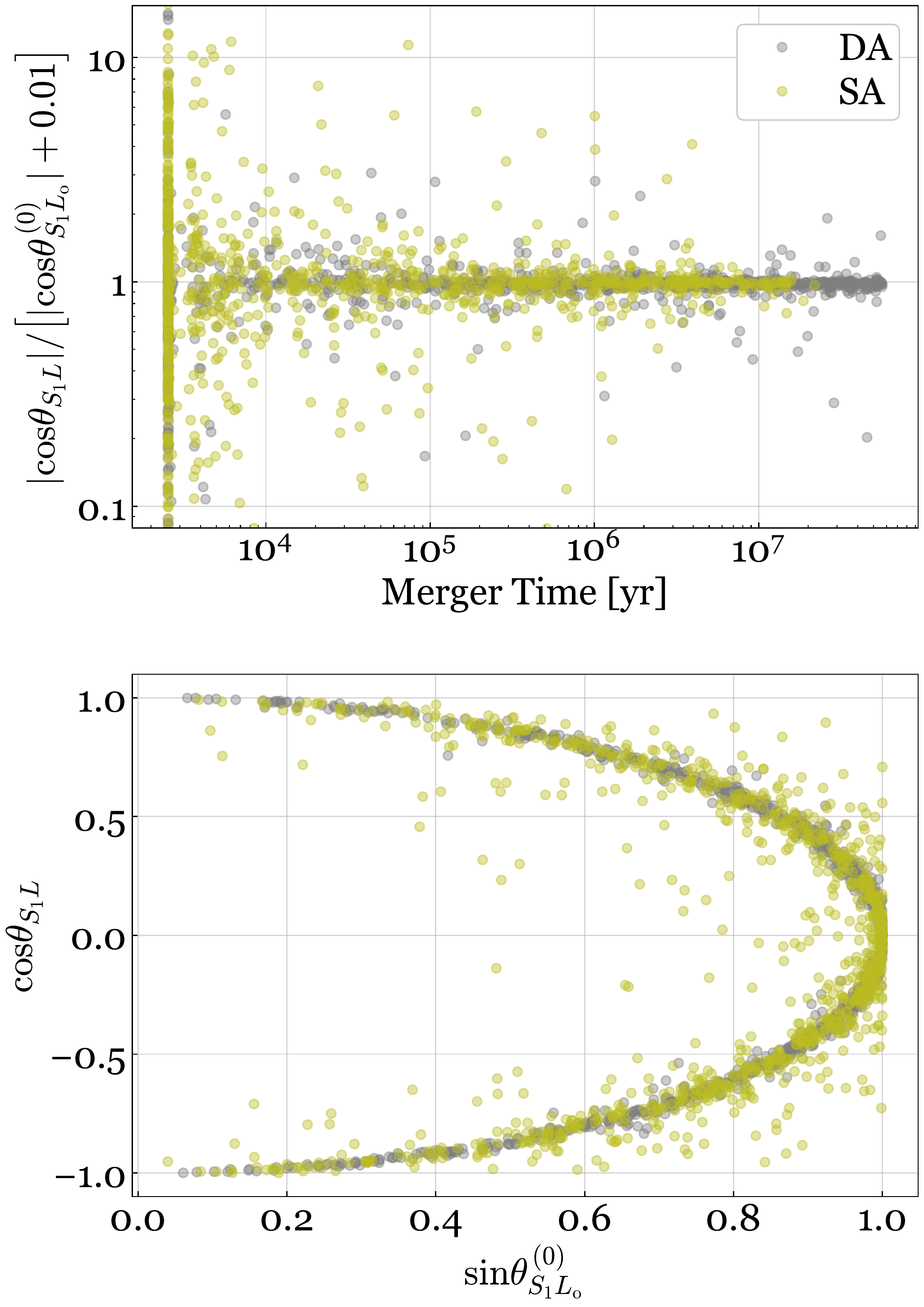}
\caption{
Top panel: the ratio between the spin-orbit alignment at the end of the LK oscillation, $|\cos\theta_{S_1 L}|$, and the initial alignment between spin and the outer orbit's angular momentum, $|\cos \theta_{S_1 L_\out}^{(0)}|$. The two quantities are nearly equal for systems experiencing multiple LK oscillations. 
Bottom panel: here we focus on only those systems with $\tau_{\rm m}\geq 5\times 10^3\, {\rm yrs}$, which are those with multiple LK oscillations, displaying a clear relationship between the final angle between $S_1$ and the binary angular momentum $L$ versus the initial angle between $S_1$ and the outer angular momentum $L_\out$. %The trend between the spin-orbit alignment One can further remove the need of taking the absolute value if one plot $\cos \theta_{S_1L}$ as a function of $\sin \theta_{S_1L_\out}^{(0)}$. In the bottom panel we show only systems with . 
%Distribution of the angle between the spin and inner angular momentum at the end of the LK evolution, $\theta_{S_{1,2}L}$ as a function of angle between the initial spin and the total angular momentum of the triple system $\theta_{S_{1,2},0}$. We have selected only systems with $T_{\rm m}\geq 2\times10^7~{\rm yrs}$.
}
\label{fig:cThSL_vs_sThS0}
\end{figure}

In fact, this attraction holds generically as shown in Fig.~\ref{fig:cThSL_vs_sThS0}. In the top panel of Fig.~\ref{fig:cThSL_vs_sThS0} we show, as a function of merger time, the ratio of $|\cos \theta_{S_1L}|$ at the end of the LK evolution to the initial value of $|\cos\theta_{S_1L_\out}^{(0)}|$. Note, in the figure we have added a small value of $0.01$ to the denominator to avoid numerical singularities. Whereas those that merge in essentially a single LK cycle ($\tau_{\rm m}\simeq 2.5\times 10^{3}\,{\rm yr}$) present a large scattering for the value of this ratio, systems with $\tau_{\rm m}\gtrsim 5 \times 10^{3}\,{\rm yr}$ (i.e., experiencing multiple LK cycles) concentrate around a value of unity. Although we limit the presentation to $\vect{S}_1$, this same relation holds true for the orientation of $\vect{S}_2$. Further, if we cast $\cos\theta_{S_1L}$ as a function of $\sin\theta_{S_1L_\out}^{(0)}$, then a clear bifurcation pattern appears, as shown in the bottom panel of Fig.~\ref{fig:cThSL_vs_sThS0}. 

% This can be easily understood if one follows the argument of Liu and Lai~\cite{Liu:18} (see their sec. 4.3) and replace their $I_0$ by $\theta_{S_{1,2}, 0}$.

Qualitatively, this may be understood by generalizing the argument given in  Ref.~\cite{Liu:18} (see their sec. 4.3). 
Specifically, in a frame that rotates together with $\vect{L}$ around $\vect{L}_\out$ (indicated by a subscript ``rot''), the evolution of $\vect{S}_1$ may be approximated as
\begin{equation}
    \frac{d\vect{S}_1}{dt}\Big{|}_{\rm rot} \simeq \vect{\Omega}_{\rm eff} \times \vect{S}_1,
\end{equation}
where 
\begin{equation}
    \vect{\Omega}_{\rm eff} = \vect{\Omega}_{\rm dS}^{(1)} + \vect{\Omega}_{\rm stdLK}.
\end{equation}
The vector $\vect{\Omega}_{\rm stdLK}$ is further given by
\begin{equation}
    \vect{\Omega}_{\rm stdLK} =\frac{3\uvect{L}\cdot\uvect{L}_\out(1+4e^2)}{4\tau_{\rm LK}} \uvect{L}_\out. 
\end{equation}
One may argue that the angle between $\vect{S}_{1}$ and $\vect{\Omega}_{\rm eff}$ is an adiabatic invariant if $|\vect{\Omega}_{\rm eff}|$ is slow varying. 
%Here $\vect{\Omega}_{\rm eff}$ characterizes the precession rate of $\vect{S}_1$ around $\vect{L}_\out$ in a frame rotates together with $\vect{L}$, and it is given by 
% \begin{equation}
%     \vect{\Omega}_{\rm eff} = \vect{\Omega}_{\rm dS}^{(1)} + \vect{\Omega}_{\rm LK},
% \end{equation}
% with 
% \begin{equation}
%     \vect{\Omega}_{\rm LK} = \frac{3\uvect{L}\cdot\uvect{L}_\out(1+4e^2)}{4\tau_{\rm LK}} \uvect{L}_\out. 
% \end{equation}
Initially $|\vect{\Omega}_{\rm stdLK}|\gg |\vect{\Omega}_{\rm dS}^{(1)}|$ when the inner binary is widely separated, but as the orbit decays, at the end of the LK cycle the opposite is true $|\vect{\Omega}_{\rm stdLK}|\ll |\vect{\Omega}_{\rm dS}^{(1)}|$. This then implies that 
\begin{equation}
  \theta_{S_1L} \simeq \theta_{S_1L_\out}^{(0)}. 
\end{equation}
Note, however, that the argument does not explain why we can also have $\theta_{S_1L}\simeq \pi - \theta_{S_1L_\out}^{(0)}$ from numerical simulations, hence a more rigorous understanding of the process is needed in a future study.

From this, we now see that the attraction to $\chieff\simeq 0$ for systems experiencing multiple ``clean'' LK cycles as reported in Refs.~\cite{Antonini:18, Liu:18, Rodriguez:18b} is a consequence of their choice of initial conditions. The aformentioned studies focus on systems whose spin vectors are initially aligned with the inner AM vector, $\theta_{\rm S_1L_\out}^{(0)} = I^{(0)}$. In order for the inner binary to be excited to a large enough eccentricity that it merges within 10\,Gyr, the inner AM vector is further required to have an initial inclination of $I^{(0)}\simeq \pi/2$ with respect to the outer orbit. 
The bottom panel of Fig.~\ref{fig:cThSL_vs_sThS0} illustrates that such systems with $\sin I^{(0)}\simeq 1$ lead to $\cos\theta_{S_1L}\simeq 0$ and consequently $\chieff\simeq 0$ at the end of the LK interaction. 

While an initial alignment between $\vect{S}_1$ and $\vect{L}$ may be expected for field triples (which are the focus of Refs.~\cite{Antonini:18, Liu:18, Rodriguez:18b}), it is unclear if this assumption holds for binaries in galactic nuclei. If the spin vectors do not have a preferred direction initially,\footnote{We note that our isotropic spin prior may be an oversimplification to the problem, as other dynamical processes, such as gas torques in the disk of an active galactic nucleus (see, e.g., Refs.~\cite{Bartos:17, Tagawa:19, McKernan:19, Graham:20}), could also affect the initial spin orientation. Here we ignore these gaseous effects, leaving this to future studies. } then the LK evolution does not lead to a preferred value of $\chieff$ (relative to the isotropic spin distribution) in general.

\begin{figure}[tb]
  \centering
  \includegraphics[width=\columnwidth]{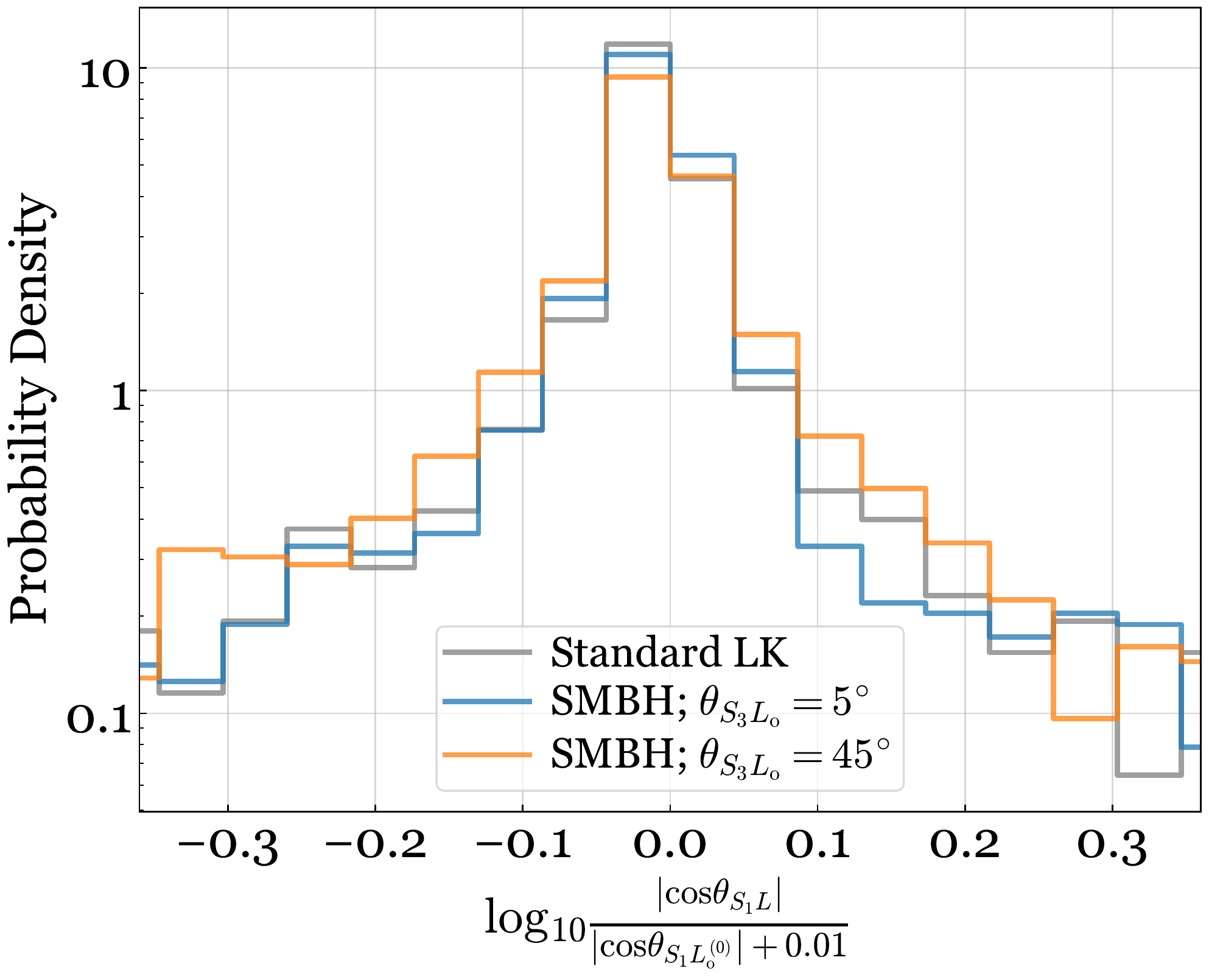}
\caption{The distribution of the ratio between $|\cos \theta_{S_1L}|$ at the end of the LK evolution and $|\cos \theta_{S_1L_\out}^{(0)}|$ initially. The grey trace corresponds the ``standard'' LK (or $\theta_{S_3L_\out}=0$), consistent with the grey dots in the upper panel of Fig.~\ref{fig:cThSL_vs_sThS0}. The blue and orange traces show the distributions when SMBH effects are included at two representative values: $\theta_{S_3L_\out}=5^\circ$ and $45^\circ$, respectively. The azimuthal angle between $\vect{L}_\out$ and $\vect{S}_3$ is randomly sampled and all data points are presented, including those merging in a single LK cycle.
}
\label{fig:cThSL_vs_cThS0_SMBH}
\end{figure}

We conclude this section by briefly examining the effects due to an SMBH~\cite{Liu:19}. As argued in Sec.~\ref{sec:smbh_effects}, we expect the effect to be mild corrections to the ``standard'' LK interactions for the set of parameters we focus on. This is demonstrated in Fig.~\ref{fig:cThSL_vs_cThS0_SMBH}, where we compare the distributions of $|\cos \theta_{S_1L}|$ and $|\cos \theta_{S_1L_\out}^{(0)}|$ with and without SMBH effects. Indeed, we see good agreement overall between the different data sets. When the $\vect{L}_\out{-}\vect{S}_3$ misalignment is significant (the orange trace with $\theta_{S_3L_\out}=45^\circ$), there is a slight hint of the attractor being degraded, as more systems experience more extreme eccentricity excitation and merge in fewer LK cycles (see also Fig.~\ref{fig:max_ecc_Tm_vs_I0}). 
Nevertheless, since the distribution of $\chieff$ is already consistent with that obtained from an isotropic spin distribution, due to the initial condition we have assumed on the spins, we do not expect SMBH effects to change this result. This is confirmed through the results presented in the bottom panel of Fig.~\ref{fig:chi_eff_dist}. This conclusion should be further strengthened for binaries that are more compact and closer to an SMBH, where its effects are more significant, as Ref.~\cite{Liu:19} showed that for a nearly fixed initial spin orientation, the final $\chieff$ distribution tends to be more broad than the isotropic-spin case. 
%Nevertheless, since the prior distribution of $\vect{S}_{1(2)}$ is already isotropic in our study, the distribution of $\chieff$ at the end of the LK evolution is not affect by the SMBH effects, as shown in the bottom panel of Fig.~\ref{fig:chi_eff_dist}. 

\section{Spin-spin evolution for binaries with arbitrary orbital eccentricity }
\label{sec:bin_evol}

In this Section, we take those binaries that have undergone LK oscillations (those we studied in Sec.~\ref{sec:LK_evol}) as the initial conditions and continue evolving the inner binaries until merger, with the goal of studying the final orientation of the spin vectors. A quantity we are particularly interested in is the angle between two spin vectors, $\theta_{S_1S_2}$.  While this angle is a subdominant effect in the inspiral GW waveform, it nonetheless plays a significant role in determining the final merger-ringdown waveform and the GW kick the system receives at the merger (see, e.g., Refs.~\cite{Campanelli:07, Kesden:10, Berti:12, Gerosa:18}). 

\begin{figure}[tb]
  \centering
  \includegraphics[width=\columnwidth]{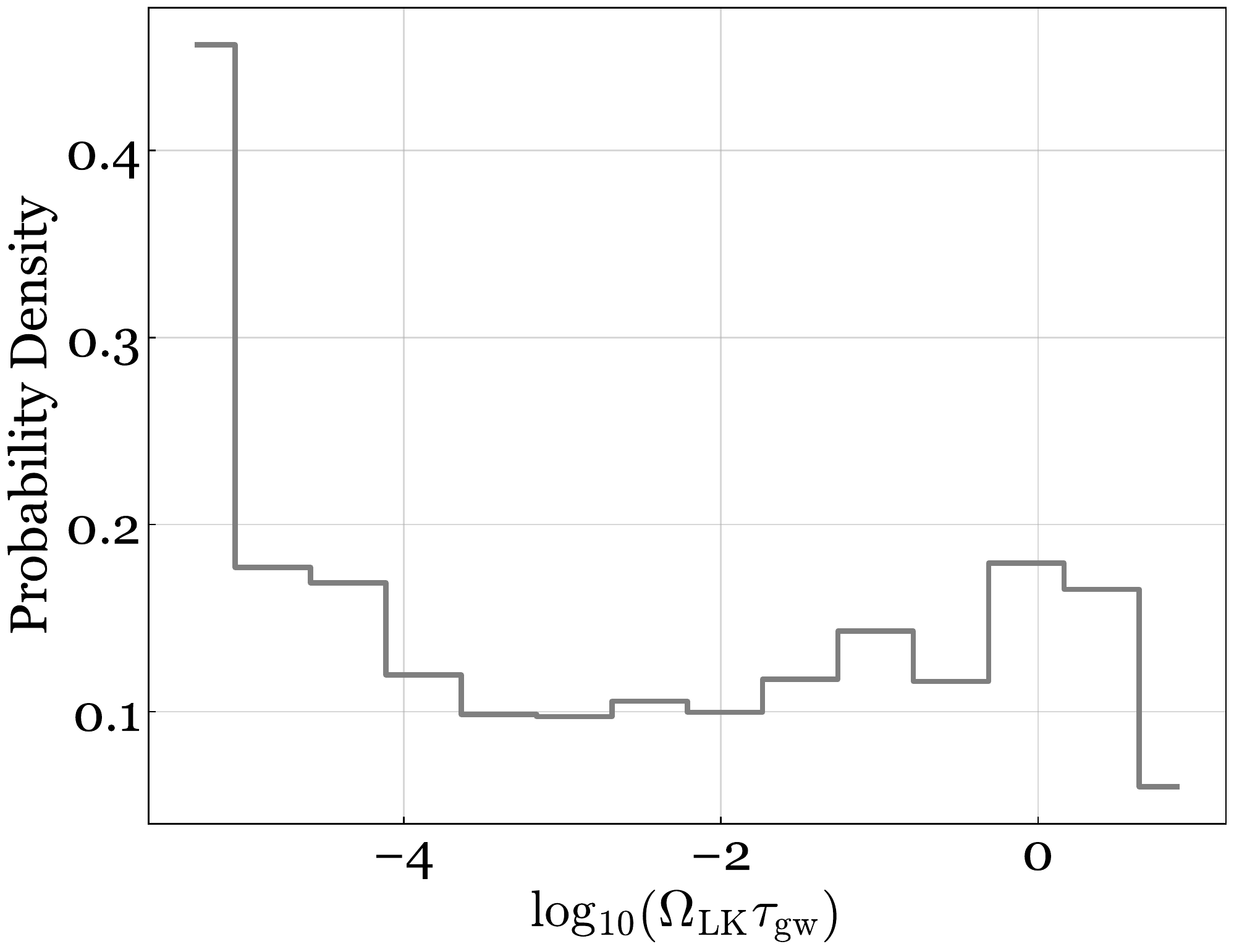}
\caption{The distribution of the product $\Omega_{\rm LK} \tau_{\rm gw}$ [Eqs.~(\ref{eq:omega_LK}) and (\ref{eq:tau_gw})] at the end of the LK evolution $a_\imag = a_\imag^{(0)}/10$ of our simulations. For the majority of the systems, we have $\Omega_{\rm LK} \tau_{\rm gw}< 1$ satisfied and therefore the inner binary is effectively decoupled from the tertiary perturber. 
}
\label{fig:omega_lk_tau_gw}
\end{figure}

Note that at this point all binaries have a separation of $a_\imag = a_\imag^{(0)}/10=0.3\,{\rm AU}$, which is the criterion for terminating the three-body LK evolution. At this point, the tidal torque for the tertiary mass to perturb is much smaller compared to the initial value. Moreover, the inner binary inspirals with an increasingly shorter timescale. As we show in Fig.~\ref{fig:omega_lk_tau_gw} (see also the bottom panel of Fig.~\ref{fig:LK_evol_sample}), for the majority of our simulations we have $\Omega_{\rm LK}\tau_{\rm gw}< 1$, and consequently, the inner binary has is decoupled from the perturber and the LK interaction terms can be safely disregarded.\footnote{We acknowledge that there are about $15\%$ of the systems shown in Fig.~\ref{fig:omega_lk_tau_gw} that do not meet the $\Omega_{\rm LK}\tau_{\rm gw}< 1$ condition because they experience a weak LK excitation and merges in more than $10^{7}$ years [cf. Eq.~(\ref{eq:tau_evap})]. We do not evolve the triple system further because that would make the majority of the systems run into the computationally expensive regime caused by the fast de Sitter precession of the inner spins. Nonetheless, one can show $\Omega_{\rm LK}\tau_{\rm gw}\propto a^3$ when $(1-e)< 1$ and by $a\simeq 0.1\,{\rm AU}\simeq 10^5 M_{\rm t}$ all the systems will satisfy $\Omega_{\rm LK}\tau_{\rm gw}< 1$. Moreover, the LK evolution only provides initial conditions for the subsequent binary evolution but will not affect any relations between various spin alignments which are the focus of Sec.~\ref{sec:bin_evol}. With or without the $\Omega_{\rm LK}\tau_{\rm gw}\propto a^3$ when $(1-e)> 1$ systems, we find the spins are consistent with an isotropic distribution at the end of the LK evolution. } 

Nonetheless, a new computational challenge appears. Note that both the de Sitter precession and the Lense-Thirring precession are of lower PN orders than the 2.5 PN GW-driven decay. In fact, we have 
\begin{equation}
    \tau_{\rm gw}\Omega_{\rm dS} \propto a^{3/2}(1-e^2)^{5/2}.
\end{equation}
One may further show that $(1-e^2)\simeq 2(1-e)\propto 1/a$ when $(1-e)\ll 1$, a condition that is typically true at the end of the LK evolution. Consequently we have
\begin{equation}
    \tau_{\rm gw}\Omega_{\rm dS} \propto 
    \begin{cases}
   a^{-1}        & \text{if } (1-e) \ll 1,\\
   a^{3/2}        & \text{if } e \ll 1.
  \end{cases}
\end{equation}
Therefore, the precession phase is largely dominant at the time when the binary has $e\sim 0.5$. This typically occurs at $a\simeq3\times 10^{-3}\,{\rm AU}$ for the binaries we consider here. A brute-force approach at evolving the set of differential equations outlined in Sec.~\ref{sec:formalism} requires a large number of precession cycles be resolved, making this approach prohibitively expensive computationally. Therefore, if we want to explore how the initial conditions affect the final spin orientation, a more efficient way of evolving the system is desired.  % On the other hand, because of the same reason\matt{which reason? too computationally expensive?}, we cannot say a priori that the initial condition plays no significant role in determining the final orientations of the spin vectors and hence ignore it\matt{what does `it' refer to here?}. 

To do so, we rely on the effective potential description and the precession-averaged orbital evolution proposed by Ref.~\cite{Kesden:15}. The derivation of Ref.~\cite{Kesden:15} is for circular orbits only, whereas the binaries considered here that merge via the LK mechanism (as well as other dynamical channels) typically have a large eccentricities.
In the following Sec.~\ref{sec:eff_spin_potential} we generalize the effective potential theory to binaries with arbitrary eccentricity. Additionally, we provide a prescription for evolving an eccentric system in a precession-averaged way. We apply this generalized theory to evolve our binaries from $0.3\,{\rm AU}$ to $300\,M_{\rm t}\simeq 3\times10^{-4}\,{\rm AU}$ in Sec.~\ref{sec:a_300}. As the binary further evolves, the precession timescale can become greater than the decay timescale and it cannot be treated in the averaged manner. In consideration of this, we evolve the full equations from $300\,M_{\rm t}$ until merger (which we define as $a=6\,M_{\rm t}$, corresponding to the inner-most stable circular orbit, or ISCO, of a Schwarzschild BH with mass $M_{\rm t}$). The final spin distribution is studied in details in Sec.~\ref{sec:fin_spin_dist}. Finally, in Sec.~\ref{sec:kick}, we demonstrate how the spin distribution affects the magnitude of the GW kick a binary receives at merger.

Before proceeding, we remind the reader that at this stage the inner binary has well decoupled from the tertiary perturber, and the LK interaction merely provides the initial conditions for the binary evolution. Therefore, in addition to studying the marginalized distributions, we also examine binaries obtained from specific slices of initial conditions. As long as a formation channel (not restricted to the LK mechanism) allows for the same slice of initial conditions, our conclusions apply generically.

\subsection{Effective spin potential and precession-averaged evolution}
\label{sec:eff_spin_potential}

We review here the effective potential theory proposed by Ref.~\cite{Kesden:15} and generalize to orbits with arbitrary eccentricity, so that the theory can be applied to eccentric binaries that dynamical formation channels (including the LK oscillation we study here) typically produce. 

To proceed, we note that the key foundation of the derivation in Ref.~\cite{Kesden:15} is that the effective spin parameter $\chieff$ [Eq.~(\ref{eq:chieff})] is preserved to at least the 2.5 PN order. In fact, this is true even for eccentric orbits (see, e.g., Ref.~\cite{Racine:08}). This, together with some geometrical relations, allows us to express the angles between different vectors as
\begin{align}
    &\cos \theta_{LJ} = \frac{J^2 + L^2 -S^2}{2JL},\label{eq:th_LJ}\\
    &\cos \theta_{S_1L} = \frac{1}{2(1-q)S_1}\left[\frac{J^2-L^2-S^2}{L} - \frac{2qM_{\rm t}^2\chieff}{1+q}\right], \label{eq:th_S1L}\\
    &\cos \theta_{S_2L} = \frac{q}{2(1-q)S_2}\left[-\frac{J^2-L^2-S^2}{L} + \frac{2M_{\rm t}^2\chieff}{1+q}\right],\label{eq:th_S2L}\\
    &\cos\theta_{S_1S_2} = \frac{S^2-S_1^2-S_2^2}{2S_1S_2},\label{eq:th_S12}\\
    &\cos\Delta\Phi = \frac{\cos\theta_{S_1S_2}-\cos\theta_{S_1L}\cos\theta_{S_2L}}{\sin\theta_{S_1L}\sin\theta_{S_2L}},\label{eq:delPhi}
\end{align}
where in the above equations $J = |\vect{L}+\vect{S}|$ is the magnitude of the total angular momentum of the binary and $S=|\vect{S}_1 + \vect{S}_2|$ is the magnitude of total spin. We use $\theta_{LJ}$ to represent the angle between $\vect{L}$ and $\vect{J}$ and $\Delta \Phi$ the angle between $\vect{S}_1$ and $\vect{S}_{2}$ in the orbital plane. Since the angles are based on geometrical relations between different vectors, they hold independent of the orbital eccentricity, as long as one uses the proper $\vect{J}$ and $\vect{L}$ for eccentric orbits. 

The effective potential is also a geometrical relation. It describes, for a given set $(J, L)$ together with constants $(M_{\rm t}, q, S_1, S_2, \chieff)$, the allowed range of the total spin magnitude $S$ can take. Specifically, the range is determined by solving the equation $\chieff^\pm(S_\pm)|_{J, L}=\chieff$, where 
\begin{align}
    &\chieff^\pm(S)|_{J, L} = \frac{1}{4qM_{\rm t}^2S^2L} \left\{\pm (1-q^2)A_1A_2A_3A_4 \right. \nonumber \\
    &\left.+\left(J^2-L^2-S^2\right)
    \left[S^2(1+q)^2-(S_1^2-S_2^2)(1-q^2) \right]  \right\},
\end{align}
with 
\begin{align*}
    & A_1 = \sqrt{J^2-(L-S)^2},\ \  
      A_2 = \sqrt{(L+S)^2-J^2}, \nonumber \\
    & A_3 = \sqrt{S^2-(S_1-S_2)^2}, \ \ 
      A_4 = \sqrt{(S_1+S_2)^2-S^2}. 
\end{align*}
The roots $S_\pm$ then defines the allowed range of $S$ as $S_-\leq S\leq S_+$. 

Within this range, the total spin magnitude varies at a rate (see Appendix~\ref{sec:dSdt} for derivation)
\begin{align}
    \frac{dS}{dt} =& -\frac{3(1-q^2)}{2q} \eta^6 \left(1-e^2\right)^{3/2}\left(\frac{M_{\rm t}^2}{L}\right)^5 \frac{S_1S_2}{M_{\rm t}S} 
    \nonumber \\
    &\times\left[1-\frac{\eta M_{\rm t}^2\chieff}{L}\right]\sin\theta_{S_1L}\sin\theta_{S_2L}\sin\Delta \Phi,\label{eq:dSdt}
\end{align}
where $\eta = M_1M_2/M_{\rm t}^2$. Note that when $e=0$, this reduces to eq.~(8) in Ref.~\cite{Kesden:15}.
%\footnote{We have also dropped an $\mathcal{O}(M^2/L)$ correction in Eq.~(\ref{eq:dSdt}) which is small in the range where this method is useful.}
Also, note that $dS/dt$ is specified in terms of $(J, L, e, S)$ and there is no explicit time dependence. 
Additionally, we define a precession timescale, $\tau_{\rm pre}$, as
\begin{equation}
    \tau_{\rm pre}(J, L, e) = 2\int_{S_-}^{S_+}\frac{dS}{|dS/dt|}. 
    \label{eq:tau_pre}
\end{equation}

We now have all the ingredients to perform the precession-averaged evolution. 
Note that $d\vect{J}/dt \propto \vect{L}$ and for the amplitudes we can write $dJ/dt = \uvect{J} \cdot d\vect{J}/dt$ and $dL/dt = \uvect{L}\cdot d\vect{J}/dt$. Thus, we have $dJ = \cos\theta_{LJ} \,dL$. Over a time $\Delta t$ with $\tau_{\rm pre}\ll\Delta t\ll\tau_{\rm GW}$, we write the precession-averaged evolution of $J$ in terms of $L$ as 
\begin{equation}
    \left\langle\frac{d J}{d L}\right\rangle = \frac{2}{\tau_{\rm pre}} \int_{S_-}^{S_+} \frac{\cos \theta_{LJ} d S}{|d S/d t|}.
    \label{eq:dJdL}
\end{equation}
Note that this is formally the same as eq. (10) in Ref.~\cite{Kesden:15}, except the precession rate $dS/dt$ now also depends on the eccentricity [Eq.~(\ref{eq:dSdt})]. The right-hand side of Eq.~(\ref{eq:dJdL}) is now fully specified in terms of $(J, L, e)$. 
% Again, the right-hand side of Eq.~(\ref{eq:dJdL}) is fully determined in terms of $(J, L, e)$, and there is no fast-oscillating (on the timescale $\tau_{\rm pre}$) quantities appearing.

%Thus, over a time $\Delta t$ with $\tau_{\rm pre}\ll\Delta t\ll\tau_{\rm GW}$, we can write the prec
%Consider a time $\Delta t$, with $\tau_{\rm pre}\ll\Delta t\ll\tau_{\rm GW}$. One may imagine that the spin-precession effect is averaged out so that the total angular momentum vector $\vect{J}$ changes by an amount $\langle\Delta \vect{J}\rangle$ that is parallel to $\vect{J}$. Meanwhile, the variation in the magnitude can be written as $dJ/dt = \uvect{J}\cdot d\vect{J}/dt \propto \cos\theta_{LJ}$. Consequently, we can write a precession-averaged evolution of $J$ in terms of $L$ as

Similarly, we cast the precession-averaged eccentricity evolution in terms of $L$ by simply dividing (the scalar version of) Eqs. (\ref{eq:dedt_gw}) and (\ref{eq:dLdt_gw}) and substitute $a$ in terms of $(L, e)$ using Eq.~(\ref{eq:L_vs_a_e}), leading to 
\begin{equation}
    \left\langle\frac{de}{dL}\right\rangle = \frac{19}{6}\frac{e}{L}\frac{1+\frac{121}{304}e^2}{1+\frac{7}{8}e^2}. \label{eq:dedL}
\end{equation}
This completes the set of precession-averaged equations. 

\begin{figure}[tb]
  \centering
  \includegraphics[width=\columnwidth]{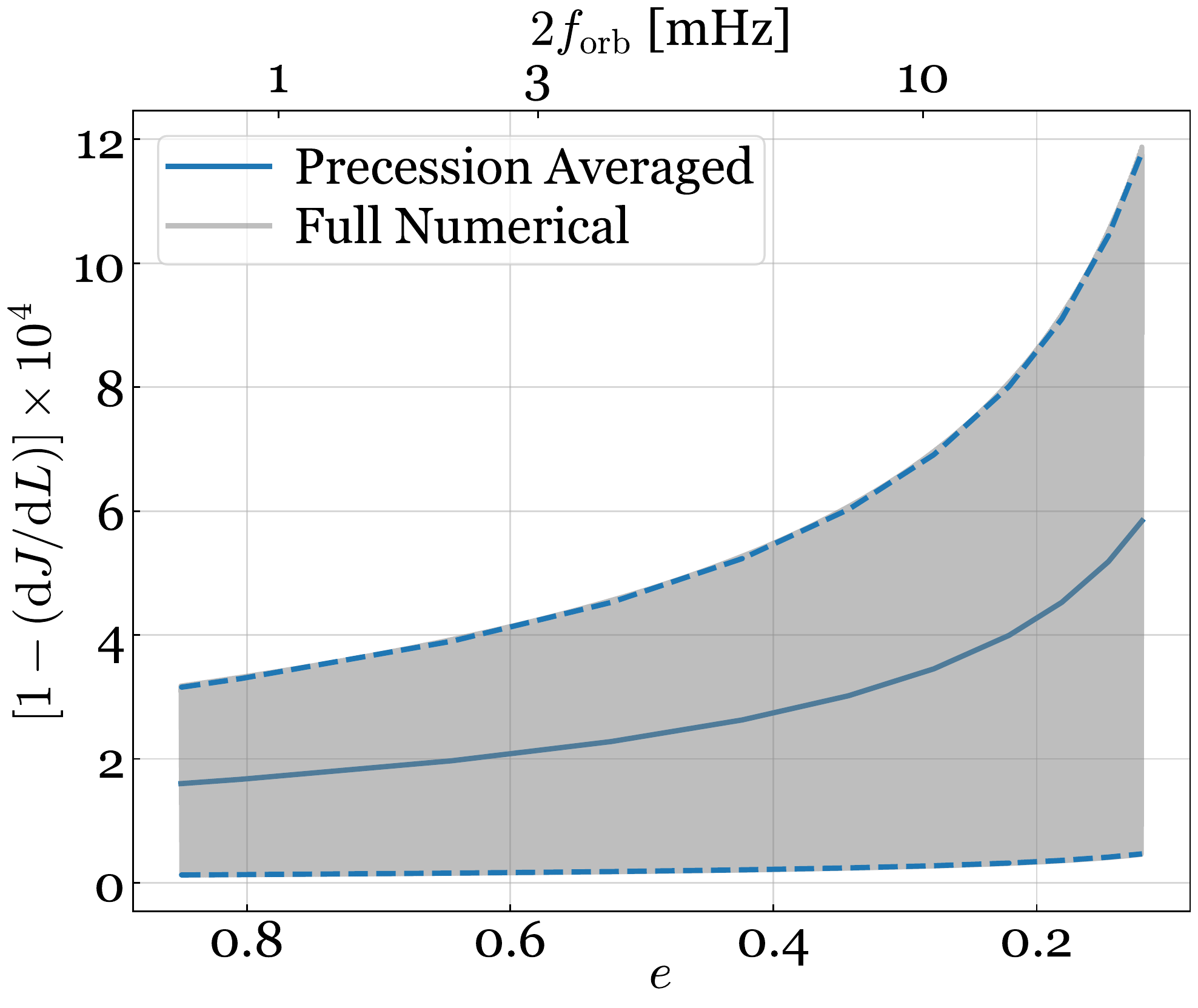}
\caption{The evolution of the total angular momentum of the inner binary $J=|\vect{L}+\vect{S}|$, with respect to the its orbital angular $L$, as a function of the orbital eccentricity. Note that we have inverted the bottom-x-axis so that the system evolves, naturally, toward smaller values of $e$.}
\label{fig:dJdL_vs_e}
\end{figure}

In Fig.~\ref{fig:dJdL_vs_e} we compare the precession-averaged evolution of $J$ (blue-solid trace) and the full numerical result (grey traces; it contains $\sim 10^4$ precession cycles in the range shown). Also shown in the blue dashed traces are the upper and lower envelopes of $d J/d L$ evaluated at $\cos \theta_{LS} (S_\mp)$. Note that in Fig.~\ref{fig:dJdL_vs_e} the x-axis corresponding to the eccentricity of the system is inverted so that left to right corresponds to a decaying orbital separation and an increasing orbital frequency. From Fig.~\ref{fig:dJdL_vs_e}, we see that the averaged evolution matches well with the full numerical result. 

We summarize the procedure for performing the precession-averaged evolution as follows. Given a set of initial conditions for $(J, L, e)$, together with a set of constant parameters $(M_{\rm t}, q, S_1, S_2, \chieff)$, one can obtain the averaged orbital evolution in terms of $L$ by solving $\left\{\langle dJ/dL \rangle, \langle de/dL \rangle\right\}$ using Eqs.~(\ref{eq:dJdL}) and (\ref{eq:dedL}). While in this process we lose track of the exact value of $S$, we nevertheless know its probability density function for each system with $(J, L, e)$ given by
\begin{equation}
    p(S|J, L, e) = \frac{2}{\tau_{\rm pre}}\frac{1}{|dS/dt|}. 
    \label{eq:p_S_vs_JLe}
\end{equation}
To get the distribution of an ensemble, we simply sum the distribution for each system together and then perform an average
\begin{equation}
    p(S) = \frac{1}{N}\sum p(S|J, L, e),
    \label{eq:p_S_summed}
\end{equation}
where $N$ is the number of systems in the ensemble.\footnote{Here, each realization of our simulation has the same weight. However, an extension that allows for different weights is straightforward to implement in this framework. } 
The probability density of any function $f$ of $S$ \big{(}and $(J, L, e)$\big{)} is described as 
\begin{equation}
    p\left[f(S)|J, L, e\right] = \frac{p(S|J, L, e)}{|df/dS|}.
    \label{eq:p_fS_vs_JLe}
\end{equation}
This allows us to, e.g., compute the distribution of different angles as shown in Eqs.~(\ref{eq:th_LJ})-(\ref{eq:delPhi}).  

In the following Section (Sec.~\ref{sec:a_300}) we apply this technique to evolve systems from the end of the LK oscillation to $a=300\,M_{\rm t}$ and study the resulting distributions.

\subsection{Evolving to $a=300\,M_{\rm t}$}\label{sec:a_300}
Among all the systems we obtain from the LK evolution, we focus specifically on those with $|\chieff|<0.1$ for the remainder of this paper (about 500 DA systems and 450 SA systems after the cut). The reasons for this restriction are as follows. First, while we have shown the LK mechanism does not provide an attractor to $\chieff=0$ once the initial spin orientation is randomized, a small $\chieff$ is nonetheless geometrically favorable for isotropic spin orientations (see Fig.~\ref{fig:chi_eff_dist}). %Additionally, a small $\chieff$ is consistent with the recent IMBH event GW190521.1\needref, so a study of this regime is useful in understanding the properties of this event.
Furthermore, spins in the orbital plane (for which $\chieff\simeq0$ is a necessary condition) is one of the conditions required to produce a particularly strong GW recoil (see, e.g., Ref.~\cite{Brugmann:08}). To further explore this configuration, we also consider a set of systems where we require not only $|\chieff|<0.1$, but also $\theta_{S_{1,2}L_\out}^{(0)}=\pi/2$ initially (including 1200 DA and 1200 SA runs in total). As the LK interaction favors $\theta_{S_{1,2}L}\simeq \theta_{S_{1,2}L_\out}^{(0)}=\pi/2$, this means each individual spin will mostly lie in the orbital plane at the end of the LK cycles. 

\begin{figure}[tb]
  \centering
  \includegraphics[width=\columnwidth]{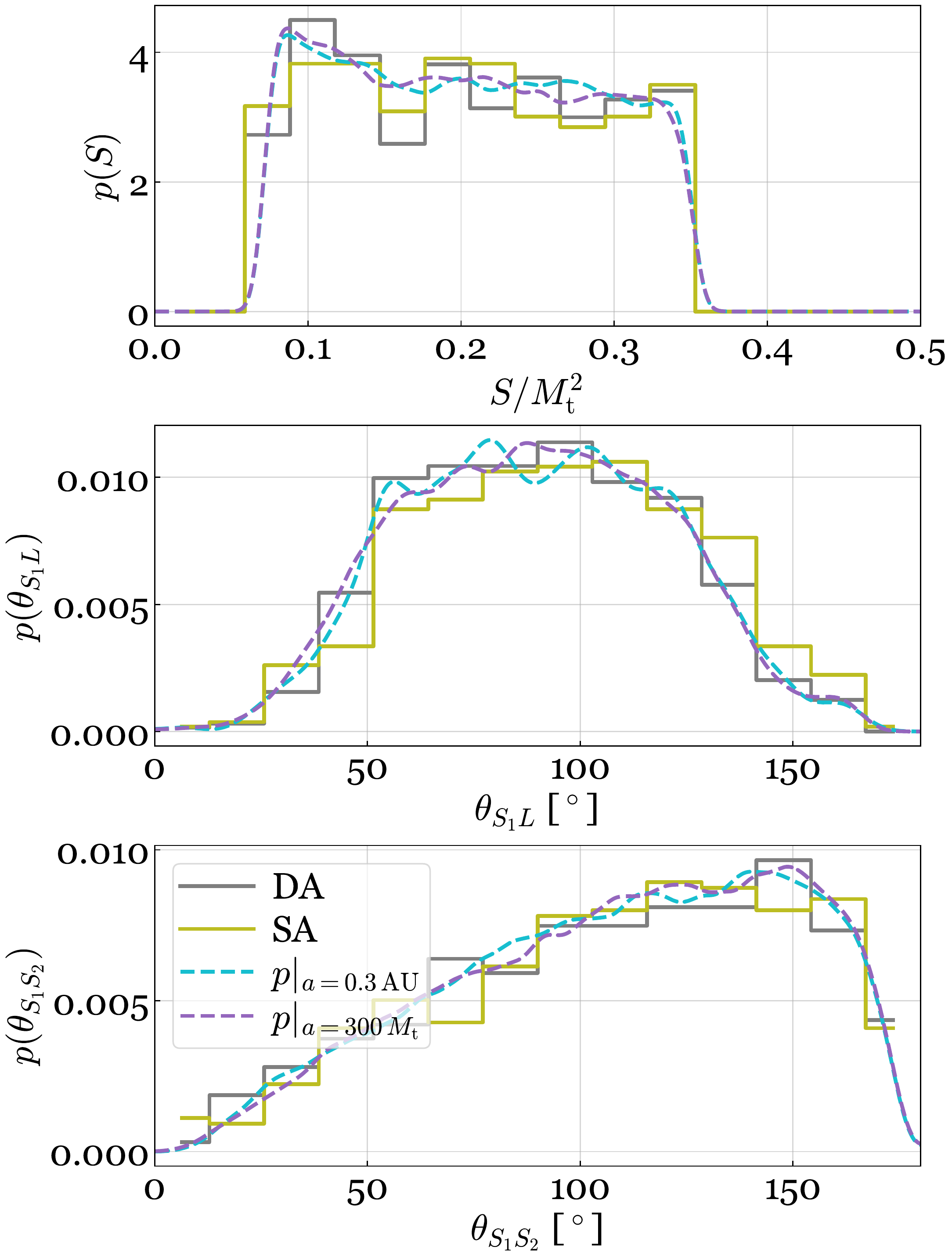}
\caption{
From top to bottom, the probability densities of the total spin magnitude $S$, the spin-orbit angle $\theta_{S_1L}$, and the spin-spin angle $\theta_{S_1S_2}$.  The solid traces are the distributions based on our numerical simulation at the end of the LK evolution ($a=0.3\,{\rm AU}$). Here we focus on those systems with $|\chi_{\rm eff}|<0.1$, which includes about 500 (450) DA (SA) systems after the cut. The dashed traces are reconstructed probability densities based on $\left\{J, L, e\right\}$ and the effective potential of $S$. The cyan traces are evaluated at $a=0.3\,{\rm AU}$ and the purple traces at $a=300\,M_{\rm t}$ (with $(J, e)$ evolved first using the precession-averaged method).}
\label{fig:chi_eff_0_S_thSS_dist_rand_thS1L}
\end{figure}

In Fig.~\ref{fig:chi_eff_0_S_thSS_dist_rand_thS1L} we show the distributions of $(S,\ \theta_{S_1L},\ \theta_{S_1S_2})$ in the (top, middle, bottom) panel, for the data set where only $|\chieff|<0.1$ is required (each individual spin vector does not necessarily lie in the orbital plane for this case). Here the solid-grey and solid-olive traces are the distributions at the end of the LK interaction (which we defined as $a=0.3\,{\rm AU}$) for those evolved numerically using the DA and SA equations. The dashed-cyan curves are the probability densities reconstructed using each individual system's $(J, L, e)$ at $a=0.3\,{\rm AU}$ according to Eqs.~(\ref{eq:p_S_vs_JLe}) and (\ref{eq:p_fS_vs_JLe}), summed together using Eq.~(\ref{eq:p_S_summed}). To get the dashed-purple traces, we first evolve the $(J, e)$ of each system as a function of $L$, using the precession-averaged method outlined in the previous Section, from $0.3\,{\rm AU}$ to $300\,M_{\rm t}\simeq 3\times10^{-4}\,{\rm AU}$, and then reconstruct the probability density. Fig.~\ref{fig:chi_eff_0_S_thSS_dist_rand_thS1L} shows that the reconstructed distribution matches well with the numerical results. Furthermore, for this data set, we do not observe a significant change in the distribution from $0.3\,{\rm AU}$ to $300\,M_{\rm t}$. Note that in the bottom panel it appears that the spins prefer to be anti-aligned. This is, however, a simple geometrical effect rather than a dynamical consequence of evolution. Intuitively, if $\vect{S}_1$ is an angle of $\alpha$ above the orbital plane, $\vect{S}_2$ needs to be at least $\alpha$ below the orbital (for $q\simeq 1$) in order to meet the $\chieff \simeq 0$ requirement. Thus the two vectors need to be at least $2\alpha$ apart, which explains why a large spin-spin angle is seemingly preferred. 

\begin{figure}[tb]
  \centering
  \includegraphics[width=\columnwidth]{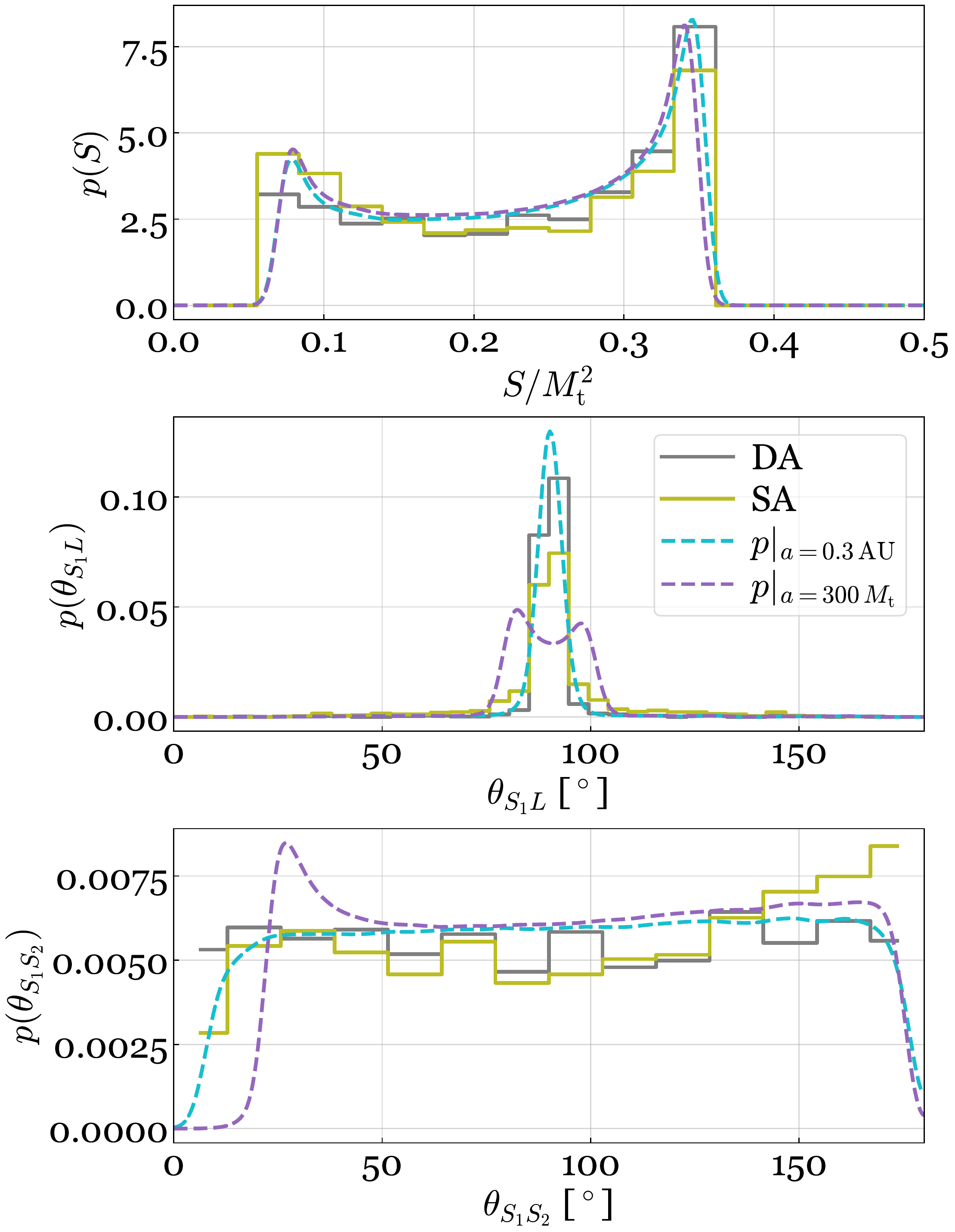}
\caption{
Similar to Fig.~\ref{fig:chi_eff_0_S_thSS_dist_rand_thS1L}, but in addition to $|\chi_{\rm eff}|<0.1$, we further require that $\theta_{S_{1(2)}L_\out}^{(0)}=\pi/2$ initially. This initial condition means that at the end of the LK evolution ($a=0.3\,{\rm AU}$), the spins vectors are approximately in the orbital plane with  $\theta_{S_{1(2)}L}\simeq \pi/2$. We specifically evolve 1200 DA and 1200 SA systems to increase the sample size here. Note that as the systems evolve from $0.3\,{\rm AU}$ to $300\,M_{\rm t}$, the distribution of $\theta_{S_1L}$ broadens and $\theta_{S_1S_2}$ begins to disfavor smaller valued angles. 
}
\label{fig:S_thSS_dist_fix_thS1L}
\end{figure}

Fig.~\ref{fig:S_thSS_dist_fix_thS1L} shows more interesting results for the evolution of the data set where we further restrict each spin to initially lie in the orbital plane. The traces of this figure retain the same definitions as those in Fig.~\ref{fig:chi_eff_0_S_thSS_dist_rand_thS1L}. As one would expect, initially $\theta_{S_1L}$ peaks at $\pi/2$ and $\theta_{S_1S_2}$ is essentially a uniform distribution. Fig.~\ref{fig:S_thSS_dist_fix_thS1L} shows that, as the system evolves, the distribution of $\theta_{S_1L}$ broadens and $\theta_{S_1S_2}$ begins to disfavor smaller values, indicating the spin-spin interaction affects the distribution. In fact, the dynamical effects are increasingly important as the inspiral continues, which we study in detail in the following Section.

\subsection{Final distribution of the spin-spin alignment}
\label{sec:fin_spin_dist}

The precession-averaged description provides an efficient way to evolve the binary when the separation is wide and we have $\tau_{\rm pre}\ll\tau_{\rm gw}$. As the orbit decays further, the separation in timescales is less well satisfied. In addition, the precession averaging ignores the spin-orbit resonances~\cite{Kesden:15, Gerosa:18, Gerosa:19}, which might become significant at small separations. As a result, from $300\,M_{\rm t}$ to $6\,M_{\rm t}$ we evolve the full set of precession equations outlined in Sec.~\ref{sec:formalism}. 

Note that as we average over precession, we do not keep track the exact value of $S$ anyone. To do the full precession-resolved evolution, we need to first reconstruct the initial conditions at $300\,M_{\rm t}$ from the averaged evolution results. This is accomplished by first randomly choosing a set $(J, L, e)$ from the numerical data at $300\,M_{\rm t}$ and sampling $S$ according to Eq.~(\ref{eq:p_S_vs_JLe}). Once $S$ is determined, we obtain the angles between different vectors according to Eqs.~(\ref{eq:th_LJ})-(\ref{eq:delPhi}), allowing us to construct the necessary vectors.\footnote{The orientation of the eccentricity vector is set by requiring $\vect{e}\cdot\vect{L}=0$. The initial angle between $\vect{e}$ and $\vect{S}$ affects only the evolution of $\uvect{e}$, not any other quantities, therefore it can be set randomly.}

\begin{figure}[tb]
  \centering
  \includegraphics[width=\columnwidth]{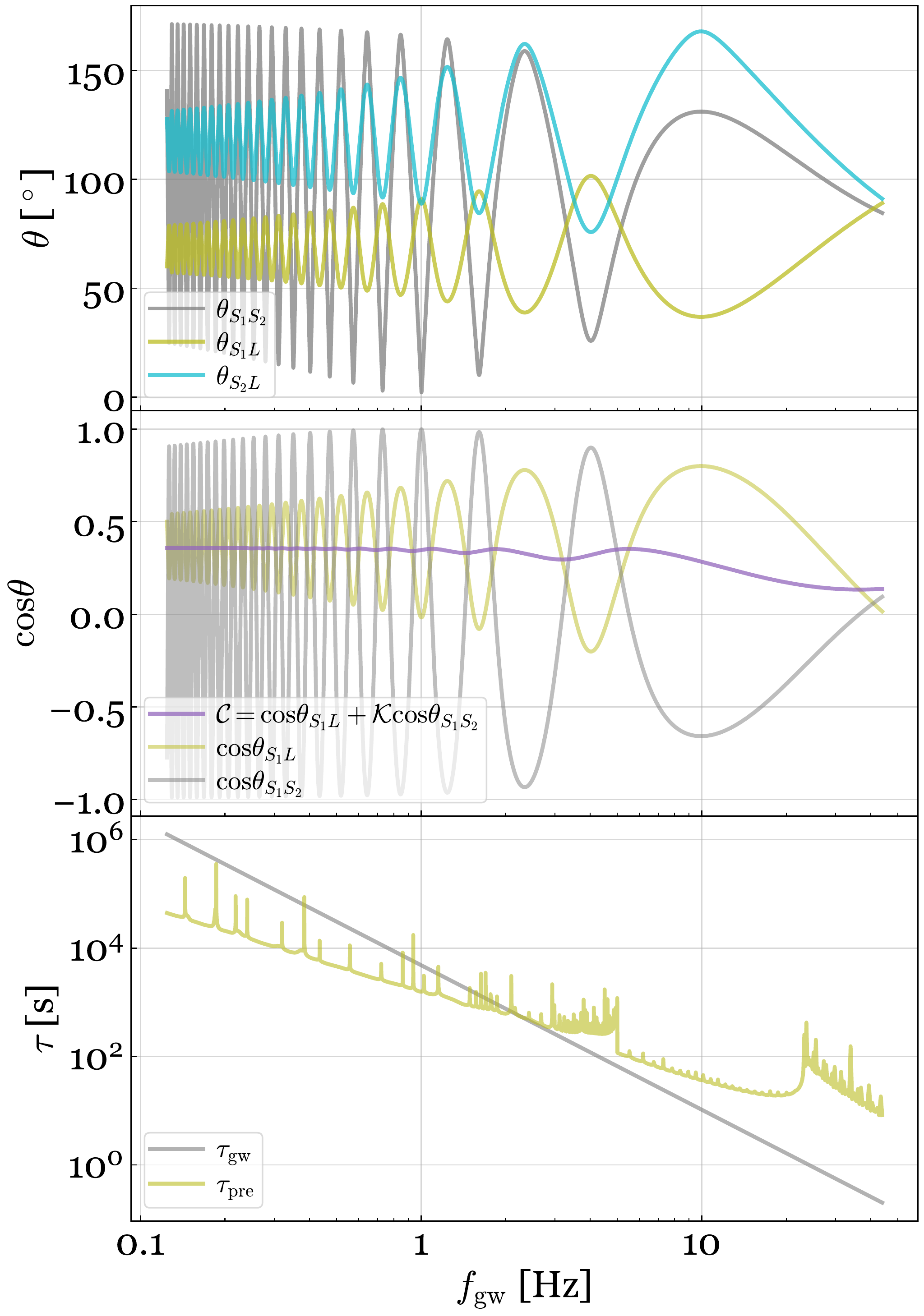}
\caption{An example of a binary evolution from $300\,M_{\rm t}$ to $6\,M_{\rm t}$ as a function of the GW frequency $f_{\rm gw}(=2f_{\rm orb}$ as the system has circularized). The top panel shows the evolution of various angles in degrees and the middle panel shows the cosine of the angles. Note that $\left(\cos \theta_{S_1L} + \mathcal{K}\cos\theta_{S_1S_2}\right)$ (purple line) stays approximately constant until a time near the merger. In the bottom panel we compare the GW decay timescale $\tau_{\rm gw}$ and the precession timescale $\tau_{\rm pre}$. }
\label{fig:bin_evol_sample}
\end{figure}

A representative evolution track from $300\,M_{\rm t}$ to $6\,M_{\rm t}$ is shown in Fig.~\ref{fig:bin_evol_sample}. In this figure, we plot different quantities as functions of the GW frequency, which is simply $f_{\rm gw}=2f_{\rm orb}$ as the eccentricity has effectively decayed away.\footnote{At $300\,M_{\rm t}$ the median eccentricity of systems in our simulation is $e=0.008$, and at $6\,M_{\rm t}$ all of the systems have $e<0.01$ . See also Figs.~\ref{fig:max_ecc_10Hz} and \ref{fig:ecc_harmonics}.} in Fig.~\ref{fig:bin_evol_sample}, from top to bottom, respectively, we show the angles between different vectors, their cosines, and the relevant timescales. Note that the precession timescale $\tau_{\rm pre}$ [Eq.~(\ref{eq:tau_pre})] can become comparable or even greater than the orbital decay timescale $\tau_{\rm gw}$ [Eq.~(\ref{eq:tau_gw})], indicating the necessity of performing a precession-resolved evolution in the last stages of the inspiral (see also Appx.~\ref{sec:bias_pS} to remove the bias that would be induced on $p(S)$ when $\tau_{\rm gw}<\tau_{\rm pre}$). In the remainder of this section, we focus in detail on the dynamics of the spin orientations.

\begin{figure*}[tb]
  \centering
  \includegraphics[width=2\columnwidth]{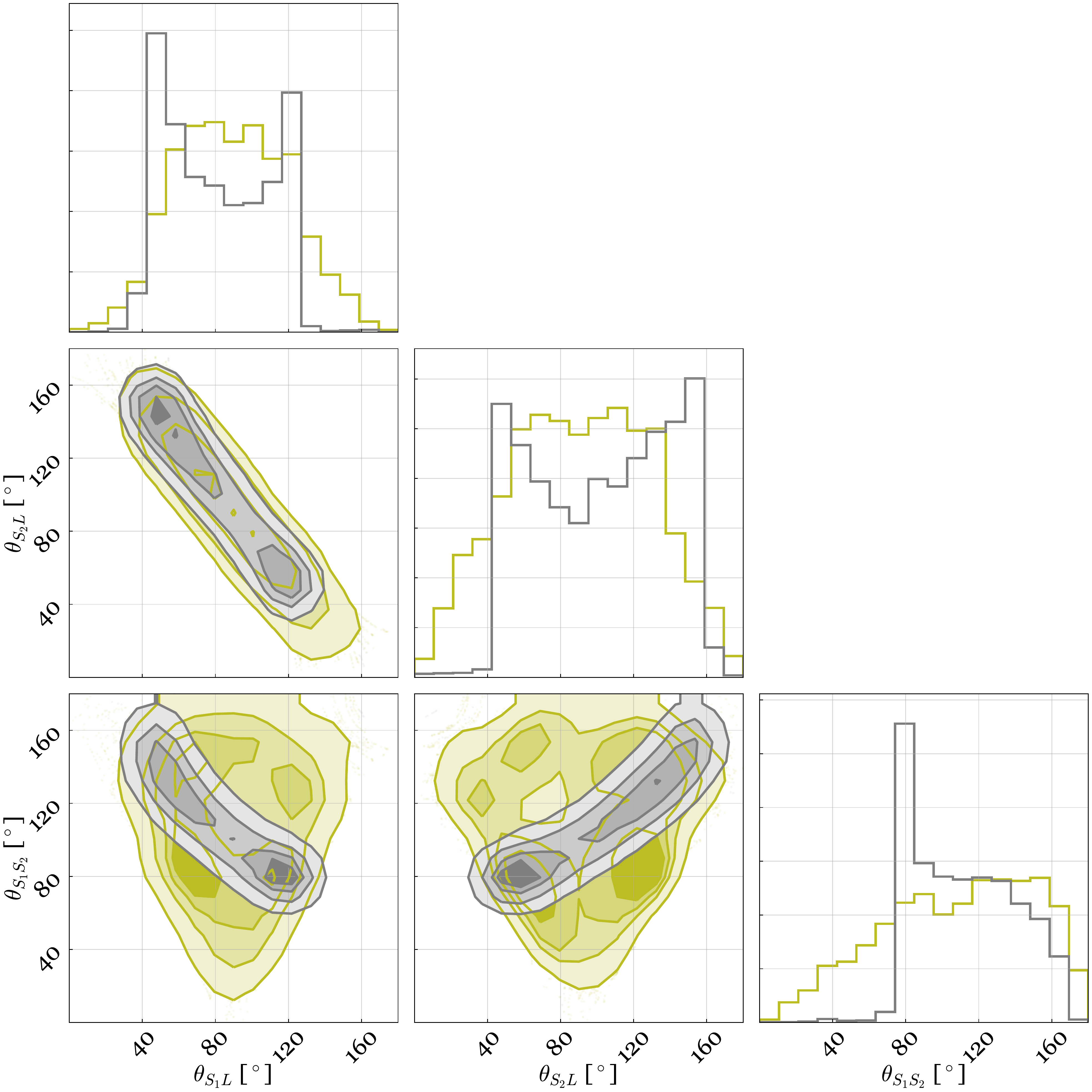}
\caption{The distribution of spin-orbit and spin-spin alignment at a separation $a=6\,M_{\rm t}$. The olive contours represent all the systems with $|\chieff| < 0.1$ after the LK evolution (the initial conditions are shown in Fig.~\ref{fig:chi_eff_0_S_thSS_dist_rand_thS1L}; including 8000 realizations in total). The grey contours further restricts the set to include only those that satisfy $\chi_{\rm 1p}\simeq \chi_{\rm 2p}\simeq 0.7$ at the end of the LK evolution (initial conditions from Fig.~\ref{fig:S_thSS_dist_fix_thS1L}; including 5000 realizations). 
%The grey traces represent the systems with $\theta_{S_{1,2}L}\simeq \pi/2$ initially (at $600\,M_\odot$) while the olive ones correspond to systems with random initial $\theta_{S_{1,2}L}$ yet still satisfy $|\chieff|<0.1$. 
Note that for the grey contours, the final spin vectors tend  with $\theta_{S_1S_2}$ disfavors strongly the aligned state, and it peaks at around $80^{\circ}$.
The grey contours also show clear correlations between different angles.
While the correlation between $\theta_{S_1L}$ and $\theta_{S_2L}$ is simply a consequence of $\chieff\simeq 0$, the interesting correlation between $\theta_{S_1L}$ and $\theta_{S_1S_2}$ is explained by the nearly conserved quantity of Eq.~(\ref{eq:c12_vs_c1_2PN}).} 
\label{fig:ang_dist_isco}
\end{figure*}

We first focus on the distributions of different angles $\theta_{S_1L}$, $\theta_{S_2L}$, and $\theta_{S_1S_2}$ at $6\,M_{\rm t}$ showm in Fig.~\ref{fig:ang_dist_isco}. There, the orange contours (including 8000 realizations) correspond to the distribution with initial conditions drawn according to Fig.~\ref{fig:chi_eff_0_S_thSS_dist_rand_thS1L}. 
%In other words, we require only $|\chieff|<0.1$ for this set of data points. 
In other words, the orange contours represent the systems starting from an isotropic spin distribution and then with the condition $|\chieff<0.1|$ imposed. 
Additionally, we show for comparison, grey contours (including 5000 realizations) corresponding to the distribution obtained from the initial condition given by Fig.~\ref{fig:S_thSS_dist_fix_thS1L}, where we further restrict the spins to be initially in the orbital plane (by setting $\theta_{S_{1}L_\out}^{(0)}=\theta_{S_{2}L_\out}^{(0)}=\pi/2$ as the initial condition for the LK evolution).

\begin{figure}[tb]
  \centering
  \includegraphics[width=\columnwidth]{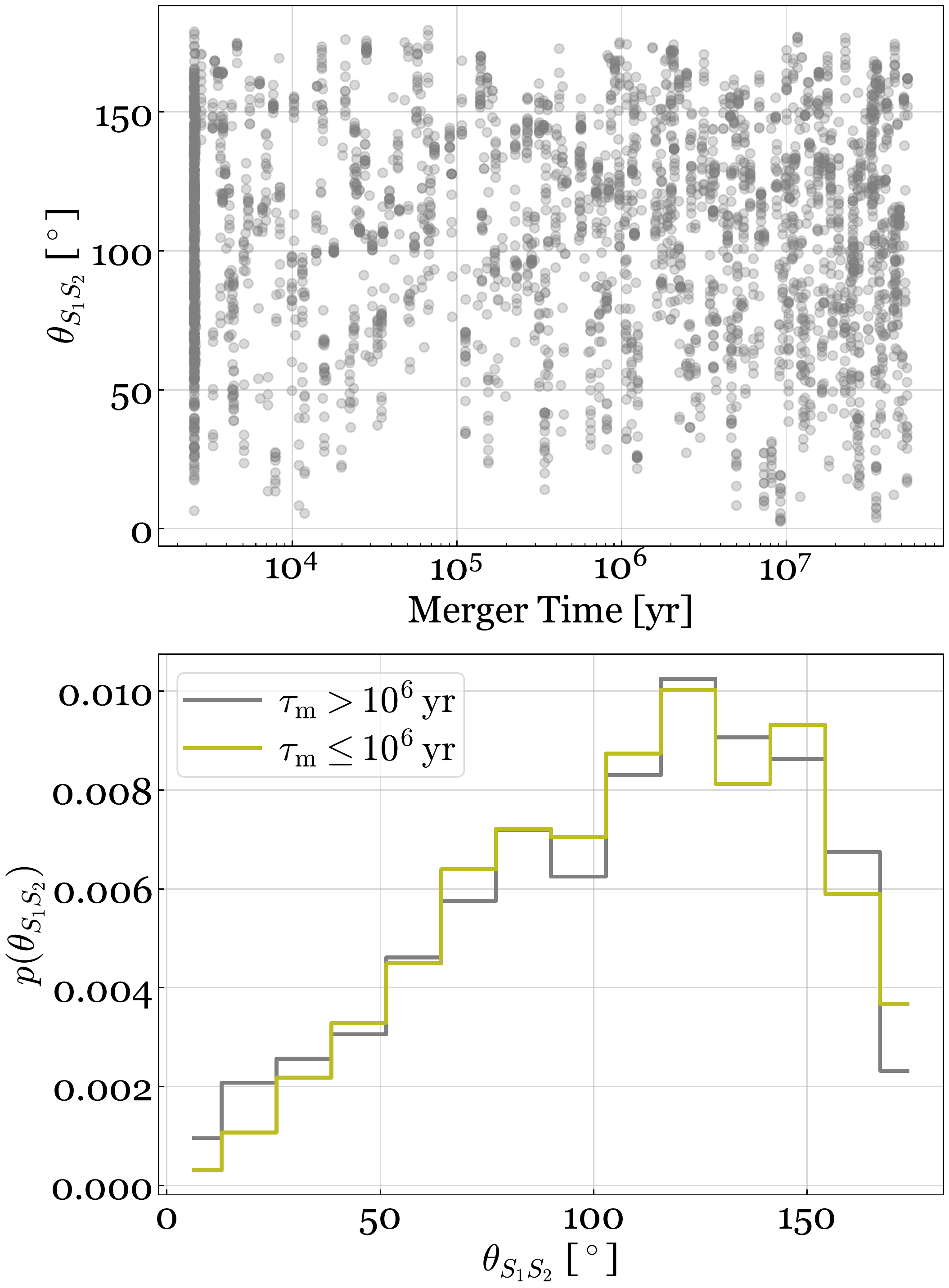}
\caption{
Top panel: scattering plot of the final angle between spin vectors $\theta_{S_1S_2}$ as a function of the merger time. 
Bottom panel: distribution of $\theta_{S_1S_2}$ for data with $\tau_{\rm m}>10^6\,{\rm yr}$ (grey) and with $\tau_{\rm m}\leq 10^6\,{\rm yr}$ (olive). 
Both plots indicate that there is no significant correlation between the spin-spin angle and the merger time. %(or maximum eccentricity the system reached during the LK evolution). 
}
\label{fig:th_12_vs_tau_m}
\end{figure}

At first glance, the orange contours appear to be similar to the initial conditions shown in Fig.~\ref{fig:chi_eff_0_S_thSS_dist_rand_thS1L}. Furthermore, we do not find a significant dependence of $\theta_{S_1S_2}$ on the merger time as shown in Fig.~\ref{fig:th_12_vs_tau_m} [see also Eq.~(\ref{eq:tau_m}). Note that the merger time is closely related to the maximum eccentricity excited by the LK mechanism]. 

\begin{figure}[tb]
  \centering
  \includegraphics[width=\columnwidth]{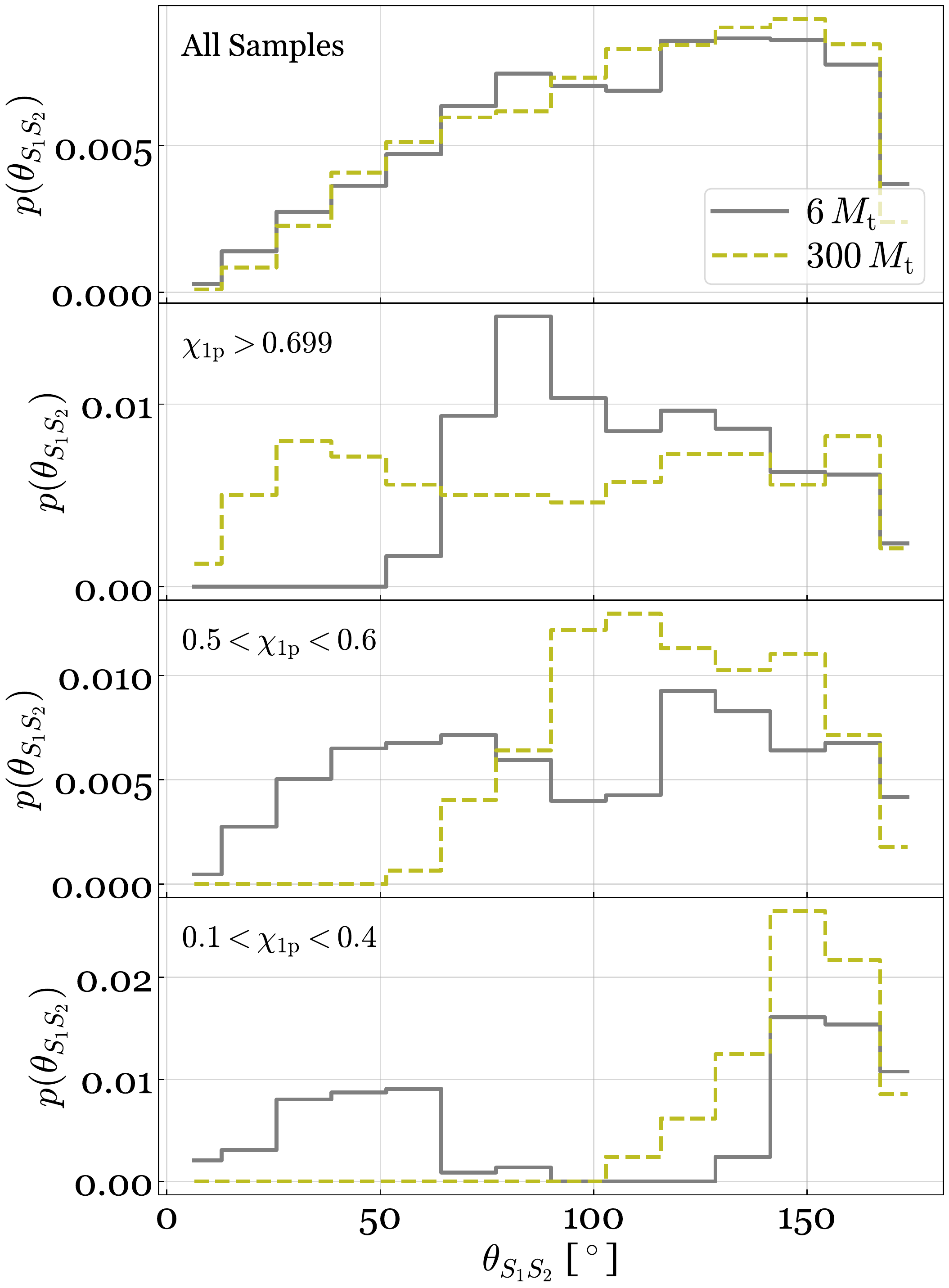}
\caption{
The distribution of $\theta_{S_1S_2}$ at $6\,M_{\rm t}$ (solid-grey traces) and at $300\,M_{\rm t}$ (dashed-olive traces). The top panel shows the distribution marginalized over $\chi_{\rm 1p}$ and the bottom three panels show the distribution corresponding to a narrow range of $\chi_{\rm 1p}$ (evaluated at $300\,M_{\rm t}$). 
}
\label{fig:th12_6M_vs_300M}
\end{figure}

Nevertheless, if we instead focus on specific slices of data, specified by a small range of values of the in-plane spin components $\chi_{\rm 1 p}\equiv \chi_1 \sin \theta_{S_1L}$,\footnote{Unlike $\chieff$ which is conserved through the evolution, the in-plane spin component $\chi_{\rm 1(2)p}$ is a time-dependent quantity. As such, we explicitly state the time at which it is evaluated whenever referring to $\chi_{\rm 1(2)p}$.} then certain evolutionary effects become clearer, as shown in Fig.~\ref{fig:th12_6M_vs_300M} (see also, e.g., Ref.~\cite{Schnittman:04}). In Fig.~\ref{fig:th12_6M_vs_300M}, we compare the angle distributions at $6\,M_{\rm t}$ (solid-grey) and at $300\,M_{\rm t}$ (dashed-olive) for different values of $\chi_{\rm 1 p}$ evaluated at $300\,M_{\rm t}$. While the olive traces are consistent with the distribution one would get by starting from an isotropic spin distribution restricted to a particular range of $\chieff$ and $\chi_{\rm 1p}$, the grey traces are nonetheless the results of dynamical interactions. Specifically, we see that for $\chi_{\rm 1p}>0.695$, the final spin vectors disfavor to be aligned, which is also demonstrated by the grey contoured data set in Fig.~\ref{fig:ang_dist_isco}. Similarly, Fig.~\ref{fig:ang_dist_isco} shows that the spin-orbit angle $\theta_{S_{1(2)L}}$ is also affected by these interactions. While the grey data set has $\theta_{S_{1(2)}L}$ peaking at $\pi/2$ initially, spins out of the orbital plane are favored at merger. More specifically, the more massive component slightly favors $\theta_{S_1L}<\pi/2$ while the less massive one favors $\theta_{S_2L}>\pi/2$.

\begin{figure}[tb]
  \centering
  \includegraphics[width=\columnwidth]{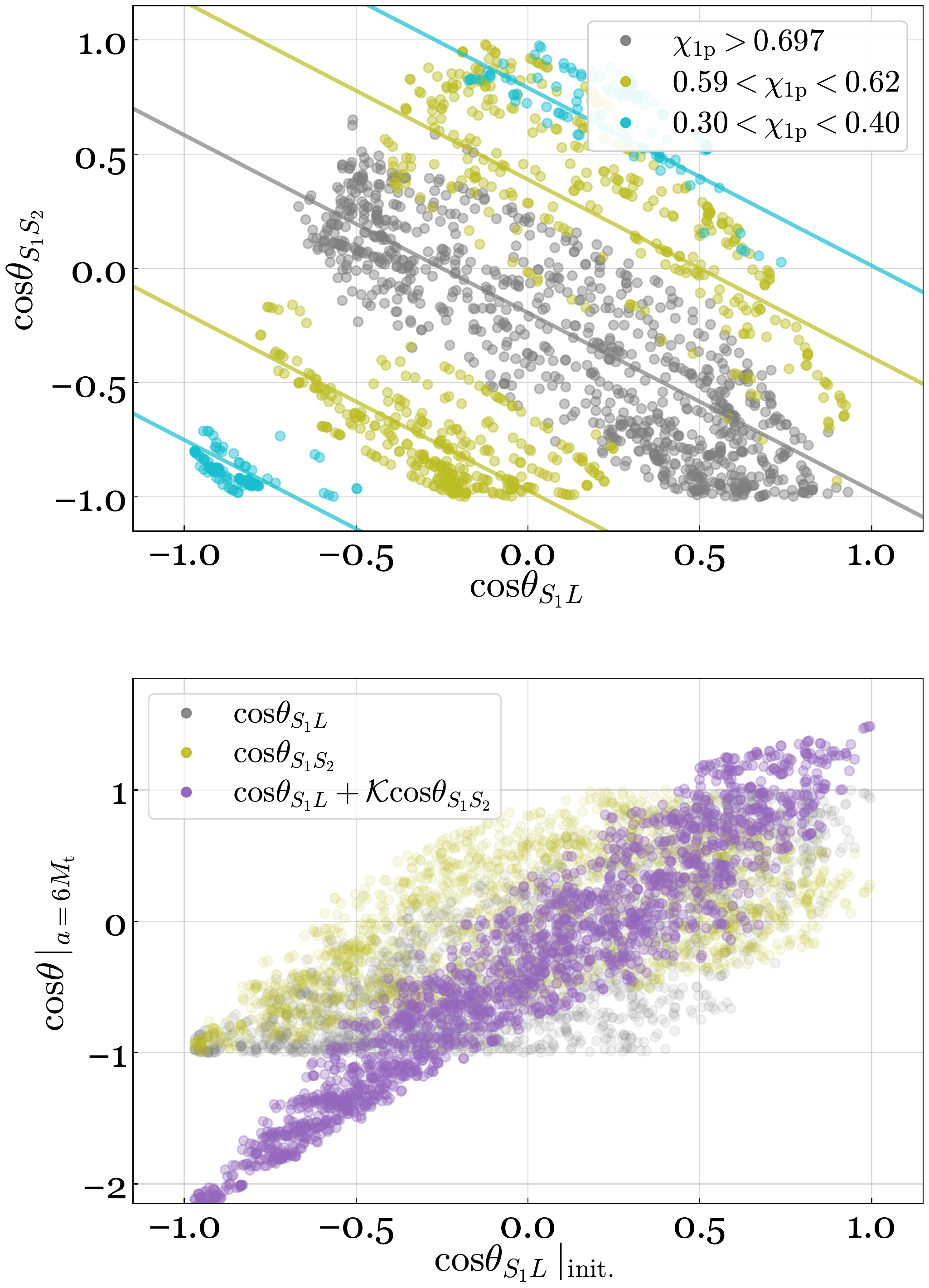}
\caption{
Top panel: a scatter plot of $\cos \theta_{S_1S_2}$ vs. $\cos \theta_{S_1L}$ at the ISCO. The points are from the olive samples in Fig.~\ref{fig:ang_dist_isco}. We color the points according to the initial values of $\chi_{\rm 1p}$ at $300\,M_{\rm t}$. Each group follows a correlation given by Eq.~(\ref{eq:c12_vs_c1_2PN}) (solid lines). 
Bottom panel: various quantities at $6\,M_{\rm t}$ as a function of the cosine of the initial ($a=300\,M_{\rm t}$) spin-orbit angle $\cos \theta_{\rm S_1L}$. Note that the quantity $\mathcal{C}$ [Eq.~(\ref{eq:c12_vs_c1_2PN})] at $6\,M_{\rm t}$ shows a clear positive correlation with respect to the initial value of $\cos\theta_{S_1L}$.
}
\label{fig:fin_vs_init_cos_dist}
\end{figure}

A closely related observation is the significant correlation between $\theta_{S_1L}$ and $\theta_{S_1S_2}$ shown by the grey contours in Fig.~\ref{fig:ang_dist_isco}. In fact, this correlation exists not only for those systems with $\chi_{\rm 1p}\simeq 0.7$, or $\theta_{S_1L}\simeq \pi/2$ initially at $300\,M_{\rm t}$, but for different values of $\chi_{\rm 1p}$ generically, as indicated in Fig.~\ref{fig:fin_vs_init_cos_dist}. 

In the top panel of Fig.~\ref{fig:fin_vs_init_cos_dist}, we show a scatter plot of $\theta_{S_1S_2}$ and $\theta_{S_1L}$. The points are colored according to the value of $\chi_{\rm 1p}$ at $300\,M_{\rm t}$. Note that each set scatters around a line corresponding to
\begin{equation}
    \mathcal{C}\equiv \cos \theta_{S_1L} + \mathcal{K} \cos\theta_{\rm S_1S_2} = {\rm Const},
    \label{eq:c12_vs_c1_2PN}
\end{equation}
where\footnote{Here we have assumed $q\neq 1$ which is the case for our simulations. The analog expression for $q=1$ is given in Appx.~\ref{sec:c12_vs_c1}, Eq.~(\ref{eq:c12_vs_c1_q1})} 
\begin{equation}
    \mathcal{K} \equiv \frac{S_2}{(1-q)L}. 
\end{equation}
Here, $\mathcal{K}$ has a well-defined value at $6\,M_{\rm t}$, as $e<10^{-2}$, and evaluates to $\mathcal{K}|_{a=6\,M_{\rm t}}=1.3$.

The above relation is a direct consequence of the fact that $J^2$ and $L^2$ are constants at 2 PN. Specifically, one may first express $\cos\theta_{S_1L}$ and $\cos\theta_{S_1S_2}$ in terms of $(J, L, S)$ using Eqs.~(\ref{eq:th_S1L}) and (\ref{eq:th_S12}), and then find a linear combination of them that eliminates $S^2$, the only variable at 2 PN. It turns out that Eq.~(\ref{eq:c12_vs_c1_2PN}) is exactly the appropriate linear combination. Hence, this relation explains the observed correlation. In fact, even when we take into account the 2.5 PN dynamics (including the decay of $J$ and $L$; see Appx.~\ref{sec:c12_vs_c1} for a detailed discussion, including the special case where $q=1$), the quantity $\mathcal{C}$ still stays approximately as a constant until the final merger.

% In fact, a similar relation holds even when we taken into account the 2.5 PN dynamics (including the decay of $J$ and $L$; see Appx.~\ref{sec:c12_vs_c1} for a detailed proof, including a discussion on the special case where $q=1$). Specifically, we have 
% \begin{align}
%     \mathcal{C} =& \left[1 + \frac{q^2}{(1+q)^2}\frac{M_{\rm t}^2\chieff}{(1-q)L}\right]\cos\theta_{S_1L}\nonumber \\ &-\frac{qS_1}{2(1-q)L}\cos^2\theta_{S_1L} +\frac{S_2}{(1-q)L}\cos\theta_{S_1S_2} \nonumber \\
%     \simeq & {\rm Const.}\label{eq:c12_vs_c1_2p5PN}
% \end{align}
% Note that when the binary is far apart with $L\gg M_{\rm t}^2\gtrsim S_{1,2}$, we have $\mathcal{C}\simeq \cos\theta_{S_1L}\simeq {\rm const}$. 

The constant nature of $\mathcal{C}$ is also demonstrated numerically in the middle panel of Fig.~\ref{fig:bin_evol_sample}, where we show Eqs.~(\ref{eq:c12_vs_c1_2PN}) in the purple trace. While both  $\cos\theta_{S_1L}$ (grey trace) and $\cos\theta_{S_1S_2}$ (olive trace) are oscillatory, the purple trace remains very well a constant until the last precession cycle ($f_{\rm gw}\gtrsim 3\,{\rm Hz}$). Close to the final merger, our assumption of Eq.~(\ref{eq:int_c1dL_approx}) breaks down, which explains the the deviation of $\mathcal{C}$ away from its constant value. 

Nevertheless, this is sufficient to explain why the top panel of Fig.~\ref{fig:fin_vs_init_cos_dist} show a clear dependence on the initial value of $\chi_{\rm 1p}$ (which determines $\sin\theta_{S_1L}$ and hence $\cos\theta_{S_1L}$). It also explains why in the bottom panel of Fig.~\ref{fig:fin_vs_init_cos_dist}, the purple dots demonstrate a clear positive correlation between $\mathcal{C}=\cos \theta_{S_1L} + \mathcal{K}\cos\theta_{S_1S_2}$ at $6\,M_{\rm t}$ and the initial value of $\cos\theta_{S_1L}$ at $300\,M_{\rm t}$.\footnote{They are not equal to each other because of the deviation shown in the middle panel of Fig.~\ref{fig:bin_evol_sample}. }

\subsection{Kick velocity distribution}
\label{sec:kick}
The angle between the two spin vectors $\theta_{S_1S_2}$ as well as its projection onto the orbital plane $\Delta \Phi$ plays a significant role in determining the final merger product. Here we consider one aspect of the merger that is influenced by spins, namely, the distribution of the GW kick velocity $v_{\rm k}$.

It has been shown that the maximum recoil velocity scales as (see, e.g, Ref.~\cite{Campanelli:07})
\begin{equation}
    \max\left[v_{\rm k, z} \right] \propto |\chi_{\rm 2p}\cos\Delta \Phi - q \chi_{\rm 1p}|,
\end{equation}
where the subscript $z$ indicates that the kick is along the direction of the orbital AM. This means an anti-aligned spin configuration (which is preferred from our spin evolution) could lead to a greater kick than the aligned case. 

To further demonstrate this point, we compute the recoil distributions for two different spin configurations. One is from our evolutionary model. Specifically, we take the olive samples from Fig.~\ref{fig:ang_dist_isco}, and further selecting those systems satisfying $\left(\chi_{\rm 1p}^2+\chi_{\rm 2p}^2\right)^2>0.68$ at $a=6\,M_{\rm t}$. As shown in the second row of Fig.~\ref{fig:th12_6M_vs_300M}, this set prefers a large angle between the two spins spins and strongly disfavors an aligned configuration. In terms of the in-plane angle $\Delta \Phi$, only $10\%$ of the systems have $\Delta \Phi<90^\circ$ after applying the $\chi_{\rm 1p}>0.68$ cut. The second set we consider is those systems with $\chieff=0$ and $\chi_{\rm 1(2)p}=\chi_{1(2)}=0.7$, and with a uniform distribution on $\Delta \Phi$. The spins are specified at $6\,M_{\rm t}$ with a randomized orbital phase and the final recoil velocity is obtained from a GW surrogate model~\cite{Varma:19, Varma:19b}.

\begin{figure}[tb]
  \centering
  \includegraphics[width=\columnwidth]{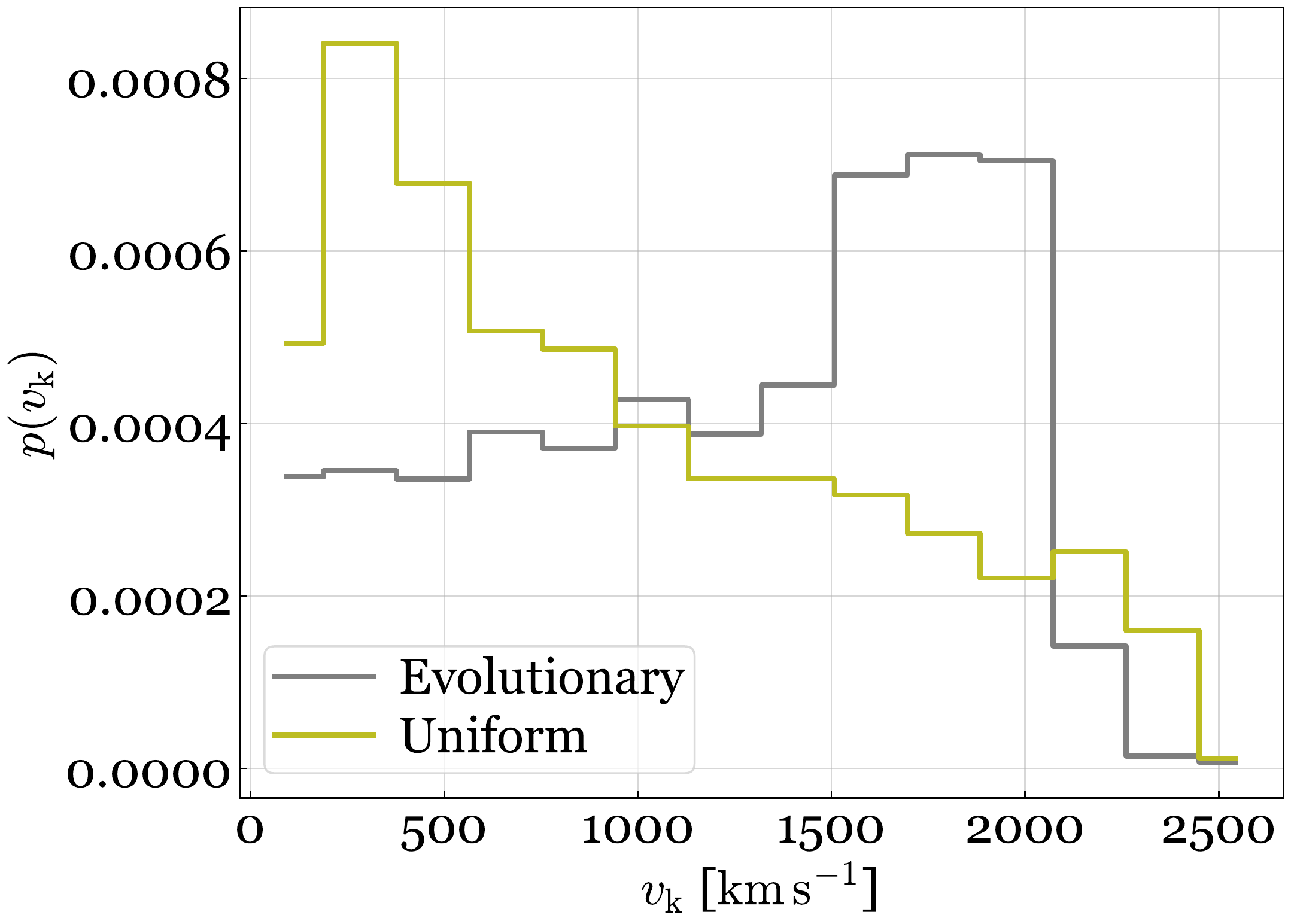}
\caption{
Distribution of the GW kick velocity $v_{\rm k}$. The grey trace is obtained from samples in Fig.~\ref{fig:ang_dist_isco} (i.e., following binary evolution), restricted to those systems with $\left(\chi_{\rm 1p}^2+\chi_{\rm 2p}^2\right)^{1/2}>0.68$ at $6\,M_{\rm t}$. For comparison, the olive trace is the kick velocity distribution for systems with the same $\chieff\simeq0$ and $\chi_{\rm p}\simeq 0.7$, but with a uniform distribution of $\Delta \Phi$ at $6\,M_{\rm t}$. While both distributions are broad and consistent with the $200\,{\rm km\,s^{-1}}$ value suggested by Ref.~\cite{Graham:20}, the evolutionary model favors a ``stronger'' kick (peaking at around $1800\,{\rm km\,s^{-1}}$) than the model with a uniform $\Delta \Phi$ prior. 
}
\label{fig:kick_dist}
\end{figure}

The result of the above procedure is shown in Fig.~\ref{fig:kick_dist}. In this figure, the grey trace corresponds to our evolutionary models and the olive trace corresponds to the reference model with uniform $\Delta \Phi$. Whereas the model with uniform $\Delta \Phi$ peaks at $v_k\simeq 250\,{\rm km\,s^{-1}}$, the evolutionary model peaks at a much higher kick velocity of $v_k\simeq 1800\,{\rm km\,s^{-1}}$. On the other hand, the evolutionary model still has a non-negligible likelihood to find a small kick velocity like the $200\,{\rm km\,s^{-1}}$ value suggested by Ref.~\cite{Graham:20}.

Lastly, we conclude this Section by re-emphasizing that whereas we focus on systems experiencing a significant LK evolution initially, the final distribution of spin-spin angle $\theta_{S_1S_2}$ holds in a more generic context. This is because the orbit has essentially circularized at $300\,M_{\rm t}$, and the final spin-spin alignment shows no obvious dependence on the eccentricity excitation (Fig.~\ref{fig:th_12_vs_tau_m}). The LK evolution simply provides an initial distribution of $\chieff$ and $\chi_{\rm 1(2)p}$. However, if certain values of $\chieff$ and $\chi_{\rm 1(2)p}$ are known (e.g., from the inspiral waveform), we can produce a relevant posterior distribution, as in Fig.~\ref{fig:kick_dist}, by restricting the systems to those consistent with the provided $\chieff$ and $\chi_{\rm 1(2)p}$ values. 

\section{Limiting eccentricity obtained during the LK oscillation}
\label{sec:max_ecc}

Having discussed the final spin distributions extensively in the previous Section, we now return to our discussions on the LK evolution, with a specific focus on the maximum achievable eccentricity. Here, we revisit the discussion in Sec.~\ref{sec:analytical_approx}, now also including the affects of dissipative GW radiation. In this Section we also examine the detectability of the orbital eccentricity by ground and space-based GW detectors.

Note that in Sec.~\ref{sec:analytical_approx} (which follows closely Refs.~\cite{Liu:15, Anderson:17, Liu:18}), we consider the limiting eccentricity for conservative systems, denoting the associated quantities with a tilde. An interesting feature of the results is that $1-\tilde{e}_{\rm lim}$ depends sensitively on the semi-major axes of both the inner and outer orbits [see Eq.~(\ref{eq:e_lim_cons})].

However, such an eccentricity is not achieved instantaneously, but instead occurs over a timescale characterized by $\tau_{\rm LK}$ [Eq.~(\ref{eq:tau_lk})]. At the same time, the eccentricity also significantly reduce the orbital decay timescale $\tau_{\rm gw}$ [Eq.~(\ref{eq:tau_gw})]. Therefore, the inner binary's eccentricity can accumulate only if $\tau_{\rm LK} < \tau_{\rm gw}$.

In fact, this timescale argument allows us to obtain the limiting eccentricity in a dissipative system by solving the equation\footnote{During the initial eccentricity excitation phase, the inner orbit's semi-major changes little and can be well approximated by its initial value $a_\imag^{(0)}$.} (see also, e.g., Ref.~\cite{Wen:03})
\begin{equation}
    \tau_{\rm gw} (e_{\rm lim}) = \tau_{\rm LK} (e_{\rm lim}). 
    \label{eq:e_lim_diss_def}
\end{equation}
In the limit $e_{\rm lim}\simeq 1$, the above equation simplifies to
\begin{align}
    % 1-e_{\rm lim}\simeq 1.9\mu^{1/3}M_{\rm t}^{5/6}M_3^{-1/3}a^{-11/6}\left(a_\out\sqrt{1-e_\out^2}\right),
    1-e_{\rm lim}&\simeq 9.1\times 10^{-5} \nonumber \\
    &\times \left(\frac{\mu}{25\,{\rm M_\odot}}\right)^{1/3}\left(\frac{M_{\rm t}}{100\,M_\odot}\right)^{5/6} \left(\frac{a_\imag^{(0)}}{3\,{\rm AU}}\right)^{-11/6} \nonumber \\
    &\times \left(\frac{M_3}{10^9\,M_\odot}\right)^{-1/3} \left(\frac{a_\out \sqrt{1-e_\out^2}}{0.06\,{\rm pc}}\right).
    \label{eq:e_lim_diss}
\end{align}
The corresponding merger timescale is now obtainable by plugging Eq.~(\ref{eq:e_lim_diss}) to Eq.~(\ref{eq:tau_m}), leading to
\begin{align}
    \tau_{\rm m, lim} &\simeq 2.5\times10^{3}\,{\rm yr}  \left(\frac{M_{\rm t}}{100\,M_\odot}\right)^{1/2}\left(\frac{a_\imag^{(0)}}{3\,{\rm AU}}\right)^{-3/2} \nonumber \\
    &\times\left(\frac{M_3}{10^9\,M_\odot}\right)^{-1} \left(\frac{a_\out \sqrt{1-e_\out^2}}{0.06\,{\rm pc}}\right)^{3}.
    \label{eq:tau_m_diss}
\end{align}
Therefore, the limiting value of $(1-e_{\rm lim})$ is now given by the maximum of Eq.~(\ref{eq:e_lim_diss}) and Eq.~(\ref{eq:e_lim_cons}). Ssimilarly, the merger timescale is given by the maximum of Eq.~(\ref{eq:tau_m_diss}) and Eq.~(\ref{eq:tau_m_lim_cons}).

\begin{figure}[htb]
  \centering
  \includegraphics[width=\columnwidth]{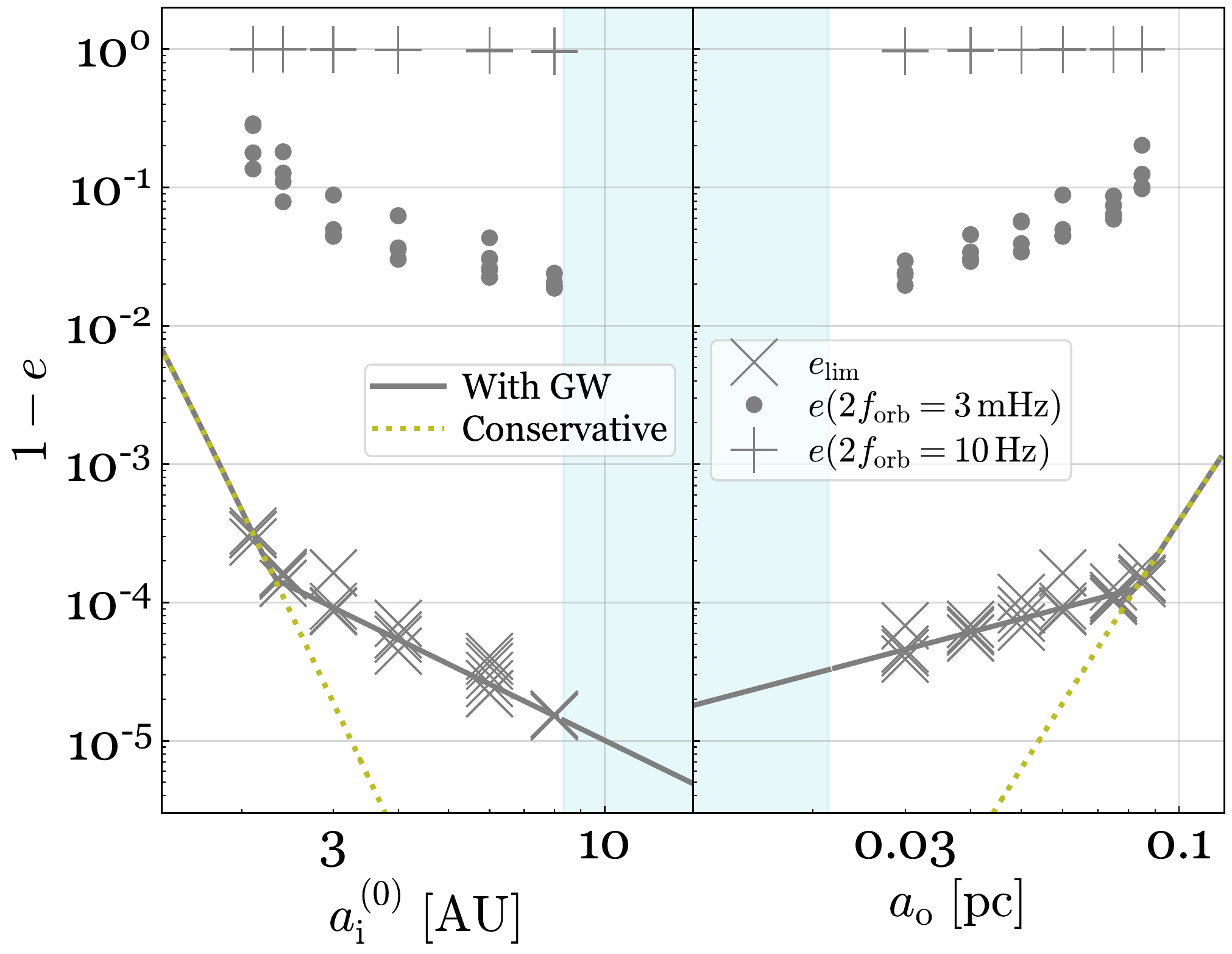}
  \caption{Limiting eccentricity achievable as a function of $a_\imag^{(0)}$ (left) and $a_\out$ (right). The limiting eccentricity during the LK process, including GW radiation, obtained numerically (crosses) are in good agreement with 
    the analytical expressions [solid-grey and dash-olive traces, corresponding respectively to Eqs.~(\ref{eq:e_lim_diss}) and (\ref{eq:e_lim_cons})]. Also shown as dots (pluses) are the eccentricity when the orbital frequency $f_{\rm orb}$ satisfies $2f_{\rm orb}=3\,{\rm mHz}$ ($2f_{\rm orb}=10\,{\rm Hz}$). The triple system has masses $(M_1, M_2, M_3)=(55, 45, 10^{9})M_\odot$, and in the left (right) plot we have fixed $a_\out=0.06\,{\rm pc}$ ($a_\imag^{(0)}=3\,{\rm AU}$). The shaded region denotes the space in which the triple system is dynamically unstable.}
\label{fig:max_ecc_vs_ai_ao}
\end{figure}

We numerically verify this result in Fig.~\ref{fig:max_ecc_vs_ai_ao} using both the DA and SA LK equations (for the SA equations, we consider 6 different initial phases of the outer orbit, each differing by $\pi/3$). The initial inclination between the inner and outer orbit is fixed at the value given by Eq.~(\ref{eq:inclination_4_max_e}). The crosses are the maximum eccentricity obtained numerically, the dotted-olive trace is the prediction for a conservative system, and the solid-grey trace corresponds to systems including GW-driven decay using Eq.~(\ref{eq:e_lim_diss_def}). Fig.~\ref{fig:max_ecc_vs_ai_ao} confirms that the timescale argument is in good agreement with the numerical results.\footnote{A caveat is that if the limiting values are set by Eqs.~(\ref{eq:e_lim_cons}) and (\ref{eq:tau_m_lim_cons}), corresponding to the cases in which the GW decay rate is always slower than the LK oscillation rate, then the use of SA equations and/or the inclusion of other effects (such as those associated with an SMBH; Sec.~\ref{sec:smbh_effects}) could exceed the bounds given by these equations. Also, for triples in the field with comparable masses, the octuple-order effects may also play a significant role. See examples from Refs.~\cite{Liu:18, Liu:19b, Liu:19}, etc.. Nonetheless, when the limiting values are set by the dissipative ones, Eq.~(\ref{eq:e_lim_diss}) and (\ref{eq:tau_m_diss}), then from the piling-up of points in, e.g., Figs.~\ref{fig:max_ecc_Tm_vs_I0} and \ref{fig:chi_eff_dist} we see that our result should still apply both when  the SA approximations is used (top panel of Fig.~\ref{fig:chi_eff_dist}) and when SMBH effects are incorporated (Fig.~\ref{fig:max_ecc_Tm_vs_I0}).\label{fn:e_caveat}}  

The limiting merger time, Eq.~(\ref{eq:tau_m_diss}), explains why in the scatter plot of Fig.~\ref{fig:chi_eff_dist} we see points piled up at a vertical line corresponding to $2.5\times 10^{3}\,{\rm yr}$ (such piling up is also seen in, e.g., fig. 3 of Ref.~\cite{Liu:18} and is explained by exactly the same reasoning). While some values of the initial inclination $I^{(0)}$ can give more extreme eccentricity excitation when the system is conservative (Eq.~\ref{eq:e_lim_cons}), once the GW decay is taken into account, the eccentricity is then limited to Eq.~(\ref{eq:e_lim_diss}). Consequently, all systems with $I^{(0)}$ in this range have the same merger time given by Eq.~(\ref{eq:tau_m_diss}). 

Note also that once the eccentricity reaches its limiting value given by Eq.~(\ref{eq:e_lim_diss_def}), the inner binary also effectively decouples from the tertiary perturber, and its eccentricity then decays monotonically according to Eq.~(\ref{eq:dedt_gw}). This allows us to explore the eccentricity at a given frequency (e.g., $2f_{\rm orb}=10\,{\rm Hz}$ with $f_{\rm orb}$ the orbital frequency) over a large range of parameter space. 

\begin{figure}[htb]
  \centering
  \includegraphics[width=\columnwidth]{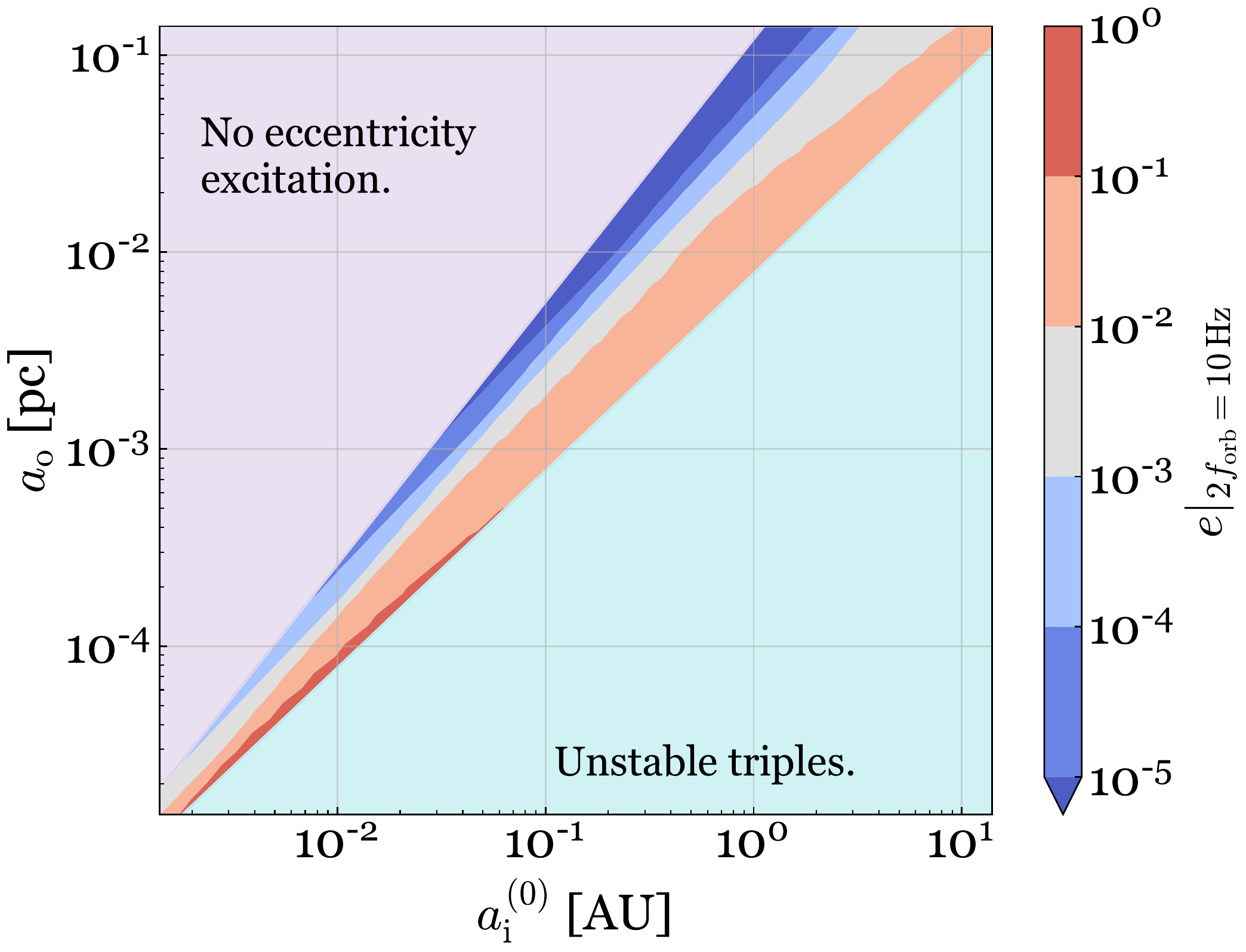}
\caption{Maximum eccentricity when the inner binary enters the LIGO band ($2f_{\rm orb}=10\,{\rm Hz}$) for a triple with $(M_1, M_2, M_3)=(55, 45, 10^{9})M_\odot$ going through the LK process.}
\label{fig:max_ecc_10Hz}
\end{figure}

One such example is shown in Fig.~\ref{fig:max_ecc_10Hz}. In Fig.~\ref{fig:max_ecc_10Hz}, we fix the triple system to have masses $(M_1, M_2, M_3)=(55, 45, 10^{9})M_\odot$ and vary the initial semi-major axes of the inner and outer orbits. We first determine the expected limiting eccentricity according to Eq.~(\ref{eq:e_lim_diss}) can be achieved through the LK process and then use $[a_\imag^{(0)}, e_{\rm lim}]$ as the initial condition for binary evolution. By solving the scalar versions of Eqs.~(\ref{eq:dLdt_gw}) and (\ref{eq:dedt_gw}) (as we do not need to follow the spin here), we can then obtain the estimated eccentricity when the inner binary enters the sensitivity band of a ground-based detector ($2f_{\rm orb}=10\,{\rm Hz}$).  

While the residual eccentricity increases as $a_\imag^{(0)}$ increases and as $a_\out$ decrease, it is unlikely to be more than $0.1$ when the binary enters LIGO's sensitivity band\footnote{Due to the caveat described in f.n.~\ref{fn:e_caveat}, we do not claim the values as absolute upper limits on the residual eccentricities. Nonetheless, they serve as decent approximations, as numerically verified in Fig.~\ref{fig:max_ecc_vs_ai_ao}.}, as to excite a greater eccentricity the triple system would be in the dynamically unstable regime~\cite{Mardling:01}. Note that this result is consistent with the pluses in Fig.~\ref{fig:max_ecc_vs_ai_ao}, where we numerically evolve the full set of equations governing the triple system. 

Furthermore, it is easy to show that the eccentricity evolution with respect to the orbital frequency, $de/df_{\rm orb}\propto e(1-e^2)/f_{\rm orb}$, is independent of the masses, yet from Eqs.~(\ref{eq:e_lim_cons}) and (\ref{eq:e_lim_diss}) we see that a massive inner binary disfavors extreme eccentricity through the LK mechanism (which is the initial condition for the binary evolution). This is also why we find smaller residual eccentricities than previous studies that focused on lighter inner binaries (see, e.g., Refs.~\cite{Wen:03, Liu:19b}). Therefore, it is unlikely for the LK mechanism to produce significant residual eccentricity for a massive binary like GW190521 when it enters the LIGO band. On the other hand, if we observe significant residual eccentricity, it would suggest the binary is likely formed via other dynamical channels (e.g., binary-single scattering~\cite{Gultekin:04, Gultekin:06, Samsing:14} or gravitational-braking~\cite{Kocsis:06, OLeary:09, Hong:15}).

Consequently, a space-based GW detector is ideal for studying the orbital eccentricity evolution at lower orbital frequencies. This idea has been studied extensively in the context of LISA (see, e.g., Ref.~\cite{Breivik:16, Giesler:2017uyu}). However, for systems reaching the limiting value [Eq.~(\ref{eq:e_lim_diss})], the eccentricity would be so high when $2f_{\rm orb}$ is in LISA's band that the orbital energy is radiated away via high-order orbital harmonics which LISA is insensitive to (see Ref.~\cite{Chen:17}). 

To demonstrate this point, we follow the approach by Ref.~\cite{Barack:04}. Specifically, we decompose the GW strain as a sum of orbital harmonics as 
\begin{equation}
    h(t) = \sum_{k=1}^{\infty} h_{k}(t),
\end{equation}
where each harmonic oscillates at $f_{\rm k} = k f_{\rm orb} + \dot{\gamma}$ with $\gamma$ the direction of the pericenter.\footnote{A circular binary only emits via the $k=2$ component, which is why we typically use $2f_{\rm orb}$ to indicate the frequency.} Each harmonic has a characteristic strain amplitude in the frequency domain, which is given by 
\begin{equation}
    h_{c, k} (f_k) = \frac{1}{\pi D_L}\sqrt{\frac{2\dot{E}_k}{\dot{f}_k}},
\end{equation}
where $\dot{E}_k$ is the GW power radiated at frequency $f_k$. We refer interested readers to Ref.~\cite{Barack:04} and references therein for the details of this calculation, while here we focus solely on the results. 

\begin{figure}[htb]
  \centering
  \includegraphics[width=\columnwidth]{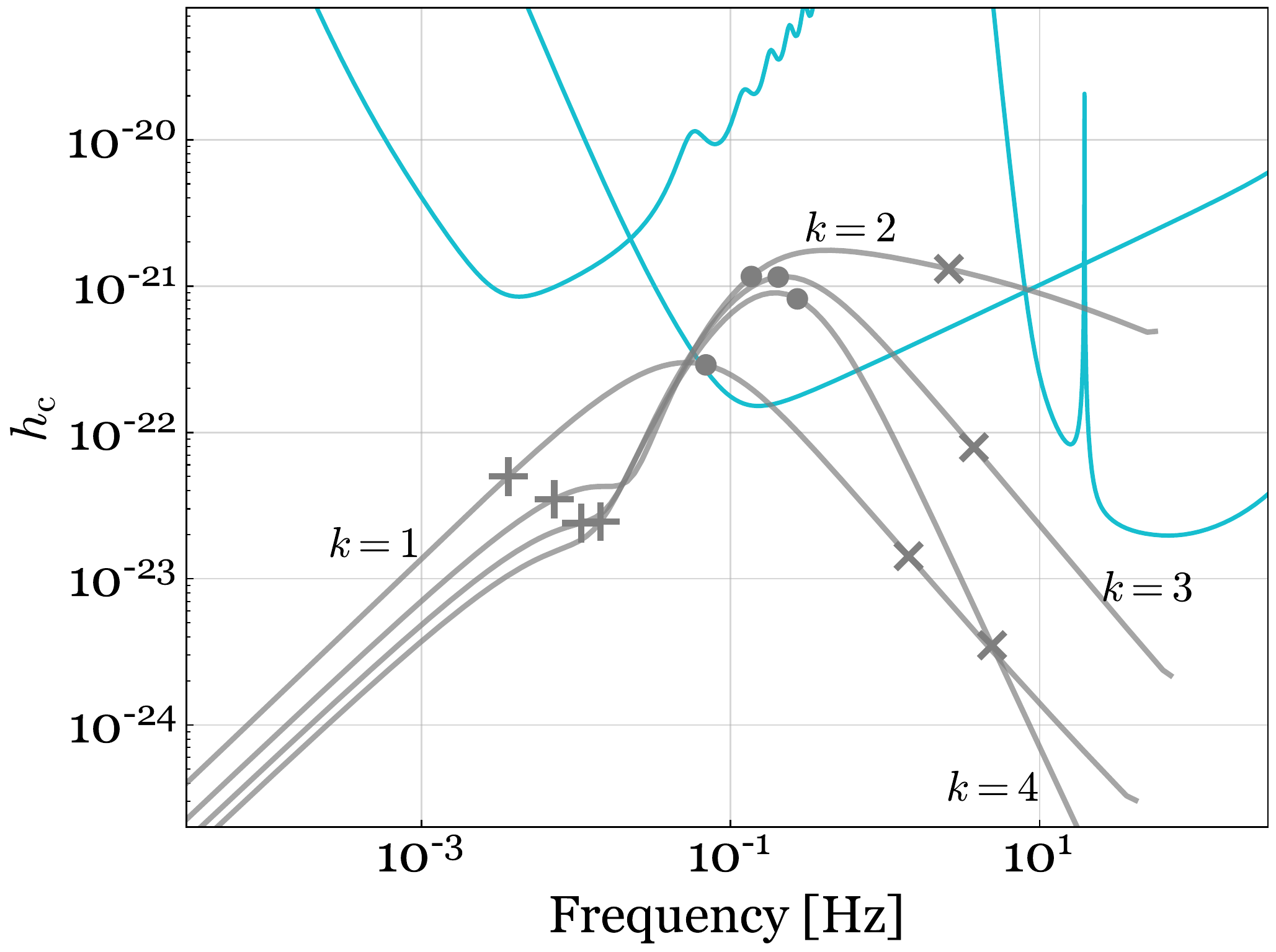}
\caption{Characteristic strain as a function of frequency for the first 4 orbital harmonics ($k=1{-}4$) for a system with $a^{(0)}=3\,{\rm AU}$, $1-e^{(0)}=10^{-4}$, and located at a cosmological redshift $z=0.44$  (corresponding to a luminosity distance $D_L\simeq2.5\,{\rm Gpc}$ assuming cosmological parameters from the Planck 2015 results~\cite{Planck:15}). The three cyan curves, from left to right, correspond to the sky-averaged sensitivities [i.e., $\sqrt{5fS_n(f)}$] of LISA, TianGO, and Voyager, respectively. We use the (plus, dot, cross) symbols to represent the instant that the binary is (1 week, 1 day, 1 hour) before the final merger.
}
\label{fig:ecc_harmonics}
\end{figure}

In Fig.~\ref{fig:ecc_harmonics} we show the evolutionary trajectories of the characteristic strain amplitudes for the first four orbital harmonics (grey traces). Here the binary is assumed to have $(M_1, M_2)=(55, 45)M_\odot$ and is at a cosmological redshift of $z=0.44$.\footnote{This is consistent with the parameters of GW190521 as reported in Ref.~\cite{Graham:20}. Note that the masses have been redshifted to $(1+z)M_{1,2}$ in the detector frame.} We further assume the binary has initial conditions of $a^{(0)}=3\,{\rm AU}$ and $1-e^{(0)}= 10^{-4}$, similar to the limiting eccentricity of the main triple system considered in this paper (see Eq.~\ref{eq:e_lim_diss}). Note that different harmonics reach the same frequency at different times, as such, we use the (plus, dot, cross) markers to represent timestamps of (1 week, 1 day, 1 hour) prior to the merger. Also shown in the plot (cyan traces), from left to right, are the sky-averaged sensitivity curves\footnote{Specifically, we plot $\sqrt{5fS_n(f)}$, where $S_n(f)$ is the power spectral density of the noise in each detector. The sky-averaged signal-to-noise ratio (SNR) for each harmonic is then ${\rm SNR}^2 = \int d\ln f \left\{h_{c, k}^2(f)/\left[5fS_n(f)\right]\right\}$. See Ref.~\cite{Barack:04}. } of LISA~\cite{Amaro-Seoane:17}, TianGO~\cite{Kuns:19}, and LIGO-Voyager~\cite{VoyagerWhite}. 

As discussed above, when the binary enters the band of a ground-based detector ($2 f_{\rm orb}\gtrsim 10\, {\rm Hz}$) only the $k=2$ harmonic has a significant amplitude, due to circularization. In the case of a milli-Hz detector, e.g. LISA, there is a potential loss of detection because when $2f_{\rm orb}$ is in the millihertz band (corresponding to the instant marked by the pluses), the GW is mostly carried away by the high-order harmonics~\cite{Farmer:03} that oscillate at frequencies above LISA's sensitivity band. However, a detector sensitive to the decihertz band, e.g. the proposed TianGO mission~\cite{Kuns:19} (middle cyan trace), could detect the evolution of these eccentric systems.

\section{Conclusion and discussions}
\label{sec:conclusion}
In this paper we studied the spin and eccentricity evolution in hierarchical triple systems via the LK mechanism, and also followed the inner binary's evolution further towards the merger. To conclude our study, we first summarize our key results in Sec.~\ref{sec:key_results}, and discuss their implications in Sec.~\ref{sec:discussion}.

\subsection{Key results}
\label{sec:key_results}

(1) We confirmed the existence of a spin attractor for systems that experience multiple ``clean'' LK cycles, as reported by Ref.~\cite{Liu:18}. However, the attraction is not towards $\chieff=0$, but it is in fact demonstrated to be $|\cos\theta_{S_{1(2)}L}| = |\cos\theta_{S_{1(2)}L_\out}^{(0)}|$ (see Fig.~\ref{fig:cThSL_vs_sThS0}). %To convert it to a preferred value of $\chieff$ would then require a detailed modeling of the initial spin orientation, which is beyond the scope of this paper. 

(2) We generalized the effective potential theory introduced by Ref.~\cite{Kesden:15} to allow for non-zero orbital eccentricity, and provided a prescription to evolve such binaries in the precession-averaged manner (Sec.~\ref{sec:eff_spin_potential}). This allows us to efficiently evolve a binary from its formation (typically with large eccentricity if the binary is formed in the dynamical channels, including the LK mechanism) to a semi-major axis of few hundred $M_{\rm t}$. 

(3) We found that the final alignment of the spin vectors are essentially independent of the maximum eccentricity excited by the LK interaction (Fig.~\ref{fig:th_12_vs_tau_m}). Instead, it depends on the initial in-plane component of the spin (Fig.~\ref{fig:th12_6M_vs_300M}). For a system with a large component spin initially lying in the orbital plane, the spin evolution significantly disfavors aligned final spins. This in fact should be true irrespective of its formation channel (whose role is to provide a prior distribution of $\chieff$ and $\chi_{\rm 1(2)p}$). 

(4) We further reported an interesting correlation between the spin-orbit and spin-spin alignments (Fig.~\ref{fig:ang_dist_isco} and \ref{fig:fin_vs_init_cos_dist}). This can be further explained by the (nearly) conserved quantities which we shown in Eq.~(\ref{eq:c12_vs_c1_2PN}) and discussed in details in Appx.~\ref{sec:c12_vs_c1}. Such a correlation could be incorporated in parameter estimation pipelines to help extract more information from detected binaries.

(5) Since the GW kick depends on the final spin-spin alignment, we found that the spin evolution may significantly affect the distribution of the kick velocity (Fig.~\ref{fig:kick_dist}). 

(6) We considered the limiting eccentricity that can be achieved by the LK mechanism in the presence of GW radiation and provided bounds derived from a timescale argument [Eq.(~\ref{eq:e_lim_diss_def}) and Fig.~\ref{fig:max_ecc_vs_ai_ao}]. For binaries in the vicinity of an SMBH, we showed that the residual eccentricity is typically small ($\lesssim 0.1$) when the binary enters a LIGO-like ground-based detector's band (Fig.~\ref{fig:max_ecc_10Hz}) for two main reasons: the triple stability requires the octuple effects to be small (Sec.~\ref{sec:formalism}), and inner binaries may be intrinsically massive [Eqs.~(\ref{eq:e_lim_cons}) and (\ref{eq:e_lim_diss})].  Furthermore, in order to capture the full orbital evolution, a decihertz detector would be necessary (Fig.~\ref{fig:ecc_harmonics}). 

\subsection{Discussion}
\label{sec:discussion}

In this study we made no attempt to predict the merger rates, given the complicated dynamics in dense stellar environments. Instead, we focused on studying the spin orientation at the end of the LK interaction. We further restricted to the leading-order (quadrupole) interactions which, according to Ref.~\cite{Liu:18}, showed the cleanest attraction of the spin vectors. If further corrections are included (see, e.g., Ref.~\cite{Liu:19}), it typically broadens the distribution of the spin-orbit angle. Nonetheless, as the spin attraction is towards the initial angle between the spin and the outer AM, we do not expect high-order corrections to significantly affect the distribution of $\chieff$ for an initially isotropic spin distribution. On the other hand, if the spins have a preferred initial orientation after takeing other astrophysical processes into account, we would then expect the LK process to shape the distribution of $\chieff$. 

While we started our discussion regarding the final spin orientations in the context of LK interactions in Sec.~\ref{sec:bin_evol}, we also considered the orientations obtained from a specific slice of $(\chieff, \chi_{1p})$. This allows our conclusions in that section to be extended to a more generic context, which hold as long as a formation channel allows for the same initial conditions. 

Specifically, the correlation between the spin-orbit and spin-spin alignments [Eq.~(\ref{eq:c12_vs_c1_2PN})] is derived based on binary PN dynamics. Such a correlation could be further used to improve parameter estimation. For example, if we could measure the angle between spin and orbit first with a space-based detector in, e.g., the decihertz band (as demonstrated in Ref.~\cite{Kuns:19}), and then again with a ground-based detector at its merger, then Eq.~(\ref{eq:c12_vs_c1_2PN}) and Fig.~\ref{fig:fin_vs_init_cos_dist} indicate the final spin-spin angle, $\theta_{S_1S_2}$, is no longer a free parameter to be inferred, but can in fact be constrained by the evolution from the lower-frequency measurement.
With a better constrained $\theta_{S_1S_2}$, it could further improve our prior on, e.g., the GW recoil velocity. These ideas provide much to explored in future studies.
%many conclusions we drew in Sec.~\ref{sec:bin_evol} in fact holds in more generic contexts, as they only require the formation channel to provide the same slice of $(\chieff, \chi_{1(2)p})$.

\section*{ACKNOWLEDGMENTS}
We thank Nathan Johnson-McDaniel for pointing out an error that the quadrupole-monopole interaction was not properly accounted for in an earlier version of the manuscript. We also thank Fabio Antonini, Linqing Wen, Ling Sun, Ka-Lok Rico Lo, Dong Lai, Yubo Su, and the referee for useful comments and discussions. HY and MG are supported by the Sherman Fairchild foundation.  SM and YC are supported by the Brinson Foundation and the Simons Foundation (Award Number 568762). SM, MG, and YC are additionally supported by the National Science Foundation (Grants PHY-1708212 and PHY-1708213). The
authors also gratefully acknowledge the computational
resources provided by the LIGO Laboratory and supported by NSF grants PHY-0757058 and PHY-0823459.

\appendix
\section{Explicit equations of motion}
\label{sec:explicit_LK_eqs}
In this section we provide the explicit equations of motion of the LK interaction for both the DA and SA approximations. Consistent with the main text, here we truncate to the quadrupole order. The octupole-order terms are available in, e.g., Refs.~\cite{Liu:15} and \cite{Liu:18} for DA and SA approximations, respectively. 

To obtain the DA LK evolution, we integrate $(\vect{L}, \vect{e}, \vect{L}_\out, \vect{e}_\out)$, 
\begin{align}
    &\frac{d\vect{L}}{dt}|_{\rm LK} = \frac{L\Omega_{\rm DA}}{\sqrt{1-e^2}}  \left[(1-e^2)\left(\uvect{L}\cdot \uvect{L}_{\out}\right)\uvect{L}\times\uvect{L}_\out \right. \nonumber \\
    &\quad\left. -5\left(\vect{e}\cdot\uvect{L}_\out\right)\vect{e}\times \uvect{L}_{\out} \right], \\
    &\frac{d\vect{e}}{dt}|_{\rm LK}  = 
    \Omega_{\rm DA}\sqrt{1-e^2}
    \left[(\uvect{L}\cdot\uvect{L}_\out)\vect{e}\times\uvect{L}_\out  \right.\nonumber \\ 
    &\quad\left. +2\uvect{L}\times \uvect{e}-5\left(\vect{e}\cdot\uvect{L}_\out\right)\uvect{L}\times\uvect{L}_\out\right],\\ 
    &\frac{d\vect{L}_\out}{dt}|_{\rm LK} = \frac{L\Omega_{\rm DA}}{\sqrt{1-e_\out^2}}\left[(1-e^2)\left(\uvect{L}\cdot \uvect{L}_{\out}\right)\uvect{L}_\out\times \uvect{L} \right. \nonumber \\
     &\quad\left. -5\left(\vect{e}\cdot\uvect{L}_\out\right)\uvect{L}_\out\times \vect{e} \right], \\
    &\frac{d\vect{e}_\out}{dt}|_{\rm LK} = \frac{L\Omega_{\rm DA}}{L_\out\sqrt{1-e_\out^2}} \nonumber \\
    &\times \left\{-5\left(\vect{e}\cdot\uvect{L}_\out\right)\vect{e}_\out\times \vect{e}  + (1-e^2)\left(\uvect{L}\cdot \uvect{L}_{\out}\right)\vect{e}_\out\times \uvect{L}\right.\nonumber \\
    &\hspace{-0.3cm}\left.-\left[\frac{1}{2}{-}3e^2{+}\frac{25}{2}\left(\vect{e}\cdot\uvect{L}_\out\right)^2 {-} \frac{5(1{-}e^2)}{2}\left(\uvect{L}\cdot\uvect{L}_\out\right)^2\right]\uvect{L}_\out{\times}\vect{e}_\out\right\},
\end{align}
where 
\begin{equation}
    \Omega_{\rm DA} = \frac{3}{4}\left(\frac{M_3}{M_1+M_2}\right)\left(\frac{a}{a_\out\sqrt{1-e_\out^2}}\right)^3\Omega_{\rm orb}, 
\end{equation}

The SA LK evolutions are solved in terms of $(\vect{L}, \vect{e}, \vect{r}_\out, d\vect{r}_\out/dt)$,
\begin{align}
    &\frac{d\vect{L}}{dt}|_{\rm LK} = \frac{L\Omega_{\rm SA}}{\sqrt{1-e^2}}  \left[- (1-e^2)\left(\uvect{L}\cdot \uvect{r}_{\out}\right)\uvect{L}\times\uvect{r}_\out \right. \nonumber \\
    &\quad\left.+5\left(\vect{e}\cdot\uvect{r}_\out\right)\vect{e}\times \uvect{r}_{\out} \right], \\
    &\frac{d\vect{e}}{dt}|_{\rm LK}  = 
    \Omega_{\rm SA}\sqrt{1-e^2}
    \left[-(\uvect{L}\cdot\uvect{r}_\out)\vect{e}\times\uvect{r}_\out  \right.\nonumber \\ 
    &\quad\left.-2\uvect{L}\times \uvect{e}+5\left(\vect{e}\cdot\uvect{r}_\out\right)\uvect{L}\times\uvect{r}_\out \right],\\ 
    &\frac{d^2\vect{r}_\out}{dt^2} =-\Phi_{\out} \left(\frac{\uvect{r}_\out}{r_\out}\right) \nonumber \\
    &-\Phi_{\rm Q}\left\{-3\left(\frac{\uvect{r}_\out}{r_\out}\right) \right.\nonumber \\
    &\left.\times \left[-1+6e^2+3(1-e^2)\left(\uvect{L}\cdot\uvect{r}_\out\right)-15\left(\vect{e}\cdot\uvect{r}_\out\right)\right]\right.\nonumber \\
    &\quad+6\frac{1-e^2}{r_\out}\left.\left(\uvect{L}\cdot\uvect{r}_\out\right) \left[\uvect{L}  -\left(\uvect{L}\cdot\uvect{r}_\out\right) \left.\uvect{r}_\out\right. \right]  \right.\nonumber \\
    &\quad-30\left.\frac{\left(\vect{e}\cdot\uvect{r}_\out\right)}{r_\out} \left[\vect{e}  - \left(\vect{e}\cdot\uvect{r}_\out\right) \left.\uvect{r}_\out\right.  \right] \right\},
\end{align}
where in the above equations we have defined 
\begin{align}
    &\Omega_{\rm SA} = \frac{3}{2}\left(\frac{M_3}{M_1+M_2}\right)\left(\frac{a}{r_\out}\right)^3\Omega_{\rm orb}, \\
    &\Phi_{\out}=\frac{(M_1+M_2+M_3)}{ r_\out}, \\
    &\Phi_{\rm Q}=\frac{1}{4}\frac{M_3}{r_\out}\left(\frac{\mu}{\mu_\out}\right)\left(\frac{a}{r_\out}\right)^2. 
\end{align}
\section{Deriving $dS/dt$ for eccentric orbits }
\label{sec:dSdt}
\begin{align}
    \frac{dS}{dt} &= \frac{1}{2S}\frac{d\left(\vect{S}\cdot \vect{S}\right)}{dt} 
    =\frac{1}{S}\frac{d\left(\vect{S}_1\cdot\vect{S}_2\right)}{dt} \nonumber \\
    & = \frac{1}{S}\left(\frac{d\vect{S}_1}{dt}\cdot \vect{S}_2 + \vect{S}_1 \cdot \frac{d\vect{S}_2}{dt} \right),
\end{align}
where we have used $\vect{S} = \vect{S}_1 + \vect{S}_2$ and the magnitudes $S_1$, $S_2$, $S_1^2$, $S_2^2$ are constants. Now plugging in Eqs.~(\ref{eq:dS1_v_dt}) and (\ref{eq:omega_dS})-(\ref{eq:omega_QM}) we have,
\begin{align}
    &\frac{d}{dt} \left(\vect{S}_1\cdot \vect{S}_2\right)\nonumber \\
    &=-\frac{3}{2a^3 (1-e^2)^{3/2}}\frac{1-q^2}{q}\left(1 - \frac{M_1M_2\chieff}{L}\right)\nonumber \\
    &\quad \times \vect{S}_1\cdot \left(\vect{S}_2 \times \vect{L}\right) \nonumber \\
    &=-\frac{3}{2}\eta^6\left(1-e^2\right)^{3/2}\left(\frac{M_{\rm t}^2}{L}\right)^5\frac{S_1S_2}{M_t}
    \frac{\left(1-q^2\right)}{q}
    \nonumber \\
    &\ \ \times \left(1-\frac{\eta M_{\rm t}^2\chieff}{L}\right)\uvect{S}_1\cdot \left(\uvect{S}_2\times \uvect{L}\right),
    \label{eq:d_S1dotS2_dt}
\end{align}
where we have replace the semi-major axis $a$ in terms of $(L, e)$. 

Further note that geometrically we have 
\begin{equation}
    \uvect{S}_1\cdot \left(\uvect{S}_2\times \uvect{L}\right) = \sin \theta_{S_1 L } \sin\theta_{S_2 L }\sin\Delta \Phi,
\end{equation}
with the angles given by Eqs.~(\ref{eq:th_S1L}), (\ref{eq:th_S2L}), and  (\ref{eq:delPhi}), and each angle is a function of $(J, L, S)$. 
Similarly, we can write $\vect{S}_1 \cdot \vect{L}$ and $\vect{S}_2 \cdot \vect{L}$ in terms of $\cos \theta_{S_1 L}(J, L, S)$ and $\cos \theta_{S_2L}(J, L, S)$. %we nonetheless dropped them in Eq.~(\ref{eq:dSdt}) as they are suppressed by a factor $S_1/L \sim S_2/L \ll (1-q^2)$ over the range we are interested in. 

\section{Bias in the spin distribution when $\tau_{\rm gw}< \tau_{\rm pre}$.}
\label{sec:bias_pS}
% \begin{figure}[tb]
%   \centering
%   \includegraphics[width=\columnwidth]{timescale_300.pdf}
% \caption{Comparison between the GW-decay timescale $\tau_{\rm gw}$ and spin-precession timescale $\tau_{\rm pre}$ at $a=300\,M_{\rm t}$. As $\tau_{\rm gw}\gg \tau_{\rm pre}$, it justifies our usage of the precession-averaged equations to evol the system to $a=300\,M_{\rm t}$.}
% \label{fig:timescale_300}
% \end{figure}

\begin{figure}[tb]
  \centering
  \includegraphics[width=\columnwidth]{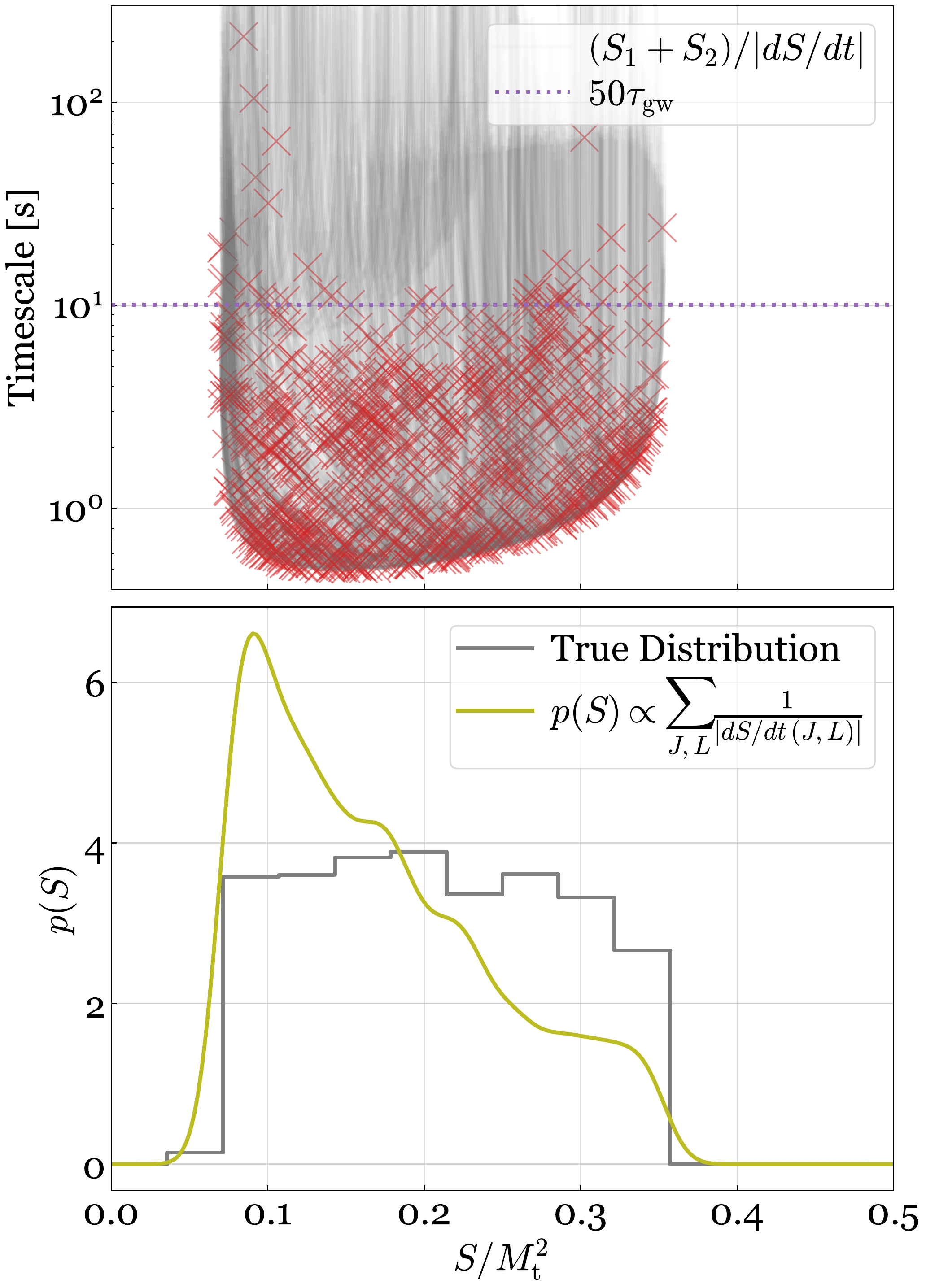}
\caption{Top panel: timescale comparison at the ISCO. Each grey trace represents the allowed instantaneous precession time based on the final $J$ and $L$ using the effective potential theory, and the red cross is the true value we obtained in the simulation. The purple-dotted line is the fifty times the GW decay timescale at $6\,M_{\rm t}$. Bottom panel: the true distribution of $S$ at the $6\,M_{\rm t}$ (grey-solid trace). If we simply assign $p(S|J, L)\propto 1/|dS/dt(J, L)|$ as we have done for Figs.~\ref{fig:chi_eff_0_S_thSS_dist_rand_thS1L} and \ref{fig:S_thSS_dist_fix_thS1L}, we would obtain a probability density function described by the olive trace, which is significantly biased relative to the true distribution.}
\label{fig:S_dist_isco}
\end{figure}

In the main text we have used Eqs.~(\ref{eq:p_S_vs_JLe})-(\ref{eq:p_fS_vs_JLe}) to generate the probability density functions of various quantities, and as shown in Figs.~\ref{fig:chi_eff_0_S_thSS_dist_rand_thS1L} and \ref{fig:S_thSS_dist_fix_thS1L}, our method reproduces well the distribution obtained from numerical simulations. However, this method can only be applied when we have $\tau_{\rm pre}\ll \tau_{\rm gw}$, and we demonstrate here the potential bias that would be induced when the timescale requirement is not satisfied. 

Specifically, we can repeat the process we have used in generating Fig.~\ref{fig:chi_eff_0_S_thSS_dist_rand_thS1L} for data at $6\,M_{\rm t}$ (the olive data in Fig.~\ref{fig:ang_dist_isco}). The reconstructed probability of the total spin magnitude $S$ is shown in the olive trace in the bottom panel of Fig.~\ref{fig:S_dist_isco}. As a comparison, the true distribution from the numerical data is shown in the grey trace. Clearly, the reconstructed probability is biased towards small $S$.

To examine things in more details, we also show the instantaneous precession time which we define as $\left(S_1+S_2\right)/|dS/dt|$ and compare it with the GW decay timescale in the top panel of Fig.~\ref{fig:S_dist_isco}. Here each grey trace is generated with a set of $(J, L, e)$ and the full range of $S$ allowed by the effective potential, and each red cross is the true value of $S$ obtained from the evolution. While the instantaneous precession time can be hundreds of times longer than $\tau_{\rm gw}$, and according to Eq.~(\ref{eq:p_S_vs_JLe}) those locations should be more likely to be sampled, we nonetheless see that the majority of the realization actually happens in the region where the precession time is less than $50\tau_{\rm gw}$ (dotted-purple line). 

Consequently, we conclude that while the effective potential theory is still valid at $6\,M_{\rm t}$, it cannot be used to directly predict the likelihood that the condition $\tau_{\rm pre}\ll \tau_{\rm gw}$ is not satisfied. To obtain a faithful distribution, a full numerical simulation over a large ensemble would thus be necessary.

\section{Understanding the correlation between $\cos \theta_{S_1 L}$ and $\cos \theta_{S_1 S_2}$}
\label{sec:c12_vs_c1}
In this section we study the dynamical relations between the spin-orbit angle $\theta_{S_1 L}$ and the spin-spin angle $\theta_{S_1S_2}$. The goal is to better understand the correlations shown in, e.g., Fig.~\ref{fig:ang_dist_isco}, and the nearly constant quantities shown in the middle panel of Fig.~\ref{fig:bin_evol_sample}.  Note that our derivations here do not assume a circular orbit, but holds generically for eccentric orbits as well. 

We have\footnote{Note that if $d\vect{L}/dt = \vect{\Omega}\times \vect{L} + \left(dL/dt\right) \uvect{L}$, then $d\uvect{L}/dt = (d\vect{L}/dt)/L - \uvect{L} \left(dL/dt\right)/L =\vect{\Omega}\times \uvect{L}$. } 
\begin{align}
    &\frac{d}{dt} \cos \theta_{S_1 L}
    = \frac{d }{dt }\left(\uvect{L}\cdot \uvect{S_1}\right) 
    = \frac{d \uvect{L}}{dt}\cdot \uvect{S}_1 +\uvect{L}\cdot\frac{d\uvect{S}_1}{dt},
%     \nonumber \\
% = & \left\{\left[ \vect{\Omega}_{\rm dS,br}^{(S_1)} + \vect{\Omega}_{\rm dS,br}^{(S_2)} + \vect{\Omega}_{\rm LT, br}\right]\times \uvect{L}\right\}\cdot \uvect{S}_1 
%   % -\frac{dL/dt}{L}\uvect{L}\cdot \uvect{S}_1
%     \nonumber \\
%     &+ \uvect{L}\cdot \left\{\left[\vect{\Omega}_{\rm dS}^{(S_1)} + \vect{\Omega}_{\rm LT}^{(S_1)} \right]\times \uvect{S}_1\right\},
    \label{eq:dc_th_S1L_def}
\end{align}
leading to
\begin{align}
    &\frac{d}{dt} \cos \theta_{S_1 L} 
    = \frac{3S_2(1+q)}{2a^3(1-e^2)^{3/2}q} \nonumber \\
    &\times\left[1 - \frac{M_1M_2\chieff}{L}\right]
    \uvect{S}_1 \cdot \left(\uvect{S}_2 \times \uvect{L}\right).
\end{align}

Meanwhile, from Eq.~(\ref{eq:d_S1dotS2_dt}) we have 
\begin{align}
    &\frac{d}{dt} \cos \theta_{S_1 S_2} 
    = \frac{-3L(1-q^2)}{2a^3(1-e^2)^{3/2}q} \nonumber \\
    &\times\left[1-\frac{M_1M_2\chieff}{L}\right] 
     \uvect{S}_1 \cdot \left(\uvect{S}_2 \times \uvect{L}\right), 
     \label{eq:dc12dt}
\end{align}
which has a similar form as $d\cos\theta_{S_1L}/dt$. 

Therefore, we have
% \begin{align}
%     &(1-q)\left(1 - \frac{\eta M_{\rm t}^2\chieff}{L}\right) \frac{d}{dt}\cos\theta_{S_1L} \nonumber\\
%     & +\frac{S_2}{L}\left(1-\frac{M_1M_2\chieff}{L}\right) \frac{d}{dt}\cos\theta_{S_1S_2} = 0. \label{eq:c12_vs_c1_d_dt}
% \end{align}
\begin{equation}
    (1-q)\frac{d}{dt}\cos\theta_{S_1L} + \frac{S_2}{L} \frac{d}{dt}\cos\theta_{S_1S_2}.
    \label{eq:c12_vs_c1_d_dt}
\end{equation}
If we treat $L$ as a constant first, we then have
\begin{equation}
    (1-q)\cos\theta_{S_1L} + \frac{S_2}{L}\cos\theta_{S_1S_2} = {\rm Const}. 
    \label{eq:c12_vs_c1_2PN_appx}
\end{equation}
This is also the relation we present in Eq.~(\ref{eq:c12_vs_c1_2PN}). 

\begin{figure}[tb]
  \centering
  \includegraphics[width=\columnwidth]{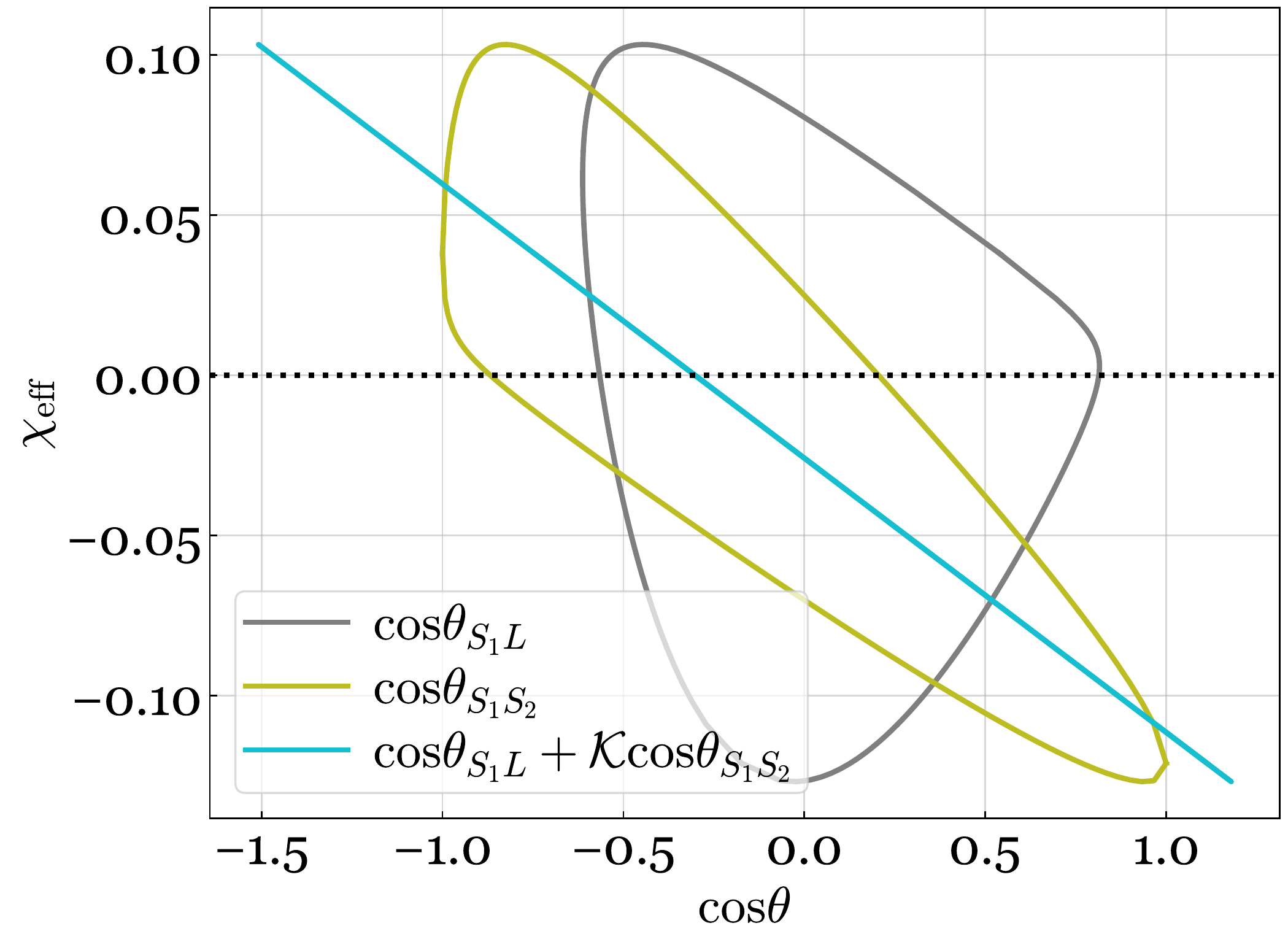}
\caption{Effective potentials at $(J, L, S, e) = (0.65\, M_{\rm t}^2, 0.61\,M_{\rm t}^2,  0.20\,M_{\rm t}^2, 0)$. }
\label{fig:eff_potential}
\end{figure}

As we argued in the main text, the 2 PN relation can also be derived from the effective potential and the fact that $J^2$ and $L^2$ are constants at 2 PN. This is illustrated in Fig.~\ref{fig:eff_potential} where we plot the contours between $\chieff$ and various $\cos\theta$ (see also Ref.~\cite{Kesden:15} on how to generate such contours). As Eq.~(\ref{eq:c12_vs_c1_2PN}) or (\ref{eq:c12_vs_c1_2PN_appx}) eliminates $S^2$, the only variable at 2 PN, it corresponds to a line in the effective potential description. Thus, once we fix the value of $\chieff$, Eq.~(\ref{eq:c12_vs_c1_2PN}) has to a fixed value (in contrast to $\cos\theta_{S_1L}$ or $\cos\theta_{S_1S_2}$ which can oscillates between the two intercepts formed by its contour and a given value of $\chieff$). 

To incorporate dynamics at higher PN orders, it is interesting to first examine the special case where $q=1$. From Eq.~(\ref{eq:c12_vs_c1_d_dt}) it is easy to see
\begin{align}
    \cos\theta_{S_1S_2} \simeq {\rm Const.}\ \ ({\rm when}\ q=1). 
    \label{eq:c12_vs_c1_q1}
\end{align}
The above equation holds at 2.5 PN order.

To obtain the more general 2.5 PN relation when $q\neq1$, it is easiest achieved by multiplying both sides of Eq.~(\ref{eq:c12_vs_c1_d_dt}) by $L$ and use $Ld\cos\theta_{S_1L}/dt = d(L \cos\theta_{S_1L})/dt - \cos\theta_{S_1L} dL/dt$. If we further define
\begin{align}
    % (1-q)L\mathcal{C} \equiv &\left[(1-q)L + \frac{q^2M_{\rm t}^2\chieff}{(1+q)^2}\right]\cos\theta_{S_1L}\nonumber \\ -&\frac{qS_1}{2}\cos^2\theta_{S_1L} +S_2\cos\theta_{S_1S_2},
    (1-q)L\mathcal{C} \equiv (1-q)L\cos\theta_{S_1L} + S_2\cos\theta_{S_1S_2},
\end{align}
we have
\begin{align}
(1-q)&\frac{d}{dt}\left(L\mathcal{C}\right) = (1-q)\cos\theta_{S_1L} \frac{dL}{dt}. \label{eq:c12_vs_c1_L_d_dt}
\end{align}
Consequently,
\begin{align}
    (1-q)\left[L\mathcal{C} - \left(L\mathcal{C}\right)^{(0)}\right] = (1-q) \int  \cos\theta_{S_1L} \frac{dL}{dt} dt. \label{eq:c12_vs_c1_appendix_ver}
\end{align}
For $q\neq 1$, we can drop the $(1-q)$ factor from both side.  We can further approximate $C^{(0)}\simeq \cos\theta_{S_1L}^{(0)}$, this leads to 
\begin{equation}
    L\mathcal{C} - \int\cos\theta_{S_1L}dL\simeq L^{(0)}\mathcal{C}^{(0)}\simeq L^{(0)}\cos\theta_{S_1L}^{(0)}. 
    \label{eq:LC_m_int_c1dL}
\end{equation}
Note that while in the second approximation we have compromised some accuracy, it nonetheless renders the right-hand side as a well-defined constant because as $L\to \infty$, $\cos\theta_{S_1L}\to{\rm constant}$ to a good approximation. 

In fact, $\cos\theta_{S_1L}$ remains a constant until 2 PN, and even when it starts to vary significantly, it oscillates around its initial value $\cos\theta_{S_1L}^{(0)}$ (see, e.g., Fig.~\ref{fig:bin_evol_sample}). Therefore we can approximate the integral as 
\begin{align}
    \int \cos\theta_{S_1L} \frac{dL}{dt} dt \simeq \cos\theta_{S_1L}^{(0)} \left[L-L^{(0)}\right].
    \label{eq:int_c1dL_approx}
\end{align}
Together with the approximation $\mathcal{C}^{(0)}\simeq \cos\theta_{S_1L}^{(0)}$, we now have (for $q\neq1$)
\begin{align}
    \mathcal{C} =& \cos\theta_{S_1L} +\frac{S_2}{(1-q)L}\cos\theta_{S_1S_2} \nonumber \\
    \simeq & \cos\theta_{S_1L}^{(0)}.
\end{align}
This means the quantity $\mathcal{C}$ can also be approximated as a constant. 

We can also write $\mathcal{C}$ in terms of $(J, L)$ as 
\begin{align}
    % &\mathcal{C} = -\frac{1}{8(1-q^2)^3 S_1 L^3}\nonumber \\
    % &\times\left\{q(1+q)\left[(1+q)(J^2-L^2-S^2)-2qLM_{\rm t}^2\chieff \right]^2 \right. \nonumber \\
    % &\ \ -4L(1-q)\left[(1-q^2)(1+q)L+q^2M_{\rm t}^2\chieff\right] \nonumber \\
    % &\left. \quad\quad \times \left[(1+q)(J^2-L^2-S^2)-2qLM_{\rm t}^2\chieff\right] \right.\nonumber \\
    % &\ \ \left.-4L^2(1-q^2)^2(1+q)\left(S^2-S_1^2-S_2^2\right)  \right\}
    &\mathcal{C} = \frac{J^2-L^2-S_1^2-S_2^2 - \left[q/(1+q)\right]\chieff M_{\rm t}^2L}{2(1-q)S_1 L}
\end{align}

If one uses the full expression of $\mathcal{C}^{(0)}$, we then have 
\begin{align}
    \mathcal{D} &\equiv \Delta_{c_1}+\frac{S_2}{(1-q)L} \Delta_{c_{12}} \nonumber \\
    &= \frac{\int \cos\theta_{S_1L} dL}{L} - \cos\theta_{S_1L}^{(0)} \left[1-\frac{L^{(0)}}{L}\right] \simeq 0,
    \label{eq:D_def}
\end{align}
where 
\begin{align*}
    &\Delta_{c_1} {=} \left[\cos\theta_{S_1L} - \cos\theta_{S_1L}^{(0)}\right], \\
    % &\Delta_{c_1^2} {=} \left[\cos^2\theta_{S_1L} - \cos^2\theta_{S_1L}^{(0)}\right], \\
    &\Delta_{c_{12}} {=} \left[\cos\theta_{S_1S_2} - \cos\theta_{S_1S_2}^{(0)}\right]\simeq \cos\theta_{S_1S_2}.
\end{align*}
Note that in practice $\cos\theta_{S_1S_2}^{(0)}$ is not a well-defined quantity at large orbital separations where spins precess faster than the orbit decays. This introduces a fundamental uncertainty of $S_2/(1-q)L$ in the value of $\mathcal{D}$. 

\begin{figure}[tb]
  \centering
  \includegraphics[width=\columnwidth]{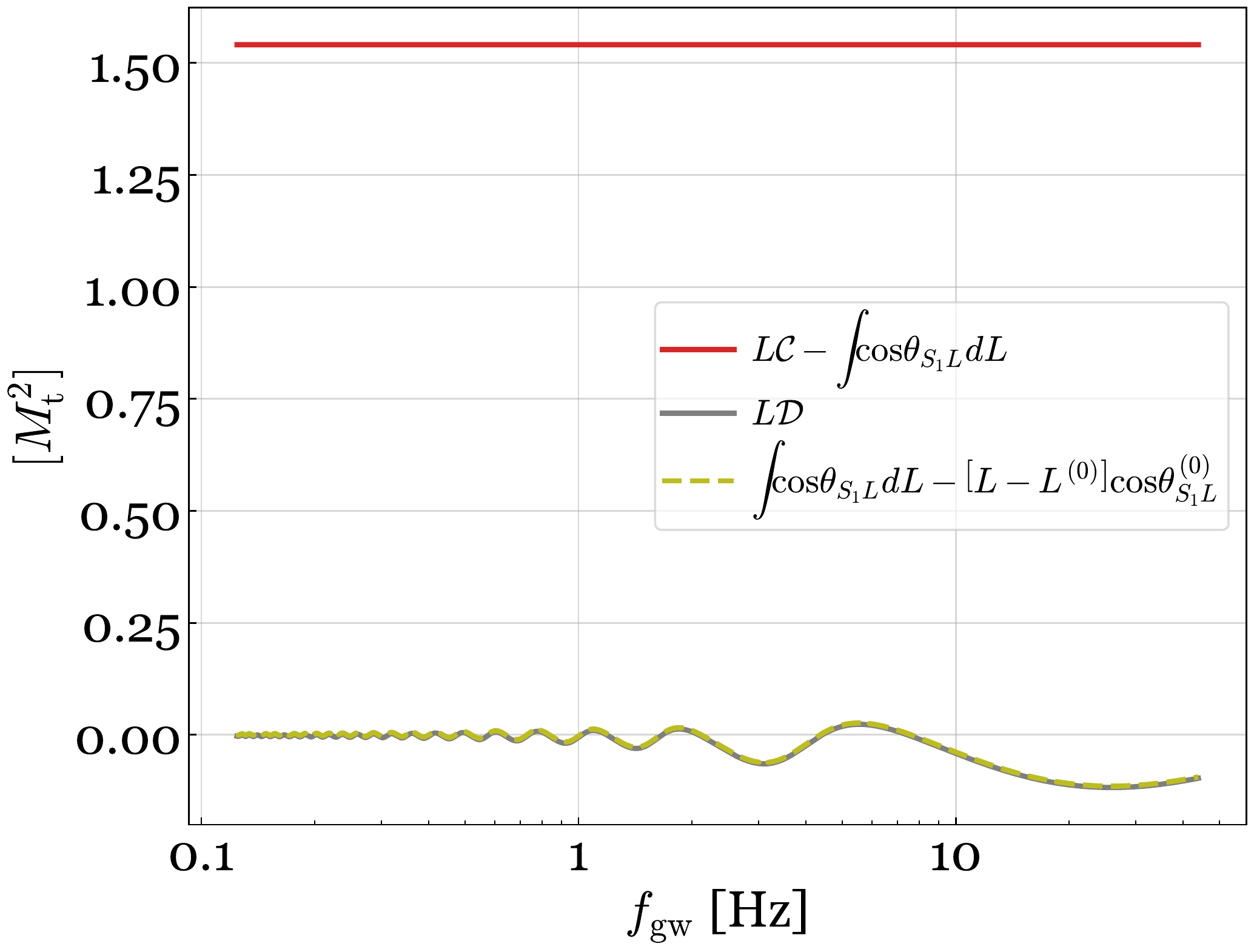}
\caption{Various quantities to show the spin dynamics at 2.5 PN. The system is the same as the one shown in Fig.~\ref{fig:bin_evol_sample}.}
\label{fig:bin_cons_sample}
\end{figure}

In Fig.~\ref{fig:bin_cons_sample}, we verify various relations we derived in this Section numerically. Specifically, the red trace corresponds to the left-hand-side of Eq.~(\ref{eq:LC_m_int_c1dL}). In the purple trace, we replace the $L\mathcal{C}$ term by its 2 PN counterpart $L(\cos\theta_{S_{1}L}+\mathcal{K}\cos\theta_{S_1S_2})$ [see Eq.~(\ref{eq:c12_vs_c1_2PN})] but still remove the secular variation piece $\int \cos\theta_{S_1L}dL$. As expected, the purple trace shows more oscillations than the red one. The dashed-olive trace is the difference between the left- and right-hand sides of Eq.~(\ref{eq:int_c1dL_approx}), whose difference should equal to $L\mathcal{D}$ (grey trace) according to Eq.~(\ref{eq:D_def}).  There is a constant offset between them because we have intentionally set $\cos\theta_{S_1S_2}^{(0)}$ to 0 when evaluating $\mathcal{D}$. Lastly, the pink-dotted trace corresponds to the last term introduced in Eq.~(\ref{eq:c12_vs_c1_L_d_dt}), which is needed to cancel the Lense-Thirring spin-spin coupling's back-reaction on the orbit. As can be seen from the plot, it is indeed a small quantity oscillating around 0 and can thus be ignored. 
 
% If $q\neq1$, we than have
% \begin{align}
%     &\left[1+\frac{q^2M_{\rm t}^2\chieff }{(1+q)^2(1-q)L} -\frac{qS_1\cos\theta_{S_1L}}{2(1-q)L}\right]\cos\theta_{S_1L} \nonumber \\
%     &+ \frac{S_2}{(1-q)L}\cos\theta_{S_1S_2} - \left[1-\frac{L^{(0)}}{L}\right]\cos\theta_{S_1L}^{(0)} \simeq \cos\theta_{S_1L}^{(0)}
% \end{align}

\clearpage
\bibliography{ref}
\end{document}